\documentclass[preprint, aip, pof, superscriptaddress]{revtex4-1}

\usepackage{times}
\usepackage{amssymb,amsmath,amsthm}
\usepackage{mathrsfs}
\usepackage{subfigure}
\usepackage{cancel}

	\usepackage[usenames,dvipsnames]{color}
	\usepackage[pdftex]{graphicx}
	\usepackage{epstopdf}

\newcommand{\ms}[1]{\mathscr{#1}}

\newcommand{\pt}{\partial_t}
\newcommand{\pth}{\partial_{\theta}}

\newcommand{\mylab}[1]{\label{#1}}

\usepackage[normalem]{ulem}

\newcommand{\bfuwe}{}

\begin{document}
\title{Bifurcation analysis of the behavior of partially wetting liquids on a rotating 
cylinder}

\author{Te-Sheng Lin}
\email{tslin@math.nctu.edu.tw}
\affiliation{Department of Applied Mathematics, National Chiao Tung University, 
1001 Ta Hsueh Road, Hsinchu 300, Taiwan}
\author{Steven Rogers}
\affiliation{Department of Mathematical Sciences, Loughborough University, 
Loughborough, Leicestershire LE11 3TU, UK}
\author{Dmitri Tseluiko}
\email{d.tseluiko@lboro.ac.uk}
\affiliation{Department of Mathematical Sciences, Loughborough University, 
Loughborough, Leicestershire LE11 3TU, UK}
\author{Uwe Thiele}
\email{u.thiele@uni-muenster.de}
\affiliation{Institut f{\"u}r Theoretische Physik, Westf{\"a}lische 
Wilhelms-Universit{\"a}t M{\"u}nster, Wilhelm Klemm Strasse 9, 
D-48149 M{\"u}nster, Germany}
\affiliation{Center of Nonlinear Science (CeNoS), Westf{\"a}lische Wilhelms Universit\"at M\"unster, Corrensstr.\ 2, 48149 M\"unster, Germany}

\date{\today}

\begin{abstract}
We discuss the behavior of partially wetting liquids on a rotating cylinder using 
\bfuwe{a model} that takes into account the effects of gravity, viscosity, rotation, 
surface tension and wettability. Such a system can be considered as a prototype 
for many other systems where the interplay of spatial heterogeneity and a lateral 
driving force in the proximity of a first- or second-order phase transition results in 
intricate behavior. So does a partially wetting drop on a rotating cylinder undergo 
a depinning transition as the rotation speed is increased, whereas for ideally wetting 
liquids the behavior \bfuwe{only changes quantitatively. We analyze the bifurcations 
that occur when the rotation speed is increased for several values of the equilibrium 
contact angle of the partially wetting liquids.  This allows us to discuss how the entire 
bifurcation structure and the flow behavior it encodes changes with changing wettability. 
We employ various numerical continuation techniques that allow us to track 
stable/unstable steady and time-periodic film and drop thickness profiles. We support 
our findings by time-dependent numerical simulations and asymptotic analyses of steady 
and time-periodic profiles for large rotation numbers.}
\end{abstract}

\maketitle

\section{Introduction} \mylab{sec:intro}

Moffatt~\cite{Moffatt1977} first studied film flow on a rotating cylinder to answer the 
question, ``How much honey can be kept on a breakfast knife, while rotating it 
about its long axis?" and similar closely related questions regarding a number of 
industrial coating and printing processes where liquid films on rotating cylinders 
play an important role. These include paper production, paint-application, 
spit-roasting, molten glass technology and even chocolate manufacturing (see the 
review by Ruschak~\cite{Ruschak1999}). The ability to control technological processes in 
these and other applications depends on our understanding of the limits that have to 
be imposed on the amount of liquid or on the rotation speed to ensure that the liquid 
remains on the cylinder.

Furthermore, Moffatt provides a long-wave model~\cite{OrDB1997rmp} in the 
overdamped limit (neglecting inertia) that incorporates gravity and viscosity effects 
but neglects surface tension~\cite{Moffatt1977}. At about the same time, 
Pukhnachov developed a model that additionally includes surface tension 
effects~\cite{Pukhnachov1977}. Various further extensions were derived that 
incorporate higher-order terms related to gravity, inertial and centrifugal effects. For 
a more detailed account see Noakes et al.~\cite{NKR06} and Kelmanson~\cite{Kelm2009jfm}. 
Note that all these models implicitly assume that the liquid ideally wets the cylinder, i.e., they do 
not incorporate terms that account for wettability.

The equation introduced by Pukhnachov~\cite{Pukhnachov1977} and later
studied, e.g., in Hinch et al.~\cite{HiKe2003prslsapes} and Karabut~\cite{Kara2007jamtp}, is amended by
Thiele~\cite{Thiele2011a} who introduces a Derjaguin (or disjoining) 
pressure term~\cite{OrDB1997rmp,StVe2009jpm}. This allows for a study 
of the influence of wettability on the drop and film behavior on a 
rotating cylinder -- an influence that becomes particularly important 
for small cylinders and thin films / small droplets.  Furthermore, 
Ref.~\onlinecite{Thiele2011a} introduces an alternative scaling that 
allows for a straightforward study of the influence of the rotation 
speed of the cylinder that represents the most important experimental 
control parameter. The governing long-wave evolution equation for the 
film thickness profile can then be presented in a form that highlights 
the analogy between the behavior of films and drops of liquid on a 
rotating horizontal cylinder and on \bfuwe{inclined substrates with 
regular one-dimensional wettability patterns, i.e., with periodic modulations 
of the equilibrium contact angle (and sometimes also the precursor film 
height)~\cite{ThKn2006prl,Herde2012,VFFP2013prl,SBAV2014pre}.}

A central result of Ref.~\onlinecite{Thiele2011a} is that the analogy holds, i.e., 
that film flow and drop motion on the exterior or interior surface of a rotating 
cylinder on one hand and on a heterogeneous substrate with lateral driving 
force on the other hand are rather similar. This implies that the bifurcation character 
of many of the occurring transitions between qualitatively different behaviors 
of films and drops is similar and corresponding results can be transferred between 
the systems. In particular, it is found that for partially wetting drops on a rotating 
cylinder there exists a counterpart \bfuwe{of the depinning transition and related 
dynamics described before for drops on heterogeneous 
substrates~\cite{Thiele2006,Herde2012,BKHT2011pre}. A depinning transition 
occurs when a driving force reaches a critical value where a steady structure 
(e.g., a drop) that is pinned by a heterogeneity (e.g., a less wettable patch or 
spatially varying gravity) starts to move. This similarity proposes that the rotating 
cylinder can serve as a model for a more general class of hydrodynamic (and other) 
depinning processes} -- a model system that naturally has periodic boundary 
conditions (BCs) facilitating its analysis. This is important, in particular, when 
comparing them to the open BCs in models for film deposition at moving contact 
lines where the transition from the deposition of a homogeneous film to the 
deposition of line patterns~\cite{FrAT2012sm,Kopf2012,DoGu2013el} can be 
interpreted as a depinning transition (for a detailed recent discussion of a number 
of such systems see end of Sec.~III of Thiele~\cite{Thie2014acis} and Sec.~4 of 
K{\"o}pf et al.~\cite{Kopf2014}).

Ref.~\onlinecite{Thiele2011a} mentions in passing that the sequence of \bfuwe{qualitative transitions} 
encountered for drops of partially wetting liquids when increasing the rotation speed 
is not present for ideally wetting liquids. There, the behavior \bfuwe{only changes quantitatively} 
with increasing the rotation speed: pendent drops are smeared out into a film in a smooth 
process, surface waves may appear but no depinning transition can occur. This poses 
the question how the entire rotation speed-related bifurcation behavior changes when 
increasing the wettability of the liquid, 
i.e., decreasing the equilibrium contact angle. \bfuwe{A particularly intriguing question is how 
the global bifurcation related to the depinning of partially wetting drops is eliminated when 
decreasing the contact angle to small values.}

The importance of the interplay of wettability and lateral driving brings the system 
as well into the context of dynamic wetting transitions that occur for drops sliding 
down an incline and for films drawn out of a 
bath~\cite{BeCP2001crassib,SZAF2008prl,ZiSE2009epjt,GTLT2014prl}. Depending 
on the particular wetting energy (or binding potential), the equilibrium wetting 
transition~\cite{BRBR2002pa,BEIM2009rmp} is a first- or second-order phase 
transition where above the critical value of temperature (or the critical value of
another control parameter) a 
homogeneous phase dominates (flat film) and below the critical value
the liquid volume separates into regions of large film height (drops) and small film
height (precursor film). At the dynamic wetting transition a thick film is drawn out of 
a finite contact angle meniscus when a lateral driving force passes a critical 
value~\cite{SZAF2008prl,GTLT2014prl}, in other words, the driving can shift the 
wetting transition.

The laid out spectrum of related effects indicates that the drop and film flow on a 
rotating cylinder is not only interesting by itself but should be seen as a model 
system not only for drops on heterogeneous substrates but, in general, for many 
more systems where (i) spatial heterogeneity and (ii) lateral driving force interact 
in the proximity of a first- or second-order phase transition in situations involving a 
conserved order parameter field. Therefore, a profound study of the rotating cylinder 
system can inform as well the investigation of many other systems. We will come 
back to this point later on.

In the present contribution, we study the transition behavior that occurs in the 
rotating cylinder model of Ref.~\onlinecite{Thiele2011a} with a special emphasis 
of the transition in the bifurcation behavior when moving from a partially wetting 
case to the completely wetting case. Besides the rotation speed of the cylinder, 
our main control parameter is the equilibrium contact angle. We begin with a 
completion of the investigation of steady film and drop profiles as their complete 
bifurcation picture is needed to understand the time-periodic depinned free-surface profiles (e.g., co-rotating droplets) and their emergence from the steady 
profiles. \bfuwe{To obtain these solutions, we employ different numerical continuation 
techniques \cite{Doedel1991,Dijkstra2014} (that allow us to track stable and unstable 
steady and time-periodic states) as well as direct numerical simulations.}

The manuscript is organized as follows: In Sec.~\ref{sec:model} we discuss the 
model, our numerical approaches and the physically relevant parameter range. 
The steady-state profiles are discussed in Sec.~\ref{sec:sss} and \bfuwe{an additional analysis of steady states without rotation is shown in Appendix~\ref{app:steady_drops}}. 
The complete picture including branches of time-periodic solutions is studied in 
Sec.~\ref{sec:complete}. The results of two accompanying asymptotic analyses 
are found in Appendix~\ref{app:asym}. In particular, Appendices~\ref{sec:B1} 
and \ref{sec:B2} analyze steady-state solutions without rotation and solutions in 
the limit of large rotation number, respectively. Finally, the concluding remarks are 
given in Sec.~\ref{sec:conc}.

\section{The model and computational methods} \mylab{sec:model}
\subsection{The governing equation}

\begin{figure}
\centering
\includegraphics[width=0.45\hsize]{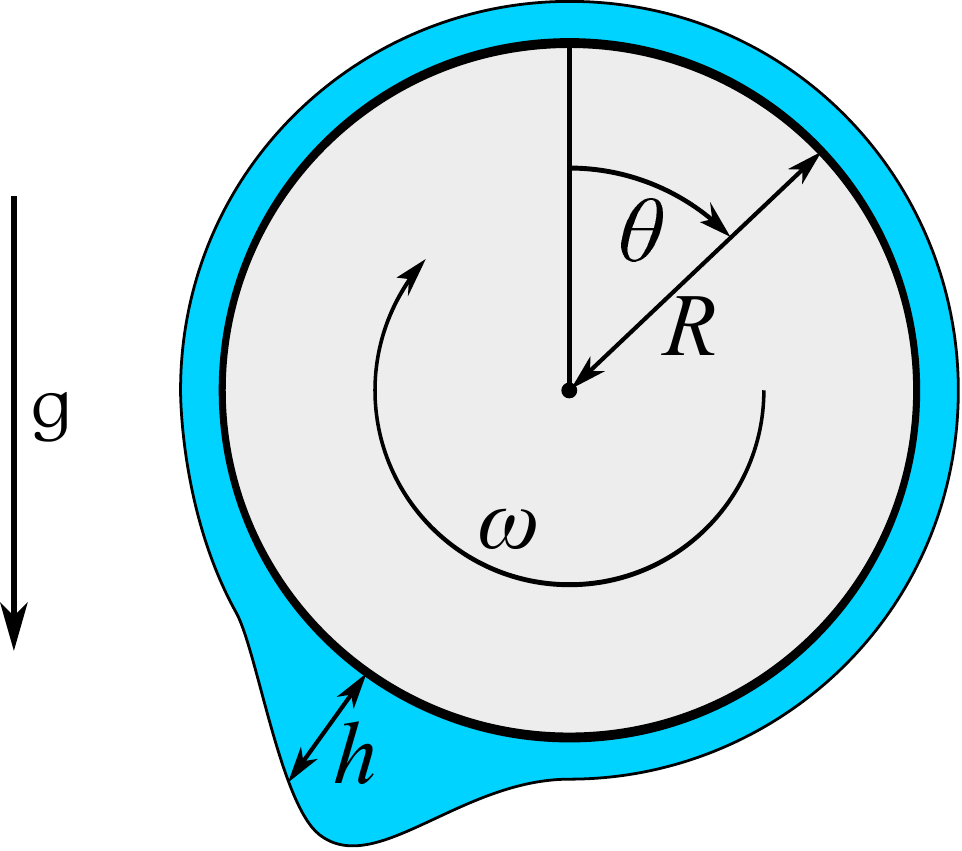}
\caption{ (Color online) \bfuwe{Sketch of a drop of a partially wetting liquid coexisting with a 
thin wetting layer on a rotating cylinder.} }
\label{Fig1}
\end{figure}

We consider a partially-wetting liquid of density $\rho$ and dynamic viscosity $\eta$ 
on a cylinder of radius $R$ that rotates about its axis at angular velocity $\omega$, 
as illustrated in Fig.~\ref{Fig1}. We assume that there are no variations in the 
direction along the cylinder axis, i.e., in practice, we consider a two-dimensional situation. Gravity $g$ acts in the vertical direction. The air-liquid surface tension coefficient is 
denoted by $\sigma$. We denote the liquid film thickness (measured along the radial 
direction) by $h(\theta,t)$, where $t$ is time and $\theta$ is the angle measured 
from the upper vertical position on the cylinder in a clockwise direction.

\renewcommand{\baselinestretch}{1}
\renewcommand{\baselinestretch}{1.5}

The evolution equation for $h(\theta,t)$ is derived employing a long-wave approximation 
that is valid in the limit $\epsilon\rightarrow 0$, where $\epsilon=\bar{h}/R$ is the 
ratio between the mean film thickness, $\bar{h}$, and the radius of the cylinder. Also, it 
is necessary to assume slow variations in the film thickness in the angular direction and 
small contact angles. With time scale $3\eta R^4/\sigma \bar{h}^3$, film thickness scale 
$\bar{h}$ and the angle $\theta$ measured in radians, the non-dimensionalized time-evolution equation is the partial differential equation (PDE)~\cite{Thiele2011a}
\begin{equation}
\pt h = - \pth\left\{h^3 \pth\left[\pth^2 h + h - B\cos\theta + \Pi(h)\right] 
+ \Omega h\right\},
\mylab{eq:timeevolution}
\end{equation}
where $B$ and $\Omega$ are the Bond and rotation numbers defined by
\bfuwe{
\begin{equation}
B=\frac{R^2\rho g}{\epsilon\sigma},\quad 
\Omega=\frac{\eta \omega R}{\epsilon^3\sigma},
\end{equation}
}
respectively. $\Pi(h)$ is the Derjaguin (disjoining) 
pressure~\cite{Genn1985rmp,StVe2009jpm,Israelachvili2011} that here combines 
long-range attraction and short-range repulsion in the form of power 
laws~\cite{Pism2001pre,PiTh2006pf}:
\begin{equation}
\Pi(h) = \frac{H}{h^3}\left(1-\frac{b}{h^3}\right),
\label{eq:djpress}
\end{equation}
where $H$ is a non-dimensional Hamaker constant. We obtain adequate values 
of $H$ and $b$ for this form of $\Pi(h)$ by relating them to the (non-dimensional) 
thickness of the equilibrium wetting layer $h_0$ and the static macroscopic equilibrium 
contact angle $\beta_0$:
\begin{equation}
b = h_0^3, \quad H = -\frac{5}{3}\,\beta_0^2\,h_0^2.
\end{equation}
Note, that $\beta_0$ is the angle in the long-wave scaling, i.e., the small physical 
equilibrium contact angle $\beta_{eq}=\epsilon\beta_0$ corresponds to the long-wave 
equilibrium contact angle $\beta_0$ of $O(1)$.

The system is $2\pi$-periodic in $\theta$, and the scaling fixes the non-dimensional 
mean film thickness to be $1$. Consequently, \bfuwe{to leading order in $\epsilon$, the volume of the film per unit length of the cylinder becomes $\int^{2\pi}_0 h(\theta, t)\,d\theta = 2\pi$.}

Note that the employed scaling introduced in Ref.~\onlinecite{Thiele2011a} allows 
one, in contrast to scalings used elsewhere in the literature, to directly relate 
individual dimensionless parameters to important physical parameters. Namely, 
the rotation number $\Omega$ is proportional to the main experimental control 
parameter, the angular velocity of the cylinder, that does not enter any other 
parameter or scale. The ratio of gravity and surface tension enters only the Bond 
number $B$ (and the overall time scale). The wettability properties control the 
dimensionless numbers contained in the Derjaguin pressure, Eq.~(\ref{eq:djpress}). 
Therefore, the scaling allows one to clearly identify transitions of behavior due to 
changes in the rotation speed or due to changes in the wetting behavior.

\subsection{Numerical methods}
\mylab{sec:numerical}

We use three different numerical approaches [(i)-(iii)] to analyze this system. 

(i) \bfuwe{With the first approach, we determine steady thickness
  profiles (steady-state solutions)} employing
continuation (or path following)
techniques \cite{Kuznetsov2010,Dijkstra2014}. By setting the time derivative in 
Eq.~(\ref{eq:timeevolution}) to zero, we obtain a fourth-order ordinary differential 
equation (ODE) for the steady film thickness profile $h(\theta)$:
\begin{equation}
- \pth\left\{h^3 \pth\left[\pth^2 h + h - B\cos\theta + \Pi(h)\right] 
+ \Omega h\right\}=0.
\mylab{eq:steadystate}
\end{equation}
We then integrate this equation once and use the substitutions $u_1 = h$, 
$u_2 = \pth h$, $u_3 = \pth^2 h$ and $u_4=\theta$ to transform the equation into 
an autonomous system of four first-order ODEs:
\begin{equation}
\left\{
\begin{array}{l}
\pth u_1 = u_2, \\
\pth u_2 = u_3, \\
\displaystyle\pth u_3 = -u_2 - B\sin(u_4) -\Pi'(u_1)\,u_2  
+ \frac{c - \Omega u_1}{u_1^3},\\
\pth u_4 = 1,
\end{array}
\right.
\mylab{eq:ODE1}
\end{equation}
where $c$ is the integration constant that represents the flux. Equations~(\ref{eq:ODE1}) 
with periodic BCs for $u_1$, $u_2$, $u_3$ and an initial condition for $u_4$ 
are solved numerically by continuation using the continuation and bifurcation software package 
Auto07p~\cite{Doedel1991,DoedelOldeman2009}. In the context of thin film equations, a 
similar approach is taken for sliding drops~\cite{TVNB2001pre,PiTh2006pf}, drawn 
films~\cite{TsGT2014epje}, self-similar solutions related to film rupture~\cite{TsBT2013ijam}, 
and for steady drops in heterogeneous systems under lateral driving~\cite{ThKn2006njp}. 
A review is given by Dijkstra et al.~\cite{Dijkstra2014} and selected tutorials can be obtained at Thiele et al.~\cite{cenosTutorial}.

(ii) Another way of looking at the problem is by treating PDE (\ref{eq:timeevolution}) 
as an infinite-dimensional system of ODEs. Then, steady states and time-periodic 
states of the PDE can be obtained as fixed points and periodic orbits of 
finite-dimensional systems of ODEs that approximate the infinite-dimensional one. 
\bfuwe{The time-periodic solutions correspond to film-thickness
  profiles of surface waves or, more importantly, of drops that rotate
  at their own speed with the cylinder (perform a stick-slip motion).}
Here, 
we take advantage of the periodicity and express the solution in the form of the 
Fourier series with time-dependent coefficients, i.e., 
$\displaystyle h(\theta, t) = \sum^{\infty}_{k=-\infty}\hat{h}_k(t)\,e^{ik\theta}$. 
Equation~(\ref{eq:timeevolution}) can be rewritten as a system of ODEs for the 
Fourier coefficients, $\hat{h}_k$, $k\in\mathbb{Z}$:
\begin{equation}
\frac{d\hat{h}_k}{dt} = F_k(\hat{h}_j), 
\mylab{eq:ODE2}
\end{equation}
where $F_k$ is a nonlinear function of all the Fourier modes
$\hat{h}_j$, $j\in\mathbb{Z}$. In practice, we truncate the number of
the Fourier modes by restricting $-N\leq k\leq N$ and solve the
finite-dimensional system, where $N$ is a large enough
integer. Alternatively, finite differences may be used for the spatial
discretization, for instance, for settings without periodic BCs~\cite{Kopf2014}. The parameter space of the obtained
finite-dimensional ODEs may then be explored with standard
  continuation techniques \cite{Kuznetsov2010,Dijkstra2014}. Here we
  employ Auto07p~\cite{Doedel1991,DoedelOldeman2009}. 
\bfuwe{The time-periodic solutions are obtained by first detecting
  Hopf-bifurcation points on the branches of steady drop profiles and
  then switching at these points to branches of time-periodic
  solutions that bifurcate at the Hopf bifurcations. We note that the
  time period is found as part of the solution, and we do not impose
  any relation between this period and the period of cylinder rotation. Earlier, the technique based on obtaining a dynamical system for the Fourier coefficients} was applied to the continuation of spiral waves in
  reaction-diffusion systems \cite{BoEn2007pd}.  The continuation of
  Fourier amplitudes can also be employed to follow steady and
  time-periodic solutions of certain integro-differential equations as
  recently done in the context of dynamical density functional theory
  (DDFT) for coarsening \cite{PoTA2014pre} and transport processes
  \cite{PAST2011pre} of interacting particles in nanopores. The
  Acknowledgement sketches the recent history of the application of
  the technique.

(iii) The third approach is to directly integrate the PDE, Eq.~(\ref{eq:timeevolution}), 
in time to obtain time-dependent solutions. We implement a \bfuwe{Fourier pseudo-spectral 
method where spatial derivatives are computed spectrally and time differentiation is performed using a second-order implicit discretization.} The resulting nonlinear 
system is solved by Newton's method at each time step until convergence is 
achieved. In this way, we obtain the time evolution for given initial conditions and 
approach steady and time-periodic states (or other more complicated behavior, 
such as solutions that are quasi-periodic in time). 
Note however, that only stable solutions can persist for long time and can be 
captured by the time-integration method. \bfuwe{Unstable solutions are
important transient states that will die out in finite 
time. They can only be captured by continuation techniques that are therefore 
indispensable if a full understanding of the drop and film hydrodynamics shall be
  achieved that is encoded in the bifurcation structure of the problem.}

We characterize steady-state and time-periodic thickness profiles by the $L^2$-norm 
and the time-averaged $L^2$-norm of \bfuwe{the difference between the thickness profile and the flat film} solution:
\begin{equation}
\|\delta h(\theta)\| \equiv \sqrt{\frac{1}{2\pi} \int^{2\pi}_0 (h(\theta)-1)^2 d\theta}
\mylab{eq:normsteady}
\end{equation}
and
\begin{equation}
\|\delta h(\theta, t)\| \equiv \sqrt{\frac{1}{2\pi T}\int^T_0\int^{2\pi}_0 (h(\theta, t)-1)^2 
d\theta dt}, 
\mylab{eq:normtime}
\end{equation}
respectively. Here, $T$ denotes the time period. 

\bfuwe{We note that, to ensure accuracy of the results presented in our study, for each of the calculations, we double the number of Fourier modes until convergence is achieved and the results do not change on the scale of the corresponding figure. Besides, we also implement a zero-padding technique to remove aliasing errors. In general, we find that taking $N=64$ is sufficient to ensure accuracy of the computations.
}

\subsection{Parameters}
\mylab{sec:parameters}

We give a brief overview of the parameters used in Ref.~\onlinecite{Thiele2011a}, 
\bfuwe{since here we investigate how wettability influences the
  basic depinning behavior discussed there}. Estimates for realistic Bond and 
rotation numbers are based on two {liquids} typically studied in the literature: (i) 
water at $25^{\circ}$C with $\sigma=0.072$N/m, $\eta=0.001$kg/ms and 
$\rho=1000$kg/m$^3$ as in Reisfeld et al.~\cite{Reisfeld1992} and a silicone oil 
with $\sigma=0.021$N/m, $\eta=1$kg/ms and $\rho=1200$kg/m$^3$ as in Evans et al.~\cite{Evans2004}. 
As experimentally feasible configurations, we assume 
angular rotation velocities $\omega=0.1$--$10$\,s$^{-1}$, cylinders of radii 
$R=10^{-3}$--$10^{-2}$ m and smallness ratios $\varepsilon=0.01$--$0.1$. 

For $\omega=1$\,s$^{-1}$, the smallness ratio $\varepsilon=0.1$ and the cylinder 
radius of $R=10^{-3}$m, we find for case (i) $B=1.36$ and $\Omega=0.014$, 
whereas for case (ii) $B=5.61$ and $\Omega=47.6$. Increasing the cylinder 
radius to $R=10^{-2}$ m, increases all the Bond numbers by a factor $10^2$ and 
all the rotation numbers by a factor $10$, whereas, decreasing the smallness ratio 
to $\varepsilon=10^{-2}$, increases all the Bond numbers by a factor $10$ and all 
the rotation numbers by a factor $10^3$. Therefore, focusing on cylinder diameters 
in the millimeter range, we mainly investigate $B\le10$, although the structure 
of steady-state solutions for higher Bond numbers is discussed. The rotation 
number can be widely varied, although we find that for $B=O(1)$, the interesting 
range for the rotation number is $\Omega=O(1)$. For instance, with $B=1$, 
one finds that depinning occurs at $\Omega\approx1.68$. 

\subsection{The aims of the present study}
\label{sec:aims}

The preliminary work in Ref.~\onlinecite{Thiele2011a} introduces the
model incorporating wettability and studies a single partially wetting
case at \bfuwe{the equilibrium contact angle $\beta_0=2$ that is rather large in the long-wave scaling}. In particular, bifurcation
diagrams are discussed for steady profiles as functions of the Bond
number and for steady pinned states and time-periodic depinned drops
as functions of the rotation number.  One particular depinning
transition -- the Saddle-Node-Infinite-PERiod (or SNIPER)
bifurcation~\cite{Strogatz1994} -- is introduced. This is contrasted
with some basic results for the completely wetting case
($\beta_0=0$). 
\bfuwe{As the focus of Ref.~\onlinecite{Thiele2011a} 
was on the main depinning transition, they only employed continuation to 
obtain branches of steady drops and film solutions directly connected to the stable 
pendent drop solution at low rotation speed or to the film solution at large rotation 
speed. In the case without rotation, the focus was on single-drop solutions. In 
consequence, the results of Ref.~\onlinecite{Thiele2011a} represent only a first 
glimpse at the system and turn out to be rather incomplete. We show here that 
the complete bifurcation diagrams are much richer even in the case without 
rotation, not to speak of the case with driving where we now discuss several 
branches of time-periodic solutions.}

\bfuwe{The present work investigates both steady and time-periodic states 
in more detail. First, this shall result in a deeper understanding of the
case with $\beta_0=2$ studied in Ref.~\onlinecite{Thiele2011a}.
The second and more important part of the present study analyzes how the
bifurcation behavior (the depinning transition) changes as the
wettability is changed. In other words, we investigate how the
complex bifurcation structure found at large contact angles emerges from the
rather trivial behavior at zero contact angle. This is done through
the consideration of several intermediate cases with  $0<\beta_0<2$.}

\bfuwe{The present detailed study of the depinning transition and 
its transformations shall also establish the present system, the hydrodynamic system 
of a rotating cylinder covered by a partially wetting liquid, as a reference system for 
depinning transitions in other hydrodynamic systems, more general soft matter systems 
and beyond. 
Such systems are characterized by a spatial heterogeneity 
(here, the inhomogeneity in the action of gravity along the cylinder circumference), 
a lateral driving (here, the rotation of the cylinder) and a cohesive force resulting 
in coherent structures (here, partial wettability resulting in coherent drops). 
A weakening of the cohesive force (here, increasing wettability, i.e., approaching the wetting transition) will change the depinning transition in all such 
systems and the present study shall reveal possible corresponding pathways in the system behavior.}

Throughout this study, we focus on positive values of the rotation
number and the Bond number, but we may also exploit the fact that all
the obtained bifurcation diagrams are symmetric about the respective
zero values of these two parameters, since a reflection in $\Omega$
corresponds to turning the cylinder in the anti-clockwise direction
and a reflection in $B$ stands for reversing the direction of
gravity. Therefore, it is perfectly acceptable to reflect solution
branches at the axes of the Bond and rotation numbers, providing a
useful technique for reducing the number of computations needed to
obtain a full set of solution branches through path following. We also
fix the thickness of the wetting layer to be $h_0=0.1$ when exploring
the other parameters.

\section{Steady-state film and drop profiles}
\mylab{sec:sss}

\subsection{Steady states in the partially wetting case}
\mylab{Partial}

\renewcommand{\baselinestretch}{1}
\begin{figure}[t]
\begin{center}
{\includegraphics[width=0.49\hsize]{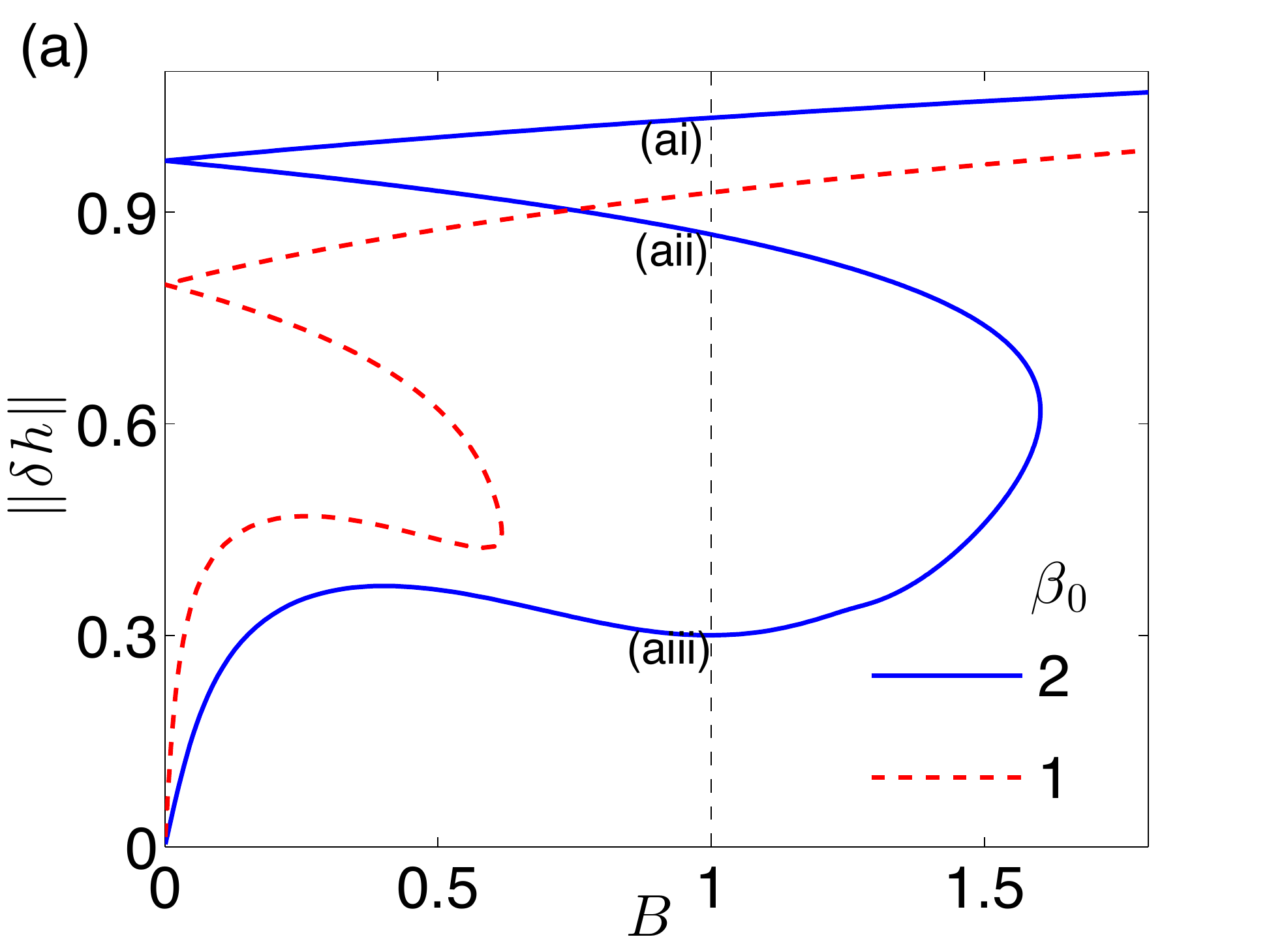}}
{\includegraphics[width=0.49\hsize]{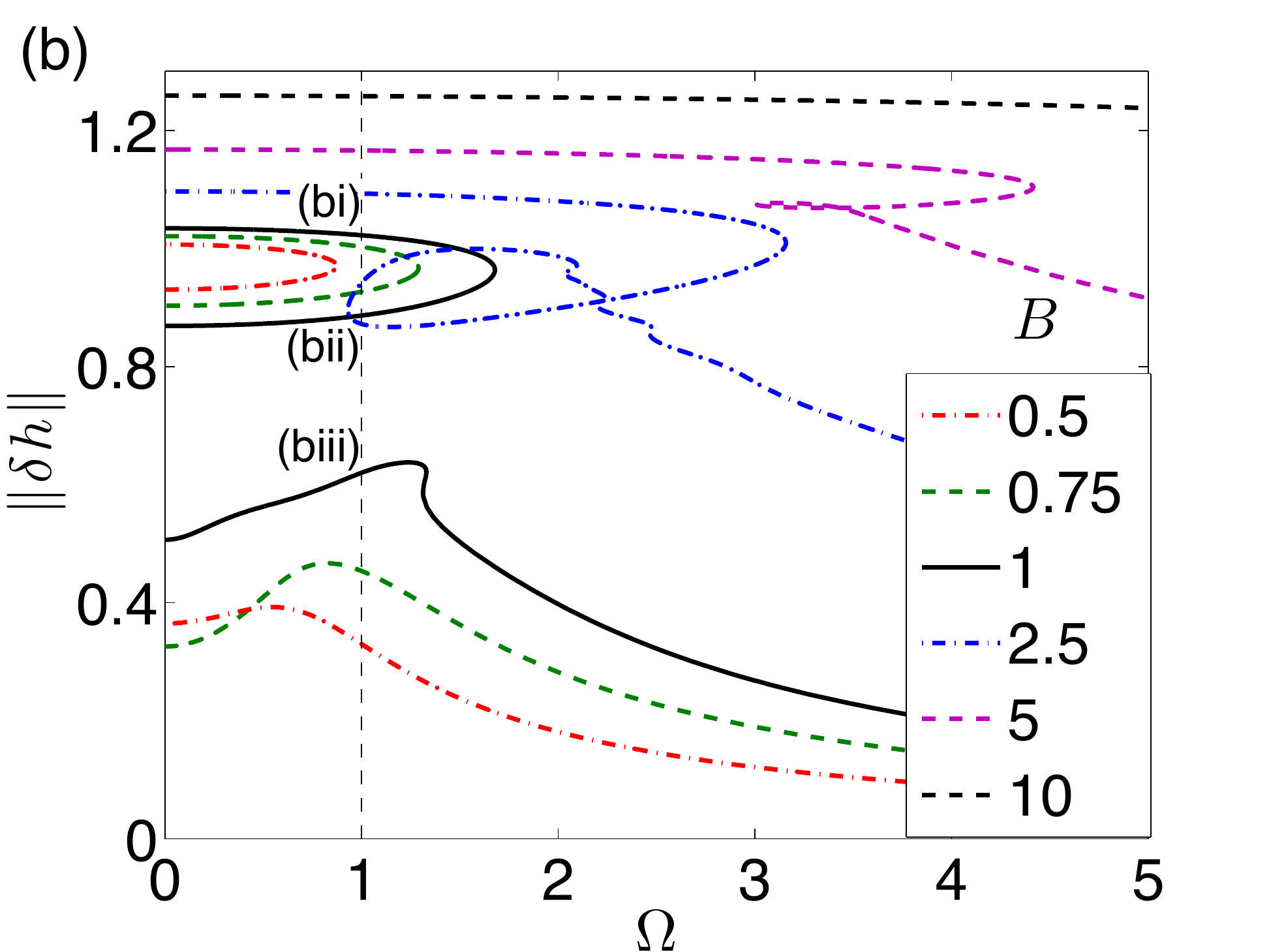}}
{\includegraphics[width=0.49\hsize]{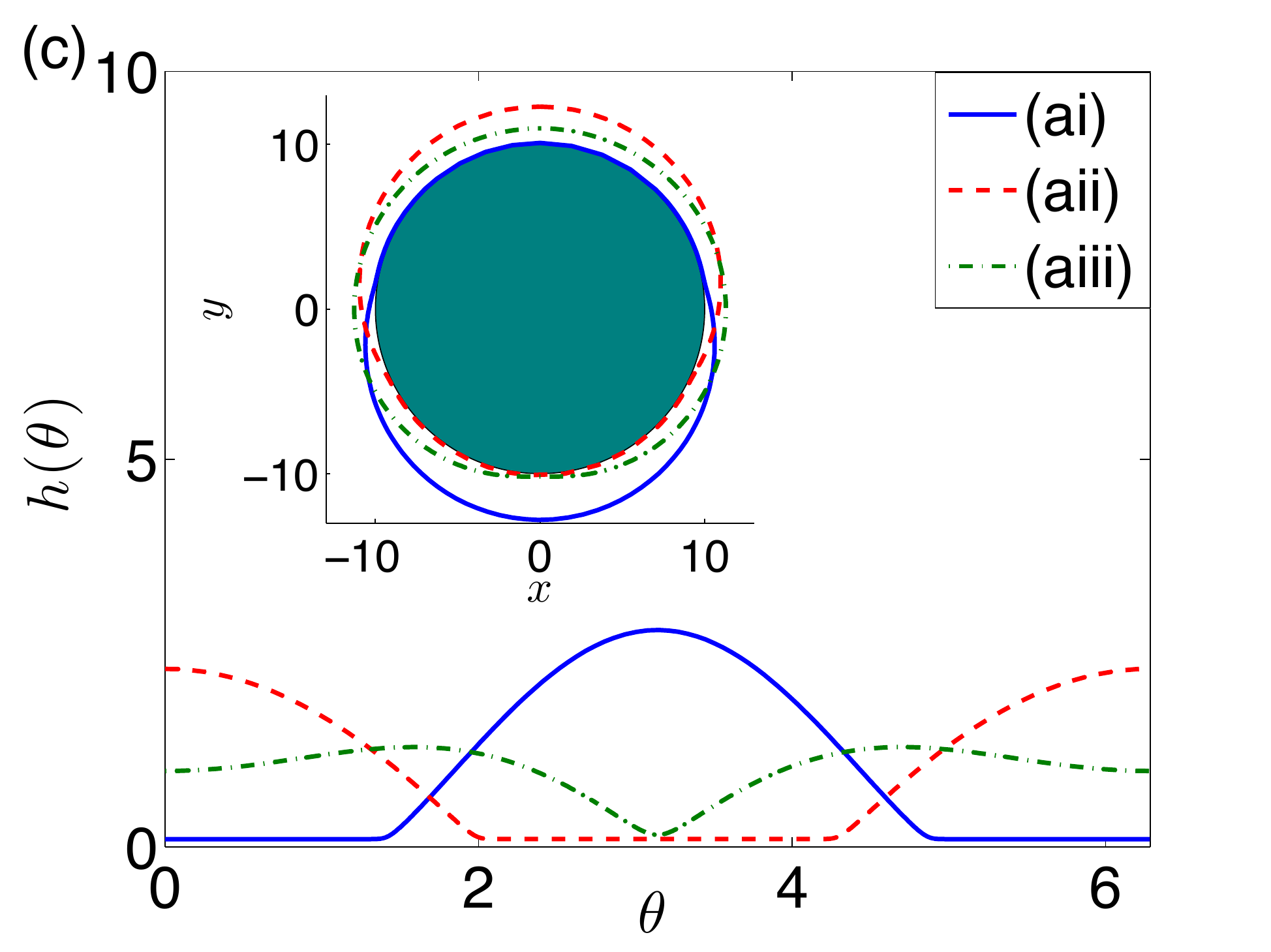}}
{\includegraphics[width=0.49\hsize]{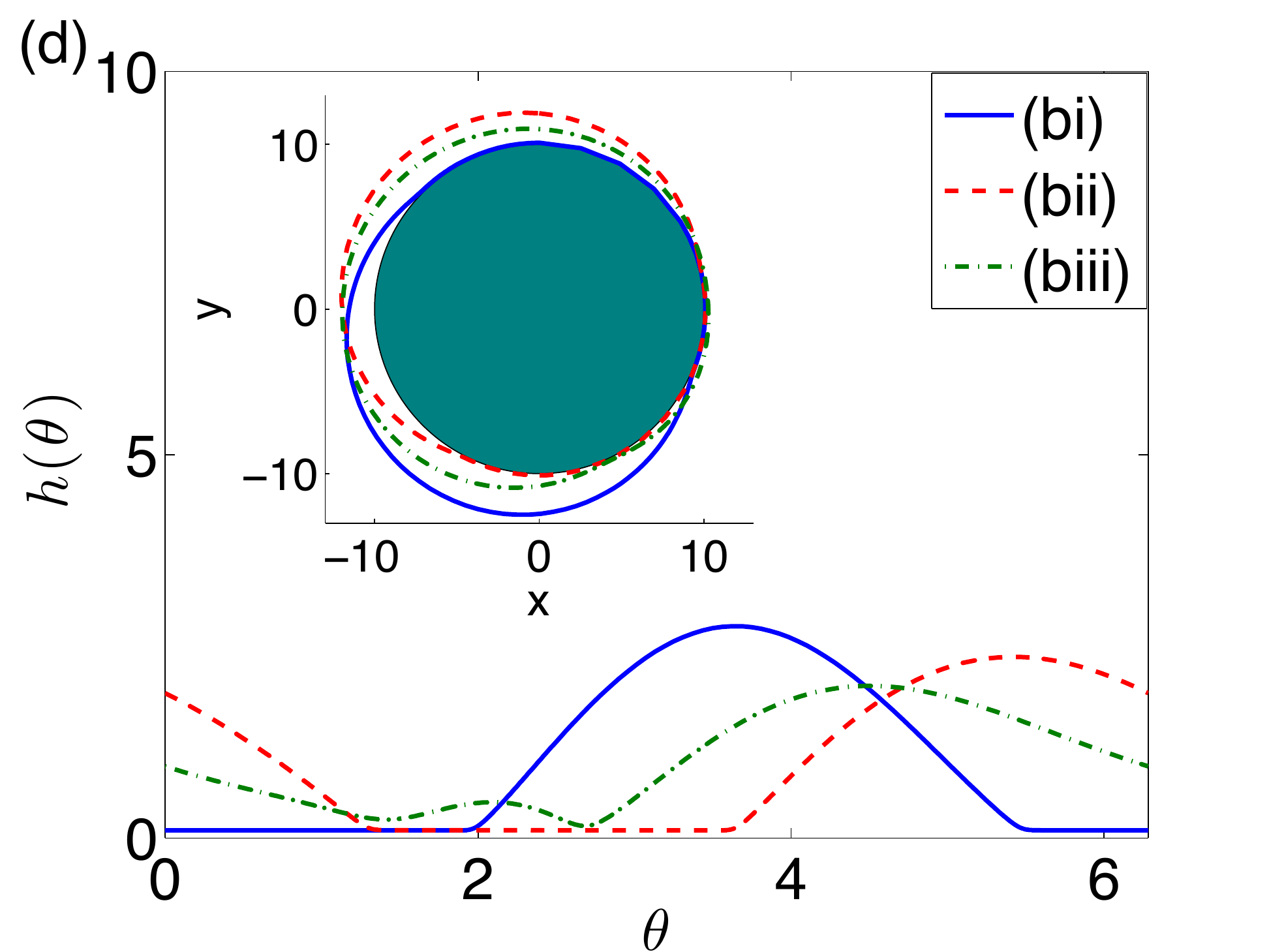}}
\vspace{-0.5cm}
\caption{(Color online) 
(a) Branches of steady drop solutions on a horizontal cylinder without
rotation ($\Omega=0$) for various equilibrium contact angles $\beta_0$
characterized by their $L^2$-norm as functions of the Bond
number. The dashed vertical line corresponds to $B=1$, and the
thickness profiles $h(\theta)$ corresponding to points (ai)--(aiii) at
$\beta_0=2$ \bfuwe{are shown in panel (c) with the inset giving the
  profiles on a cylinder assuming a radius $R=10$.}  (b) Branches of
steady drop solutions on a horizontal rotating cylinder for  $\beta_0
= 2$ as functions of the rotation number $\Omega$ for various Bond
numbers $B$ as given in the legend. \bfuwe{The dashed vertical line
  corresponds to $\Omega=1$, and the thickness profiles corresponding
  to points (bi)--(biii) at $B=1$ are shown in panel (d) with the
  inset giving the profiles on a cylinder assuming a radius $R=10$.
The curves in panels (a) and (b) agree with Figs.~2 and 4 of
Ref.~\onlinecite{Thiele2011a}, respectively. However, here they are
calculated with numerical methods (i) and (ii) (see
Sec.~\ref{sec:numerical}) that perfectly agree on the presented scale. This also serves as a validation of the novel method (ii).}
}
\vspace{-0.5cm}
\label{Fig2}
\end{center}
\end{figure}
\renewcommand{\baselinestretch}{1.5}

First, we consider steady drop and film solutions on the resting
cylinder, i.e., at $\Omega = 0$, for the two particular values
$\beta_0=1$ and $2$ of the equilibrium contact angle that correspond
to partial wetting. Changing the Bond number, $B$, we obtain families
of steady-state solutions as displayed in Fig.~\ref{Fig2}(a).  One
observes that the layout of the branches (the bifurcation structure)
is very similar in the two cases, only the size of the structure is
smaller for smaller $\beta_0$. Without gravity ($B=0$) there are only
two solutions: \bfuwe{an unstable flat film (the $L^2$-norm, defined
  by Eq.~(\ref{eq:normsteady}), is zero) and a stable drop solution}
(the $L^2$-norm is slightly smaller than one) that is invariant with
respect to translation in $\theta$ as for $B=0$ the system is
homogeneous. Once $B>0$, only two positions survive and two individual
branches emerge from the drop solution at $B=0$ that correspond to
unstable drops sitting symmetrically on top of the cylinder (smaller
$L^2$-norm) and to stable pendent drops hanging
symmetrically beneath the cylinder (larger $L^2$-norm),
  respectively. The unstable flat film also becomes modulated for
$B>0$. In consequence, for small $B$, three steady profiles exist
which correspond to different distributions of the liquid on the
cylinder. The three profiles on the dashed vertical line
corresponding to points (ai)--(aiii) in Fig.~\ref{Fig2}(a)
($\beta_0=2$ and $B= 1$) can be found in Fig.~3 of
Ref.~\onlinecite{Thiele2011a} \bfuwe{and are also shown in
  Fig.~\ref{Fig2}(c) with the inset giving the profiles on a cylinder
  assuming a radius $R=10$.} Above a critical $B=B_\mathrm{sn}$, only
the stable pendent drop remains while the other two solutions
annihilate at $B_\mathrm{sn}$ in a saddle-node (SN) bifurcation.
$B_\mathrm{sn}$ monotonically increases with the contact angle
$\beta_0$, e.g., it reaches the value $B\approx 43.3$ for
$\beta_0=10$. \bfuwe{We note that a pendent drop exists at any 
$B$ in the employed long-wave framework. However, at large $B$ the 
long-wave approximation itself becomes invalid. A full hydrodynamic 
description then shows drops that drip off the cylinder \cite{hazel}.}

Next, we discuss the changes in the observed steady profiles when the
cylinder is rotated. We increase the rotation number $\Omega$ from
zero and obtain solution branches as shown in Fig.~\ref{Fig2}(b) for
various Bond numbers and $\beta_0 = 2$. First, we consider the case
$B=1$ (solid black lines) and note that although one finds 3 solutions
for $\Omega=0$ as expected from the crossings of the dashed
line and the $\beta_0 = 2$ curves in Fig.~\ref{Fig2}(a), a close
inspection shows that the solutions of the smallest
$L^2$-norms do actually not correspond to each other. This
mismatch that was already pointed out in Ref.~\onlinecite{Thiele2011a}
indicates that the branch structures in Fig.~\ref{Fig2} are not
complete \bfuwe{(see the discussion in Sec.~\ref{sec:aims}).}  It will
be completed below after discussing Fig.~\ref{Fig2}(b). \bfuwe{The
  vertical dashed line in Fig.~\ref{Fig2}(b) corresponds to $\Omega=1$
  and the profiles (bi)--(biii) for $B=1$ are shown in
  Fig.~\ref{Fig2}(d).}

Increasing $\Omega$ from zero, the \bfuwe{stable pendent} drop (top branch in 
Fig.~\ref{Fig2}(b)) is shifted towards the left-hand side of the cylinder, i.e., 
towards larger $\theta$. Furthermore, the drop widens marginally, thereby decreasing 
its $L^2$-norm. This branch annihilates with the closest unstable branch (drops on top of the 
cylinder) in an SN bifurcation at $\Omega_\mathrm{sn}\approx1.68$ where a depinning 
transition occurs \cite{Thiele2011a}.  For $\Omega>\Omega_\mathrm{sn}$ the drop 
overcomes the downward pull of gravity, due to the frictional forces exerted by the 
rotating cylinder. As a result, the solution becomes time-periodic and the drops move 
around the cylinder with a position-dependent velocity. For space-time plots, see 
Fig.~7 of Ref.~\onlinecite{Thiele2011a}.

\noindent Close to the depinning transition, the difference 
in the time scales of slow and fast phases diverges and the drop motion strongly 
resembles stick-slip motion as also discussed in the context of drop motion on 
heterogeneous substrates~\cite{Thiele2006}. The square-root dependence of the 
frequency $1/T$ of the drop motion with time period $T$ on the driving 
$\Omega - \Omega_\mathrm{sn}$ indicates that at $\Omega_\mathrm{sn}$ one has 
a SNIPER bifurcation.

The mismatch observed above indicates that at $\Omega=0$ 
Figs.~\ref{Fig2}(a) and (b) miss at least a steady solution with an $L^2$-norm of $0.5$ 
and $0.3$, respectively. The starting point of our present analysis is a continuation of 
the steady profiles that are not consistent between the figures in the respective required
parameter. Note that this alone does not necessarily provide a
complete set of solutions since it is possible that further solutions
exist that are not connected in this way to the already known
solutions. Therefore, we also perform two-parameter continuations of
the loci of the observed SN bifurcations (fold continuation),
starting at several values of $B$ and continually check the
consistency between the figures. Once all the figures are consistent, all
the steady solution branches should be present in the studied range
of parameters. 

\begin{figure}
\begin{center}
{\includegraphics[width=0.49\hsize]{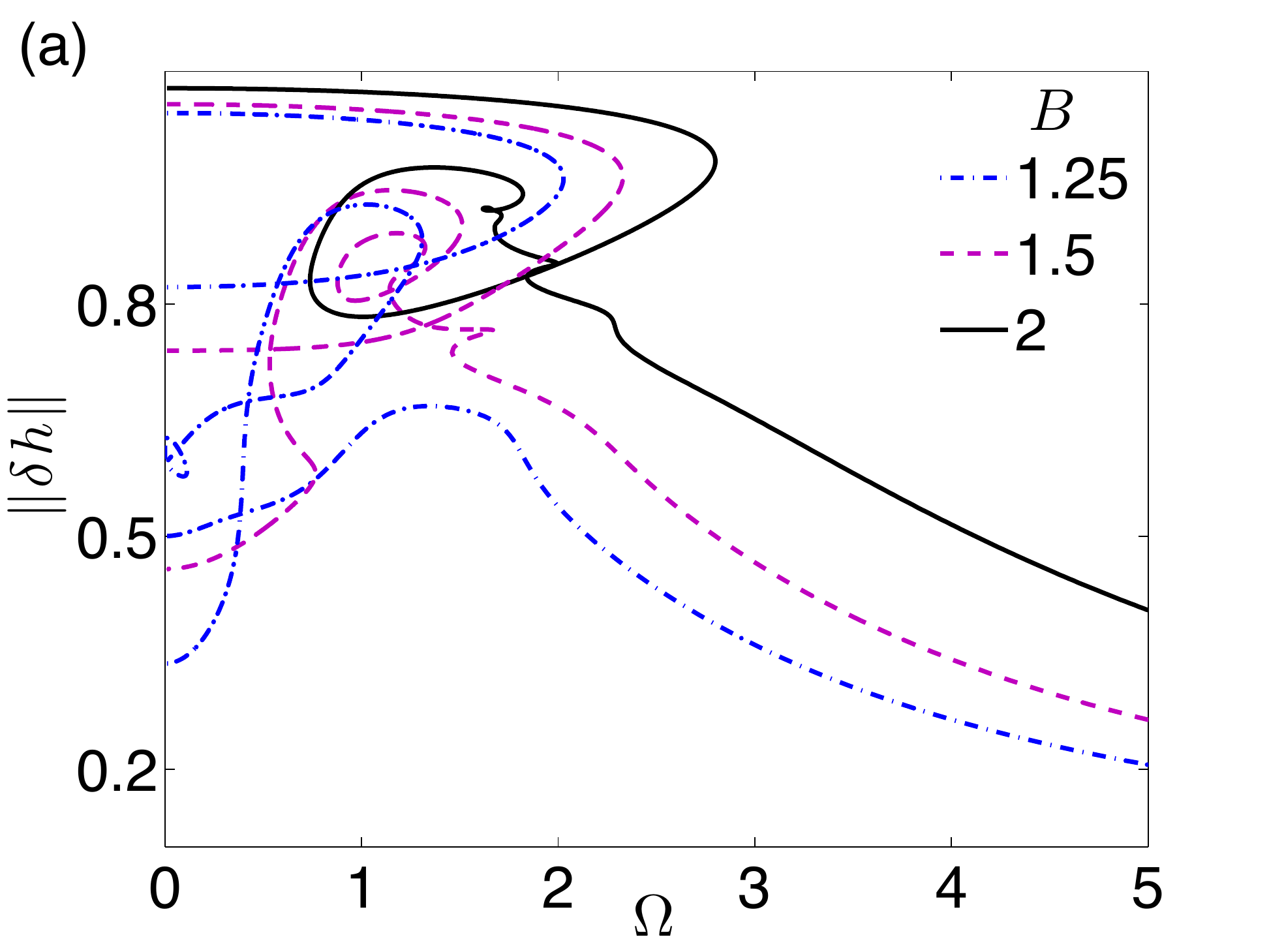}}
{\includegraphics[width=0.49\hsize]{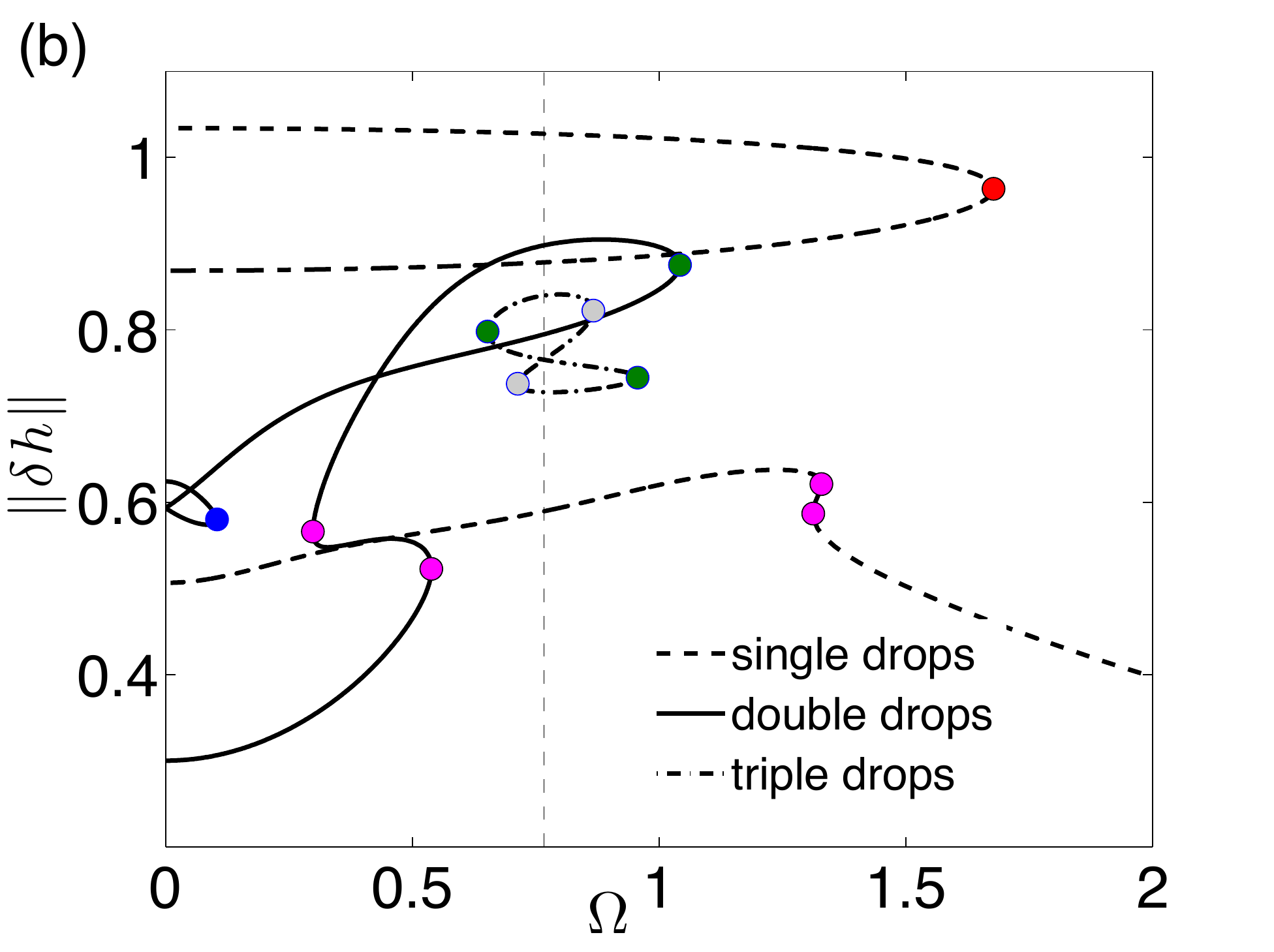}}
\caption{(Color online) 
(a) Extension of Fig.~\ref{Fig2}(b) for intermediate Bond numbers displaying 
the complex structure of the branches that involve several SN bifurcations. 
(b) Extension of Fig.~\ref{Fig2}(b) for $B=1$ and $\beta_0=2$, representing 
the steady solutions already present in Fig.~\ref{Fig2} as dashed
lines \bfuwe{(stable/unstable single drops and films)} 
and the additionally determined solutions as the solid lines
\bfuwe{(unstable double drops) and dot-dashed lines (unstable triple drops).} Saddle-node 
bifurcations (folds) are marked with solid circles with colors corresponding to the 
respective fold-continuation curves in Fig.~\ref{Fig4}. Profiles for solutions on the 
vertical thin dashed line \bfuwe{that corresponds to $\Omega=0.767$} are shown in Fig.~\ref{Fig5}.}
\label{Fig3}
\end{center}
\end{figure}

First, we expand in Fig.~\ref{Fig3}(a) the results of Fig.~\ref{Fig2}(b) for 
intermediate values $B=1.25$, $1.5$ and $2$ and notice that there exists a rather 
complex structure involving multiple solutions in an intermediate $\Omega$-range: 
for $B=1.25$ there are three separate curves that for $B=1.5$ merge into $2$ 
curves and then for $B=2$ into a single-loop structure. 
\renewcommand{\baselinestretch}{1}
\renewcommand{\baselinestretch}{1.5}

\renewcommand{\baselinestretch}{1}
\begin{figure}
\begin{center}
\subfigure[]{\includegraphics[width=0.45\hsize,height=5.5cm]{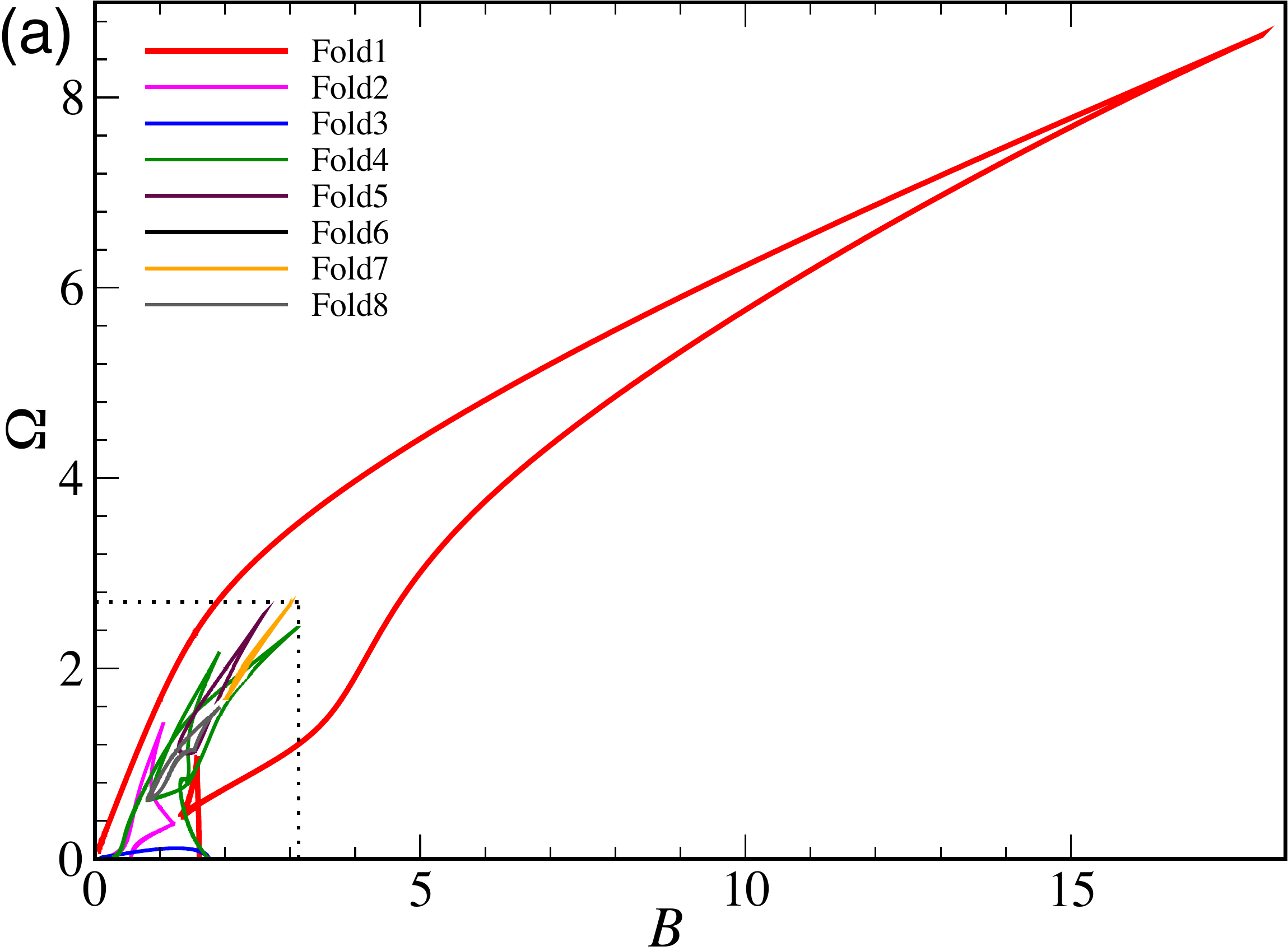}}\hspace{0.5cm}
\subfigure[]{\includegraphics[width=0.45\hsize,height=5.5cm]{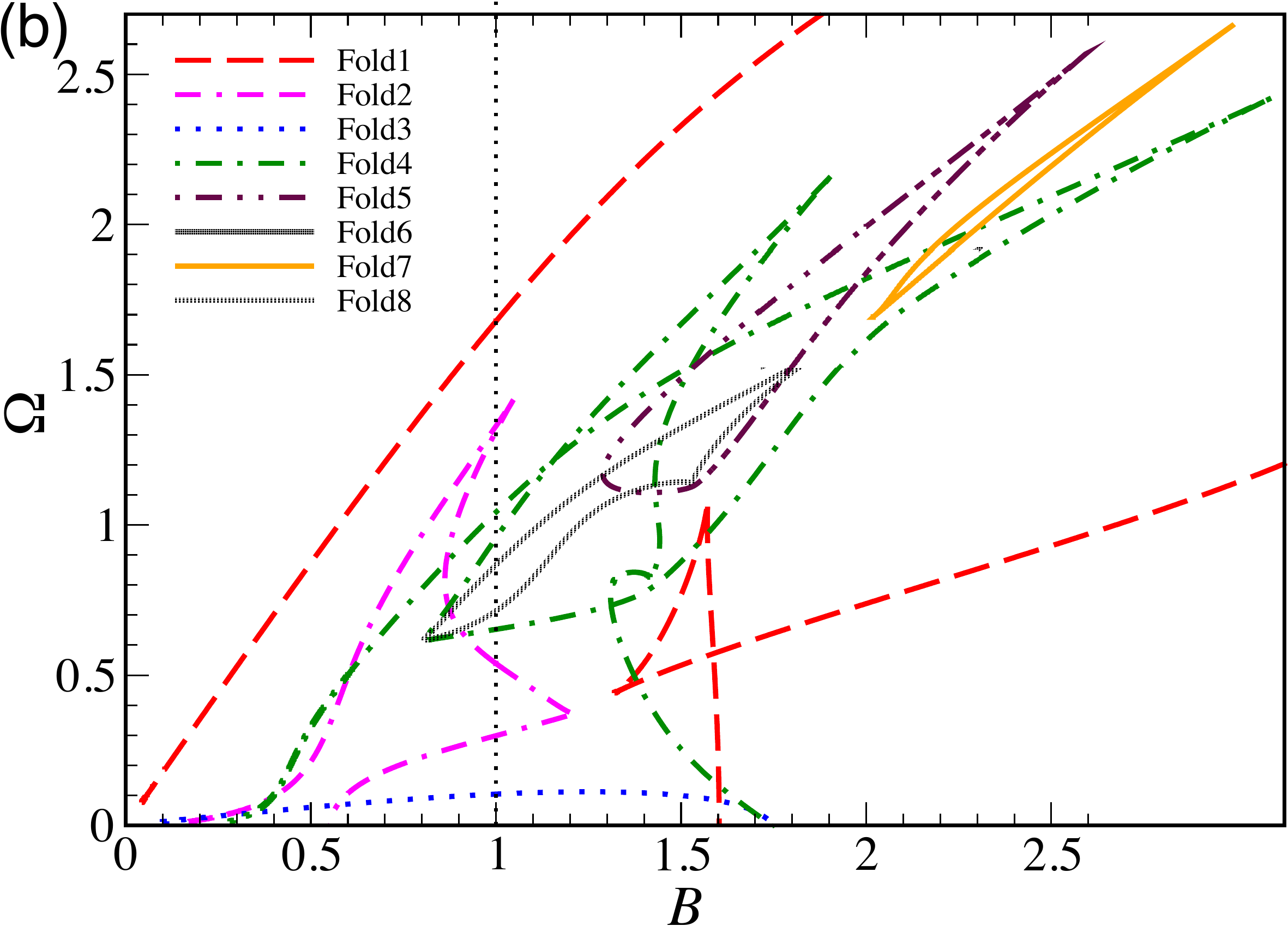}}
\caption{(Color online) 
Shown are the loci of saddle-node bifurcations in the ($B, \Omega$)-plane for 
$\beta_0=2$. Panel (a) gives a complete view while panel (b) gives a zoom into the 
region where most folds are located. The vertical thin dotted line indicates the 
case $B=1$ that \bfuwe{is consistent with Fig.~\ref{Fig3}(b) showing} eleven folds with corresponding 
line colors.
}
\label{Fig4}
\end{center}
\end{figure}
\renewcommand{\baselinestretch}{1.5}

To investigate the reconnections and the related additional branches, we continue 
the loci of all encountered SN bifurcations in the ($B, \Omega$)-plane. This includes 
codimension-two bifurcations where pairs of SN bifurcations are created/annihilated. 
The three observed processes are a hysteresis bifurcation, the destruction/creation of 
an isola of solutions and a necking bifurcation (cf.,~e.g., Thiele et al.~\cite{Thiele2004}). 
The resulting full fold-tracking diagram shown in Fig.~\ref{Fig4}(a) with a zoom into 
the most relevant region in Fig.~\ref{Fig4}(b) determines the folds that exist for any 
given Bond number. For example, at $B=1$, eleven folds at particular $\Omega$ are 
found, yet, Fig.~\ref{Fig2}(b) only displayed three. Selecting the ``new'' folds, one 
may again employ continuation in $\Omega$ to obtain the complete bifurcation diagram 
of steady profiles in Fig.~\ref{Fig3}(b). 

\begin{figure}
\begin{center}
{\includegraphics[width=0.49\hsize]{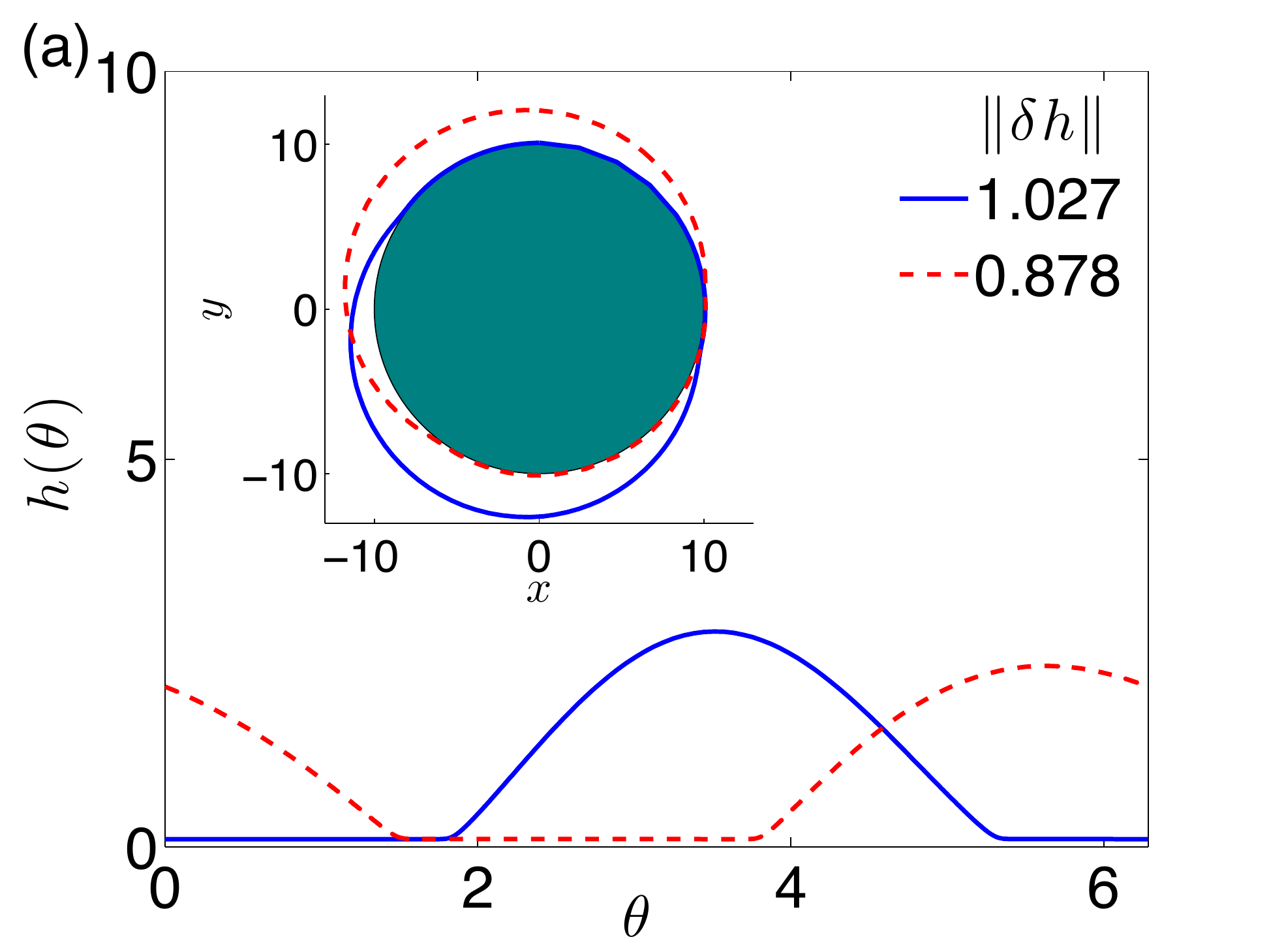}}
{\includegraphics[width=0.49\hsize]{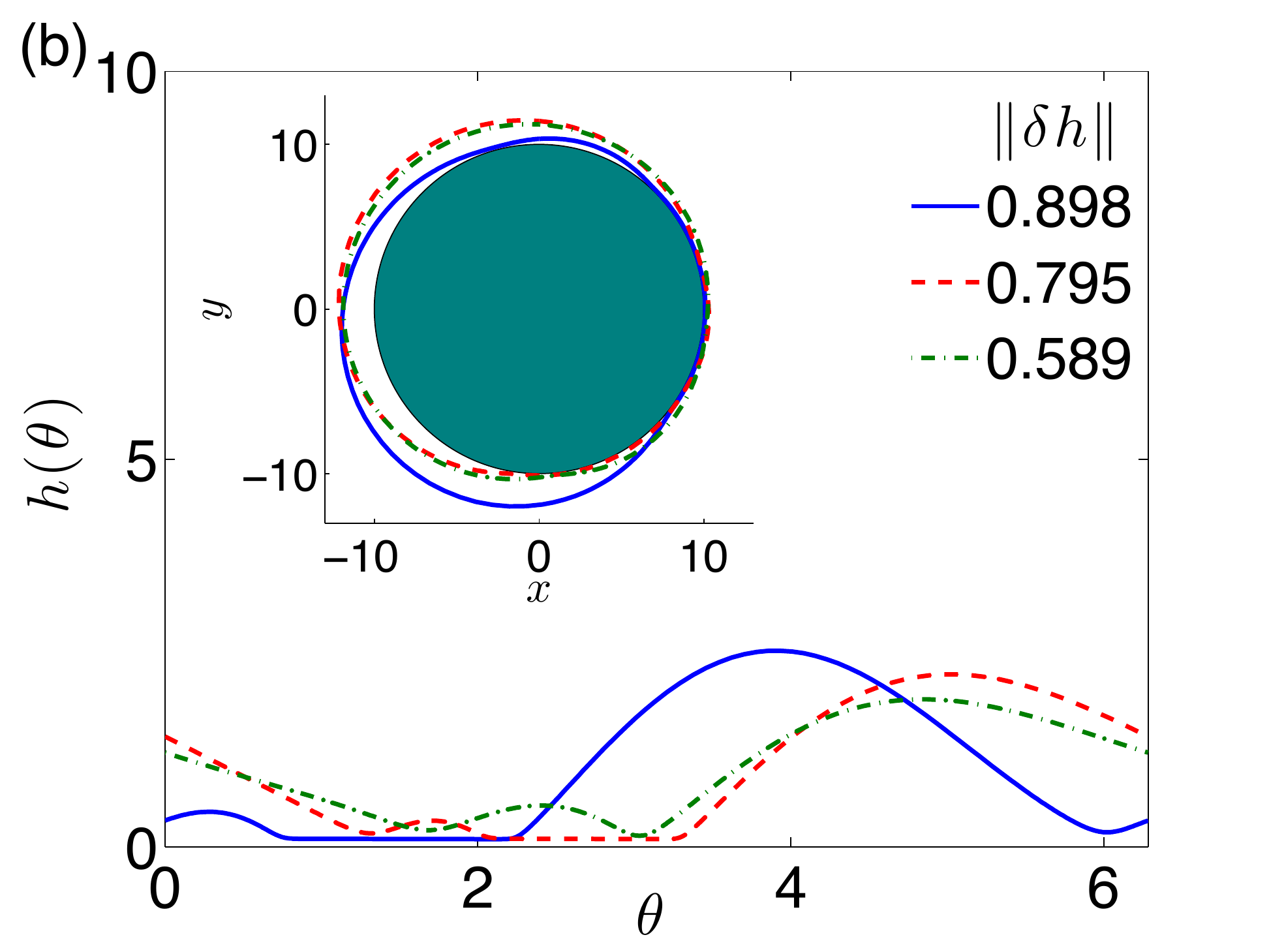}}
{\includegraphics[width=0.49\hsize]{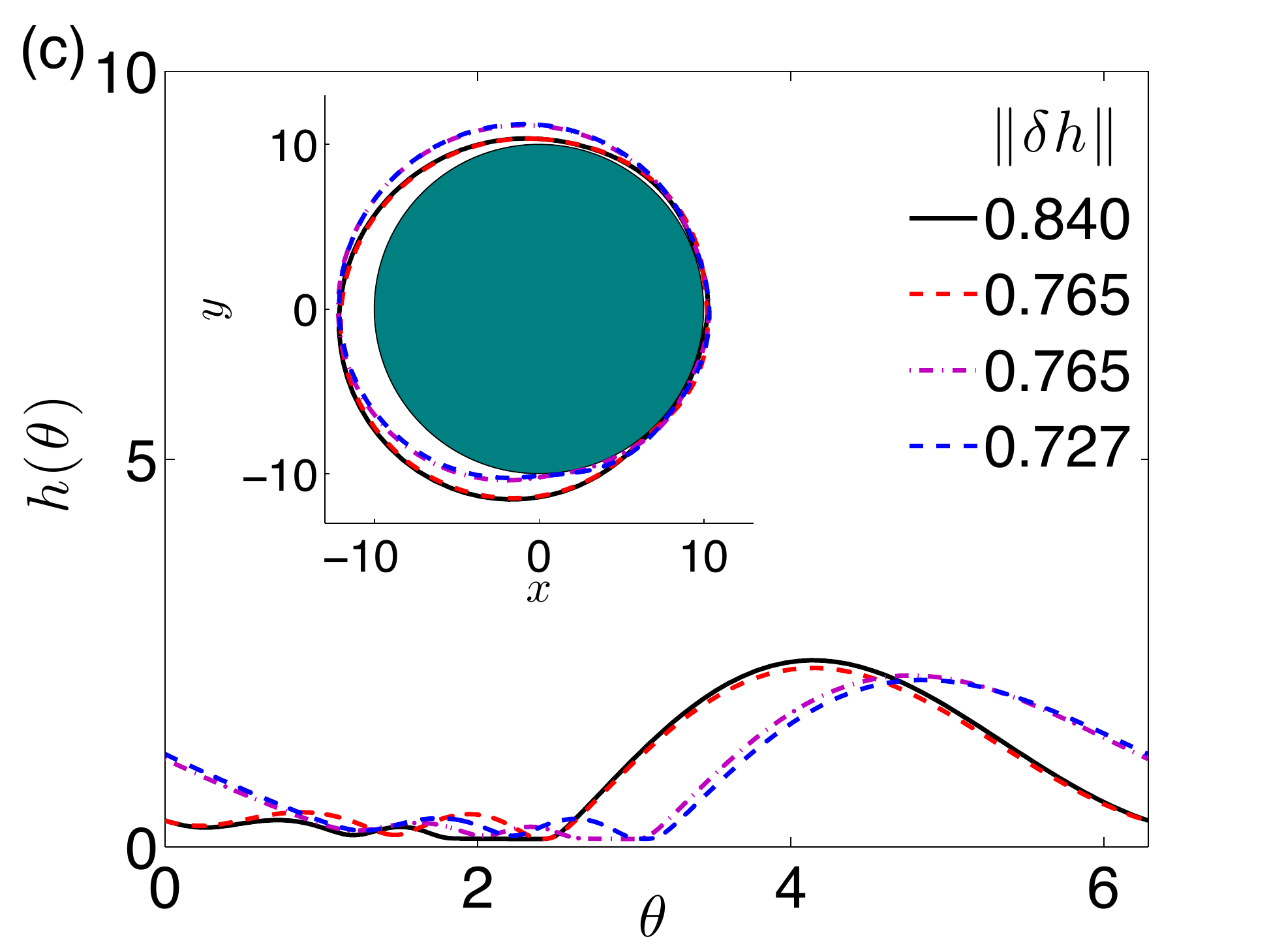}}
\caption{(Color online) 
Steady drop profiles on a horizontal cylinder for $\Omega =0.767$, 
$\beta_0=2$ and $B=1$, i.e., at the loci indicated by the
vertical thin dashed line in Fig.~\ref{Fig3}(b). 
The individual panels give
(a) single-, (b) double- and (c) triple-drop solutions, with the $L^2$-norms 
$\|\delta h\|$ as given in the legends. \bfuwe{The insets show the profiles on a cylinder assuming a radius $R=10$.}}
\label{Fig5}
\end{center}
\end{figure}

We observe that around $\Omega=0.8$ there exists a ``figure-eight"
isola of solutions.  What makes these solutions particularly
interesting is that they are not related to any equilibrium solutions
that may exist at $\Omega=0$ for otherwise the same parameters.  In
particular, the solutions correspond to \bfuwe{unstable}
triple-drop/hump solutions, that for $B=1$ and $\beta_0=2$ do not
exist without rotation, i.e., without non-equilbrium
driving. In other words, a finite driving brings into
existence solutions expected at other parameter values. Typical
single-, double- and triple-drop profiles located on the vertical
dotted line in Fig.~\ref{Fig3}(b) are illustrated in
Fig.~\ref{Fig5}. As expected, all the profiles are off center, i.e.,
rotation drags profiles towards the \bfuwe{left-hand side of the
  cylinder. Although all double- and triple-drop solutions are
  unstable, their understanding is crucial for the understanding of the
  behavior of the hydrodynamic system: they provide the phase space
  of the system with a rich structure that allows for complex
  dynamical states (see Sec.~\ref{sec:complete}).}

\renewcommand{\baselinestretch}{1}
\renewcommand{\baselinestretch}{1.5}

The remaining part of the new structure in Fig.~\ref{Fig3}(b) intersects
$\Omega=0$ four times, with the $L^2$-norms
$\|\delta h\|\approx0.30$, $0.62$ and a double intersection at
$0.59$. This indicates that an entire system of branches is missing in
Fig.~\ref{Fig2}(b) showing the equilibrium case. The solution with the
\bfuwe{$L^2$-norm $0.30$ is the solution missing in
  Fig.~\ref{Fig2}(b). As the additionally encountered branches cross
  $\Omega=0$, also the system of equilibrium profiles needs
  completion. This is achieved through continuation in $B$ at
  $\Omega=0$.}

\renewcommand{\baselinestretch}{1}
\begin{figure}
\begin{center}
\includegraphics[width=0.7\hsize]{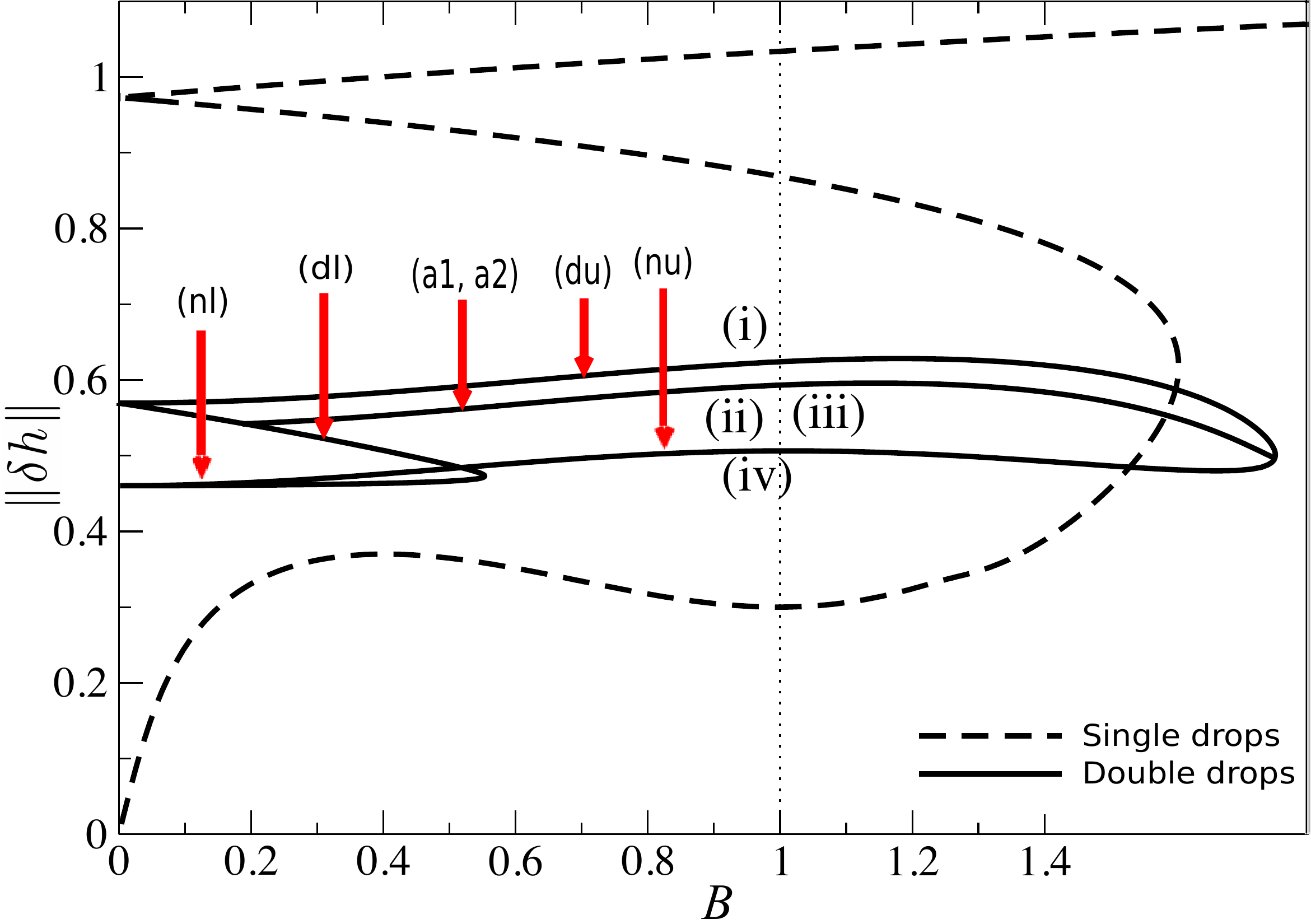}
\caption{(Color online) 
Branches of steady profiles for $\beta_0=2$. The $L^2$-norms
are given as functions of the Bond number $B$. The dashed lines represent single-drop solutions
also present in Fig.~\ref{Fig2}(a) while the solid lines give
the bifurcation structure of \bfuwe{unstable} double-drop solutions. There we distinguish
symmetric drop (d$\mathrm{u}$, d$\mathrm{l}$) and nucleation or hole (n$\mathrm{u}$, n$\mathrm{l}$) and asymetric (a$\mathrm{1}$,
a$\mathrm{2}$) solutions. Subscripts ``u'' and ``l'' refer to the respective upper and
lower solution. The profiles at $B=1$ (indicated by the vertical dotted line and
marked by (i)--(iv)) are displayed in Fig.~\ref{Fig7}. 
\bfuwe{We also note that only the branch that corresponds to pendent drop solutions is stable (the branch with the largest $L^2$-norms). All the other branches are unstable.}
}
\label{Fig6}
\end{center}

\end{figure}
\renewcommand{\baselinestretch}{1.5}
\renewcommand{\baselinestretch}{1}
\begin{figure}
\begin{center}
\includegraphics[width=0.6\hsize]{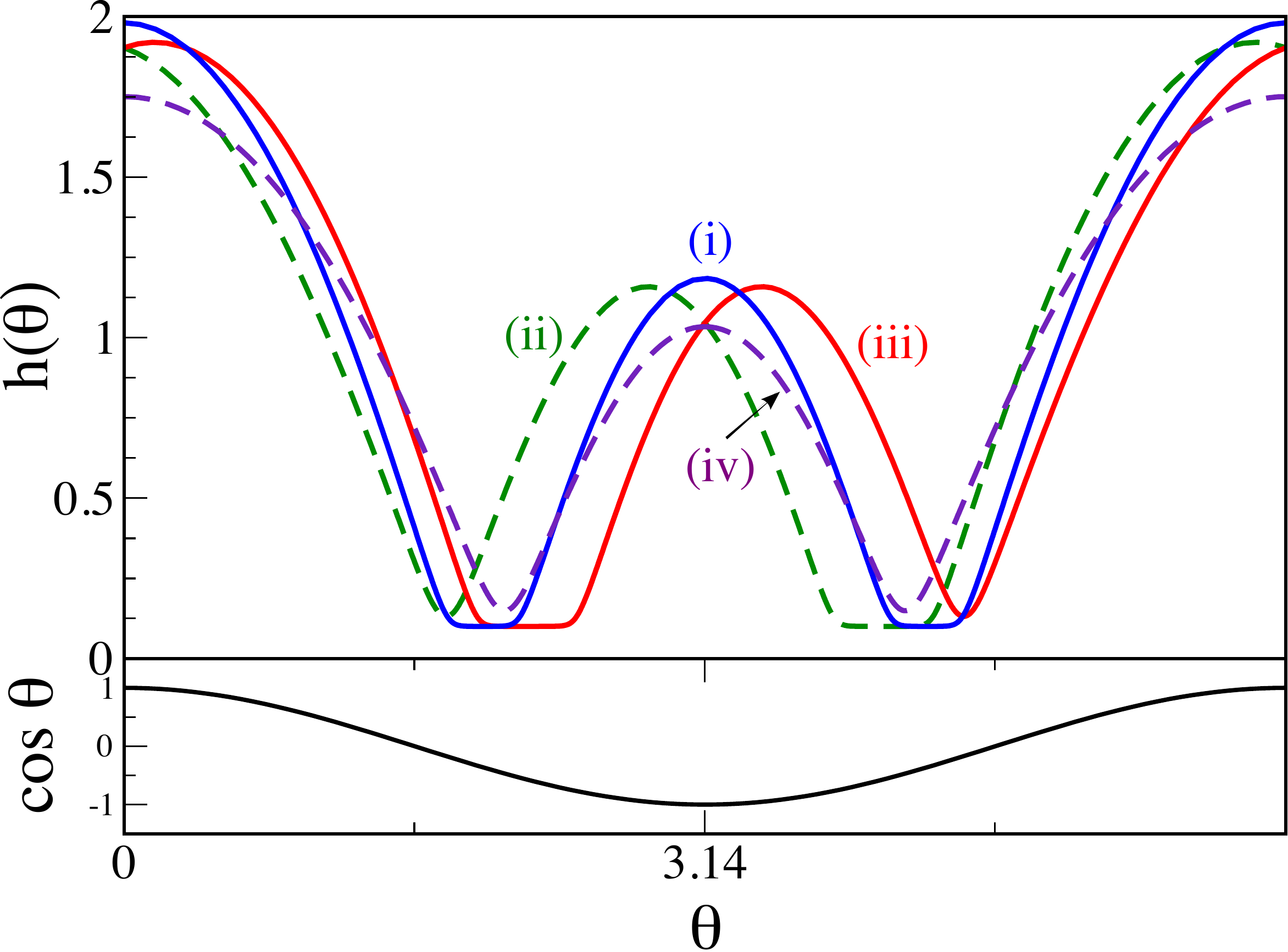}
\caption{(Color online) 
Selected profiles of steady \bfuwe{unstable} double-drop solutions on a horizontal cylinder, without rotation 
($\Omega = 0$), at $B= 1$ and $\beta_0 = 2$ corresponding  in
Fig.~\ref{Fig6} to the crossing points of the vertical dotted line and
the solid lines (with the labels (i)--(iv)). They present examples of
profiles from branches d$\mathrm{u}$, a$\mathrm{1}$, a$\mathrm{2}$ and
n$\mathrm{u}$, respectively. The symmetries are discussed in the main text.
}
\label{Fig7}
\end{center}
\end{figure}
\renewcommand{\baselinestretch}{1.5}

All the resulting solutions \bfuwe{corresponding} to
\bfuwe{unstable} double-drop solutions are included in Fig.~\ref{Fig6} 
as solid black lines. The profiles indicated at $B=1$ by the vertical dotted line and 
marked by (i)--(iv) are displayed in Fig.~\ref{Fig7}. As they also correspond to the 
$\Omega=0$ states in Fig.~\ref{Fig3}(b), we deduce that the central line in 
Fig.~\ref{Fig6} is actually a double branch since in Fig.~\ref{Fig3}(b) two branches 
meet at $\Omega=0$, with the $L^2$-norm of $0.59$. As these are the only branches in 
Fig.~\ref{Fig3}(b) which do not have zero slope at $\Omega=0$, they correspond to 
the two mirror images with respect to $\theta = \pi$ of an asymmetric solution [profiles (ii) and
(iii) in Fig.~\ref{Fig7}]. We name these branches a$\mathrm{1}$ and 
a$\mathrm{2}$ (where `a' refers to `asymmetric'). They emerge at pitchfork bifurcations 
from the other branches that correspond to drop and hole (or nucleation) solutions 
that are themselves symmetric with respect to reflection at $\theta = \pi$. We name 
them d$\mathrm{u}$, d$\mathrm{l}$, n$\mathrm{u}$, and n$\mathrm{l}$ (where `d' 
and `n' refer to `drop' and `nucleation', respectively, and `u' and `l' refer to respective 
upper and lower solutions), see Fig.~\ref{Fig6}. Figure~\ref{Fig7} gives examples for 
d$\mathrm{u}$ and n$\mathrm{u}$ solutions, see profiles (i) and (iv), respectively. For the 
d$\mathrm{u}$ and n$\mathrm{u}$ solutions shown in Fig.~\ref{Fig7}, there is one 
large drop on top of the cylinder and another smaller one underneath. In contrast, 
the d$\mathrm{l}$ and n$\mathrm{l}$ branches consist of profiles where the two 
holes are on top and underneath of the cylinder (not shown).

Note, that \bfuwe{the unstable double-drop solutions exist at larger Bond
  numbers than the unstable single-drop solutions and may, therefore,
  influence the dynamics up to larger values of $B$. Yet, as before,
  above a critical value ($B\approx 1.749$), only the stable pendent
  drop remains.}

\bfuwe{To understand why Fig.~\ref{Fig6} has the two double-drop
  solutions at $\Omega=0$ and $B=0$ that give rise to the entire
  double-drop branch structure at $B>0$ even without driving, we
  consider in Appendix~\ref{sec:steady_drops_1} the case $\Omega=0$,
  $B=0$ in an extended parameter space. Also,
  Appendix~\ref{sec:steady_drops_2} gives some further details of the
  behavior of the double-drop solutions of Fig.~\ref{Fig6}.}

\subsection{Transition to the completely wetting case}
\mylab{sec:completewetting}

The rich solution structure at equilibrium ($\Omega=0$) and, in consequence, out of 
equilibrium ($\Omega>0$) is mainly due to partial wettability. This implies that most of 
this structure \bfuwe{has to disappear when wettability is increased, i.e., when the 
equilibrium contact angle $\beta_0$ is decreased.} Before we move on to the full bifurcation 
structure including time-periodic thickness profiles, here, we briefly emphasize the 
contrasting structure of steady profiles found for completely wetting liquids, when 
$\beta_0$ approaches zero. As observed in Fig.~\ref{Fig2}(a), for a resting horizontal 
cylinder without rotation the critical Bond number $B_\mathrm{sn}$ \bfuwe{where the two 
unstable drop solutions annihilate, i.e., where a qualitative transition 
occurs, decreases with $\beta_0$. In parallel, the modulated film transforms into a 
pendent drop that is at $\beta_0=0$ the only solution for any non-zero Bond number.}

For a completely wetting liquid, there exists no force beside gravity that favors
drops as opposed to a flat film. Gravity can still produce pendent drops. However, 
for non-zero Bond numbers at $\Omega=0$, the model, Eq.~(\ref{eq:timeevolution}), 
does not have continuous $2\pi$-periodic steady-state solutions, but only 
solutions with a finite support, i.e., \bfuwe{pendent drops on a dry substrate.} They can be studied 
employing a weak formulation.  See Appendix~\ref{sec:B1} and 
Refs.~\onlinecite{Reisfeld1992, Evans2004} for further details.

\begin{figure}
{\includegraphics[width=0.49\hsize]{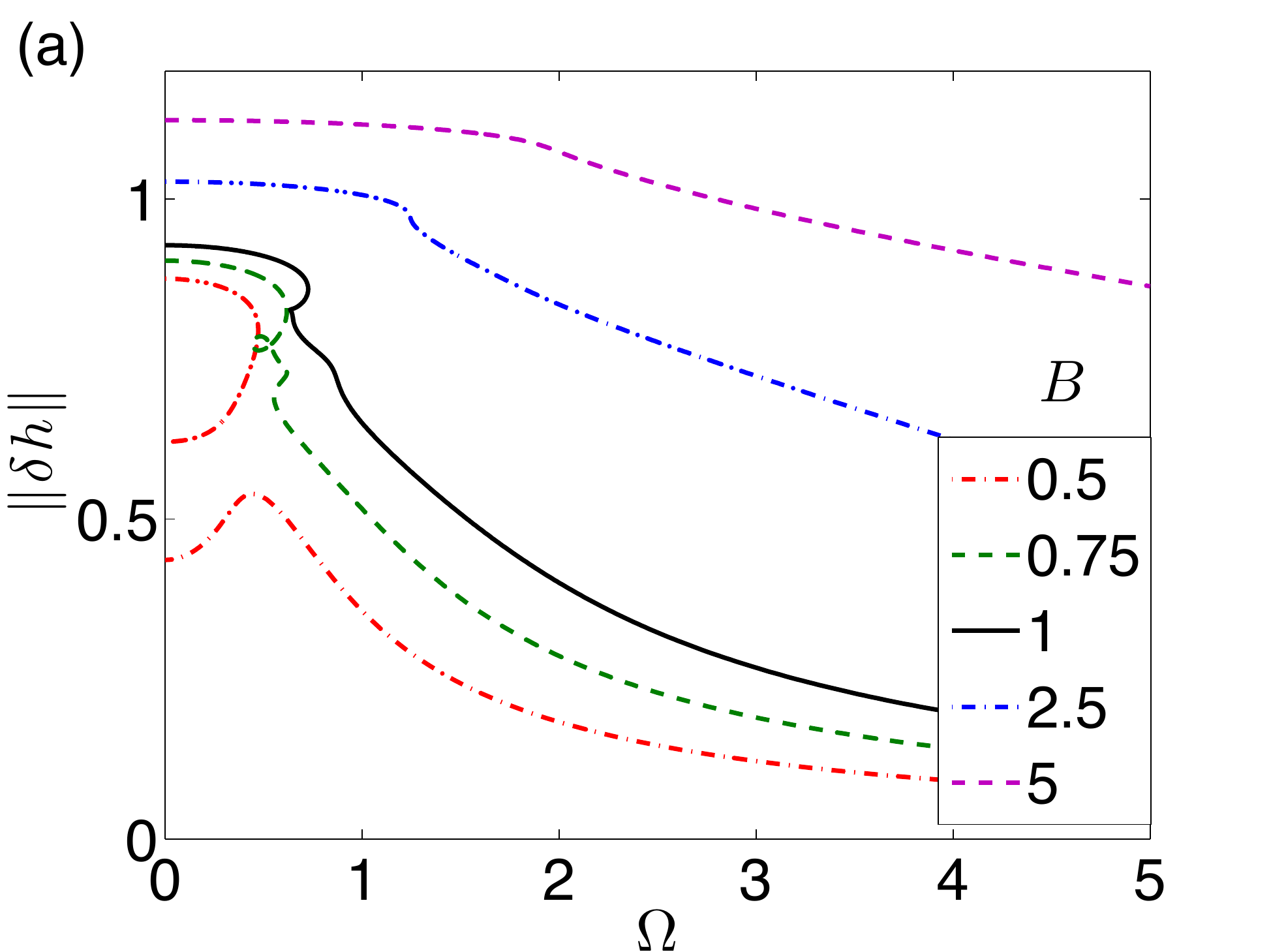}}
{\includegraphics[width=0.49\hsize]{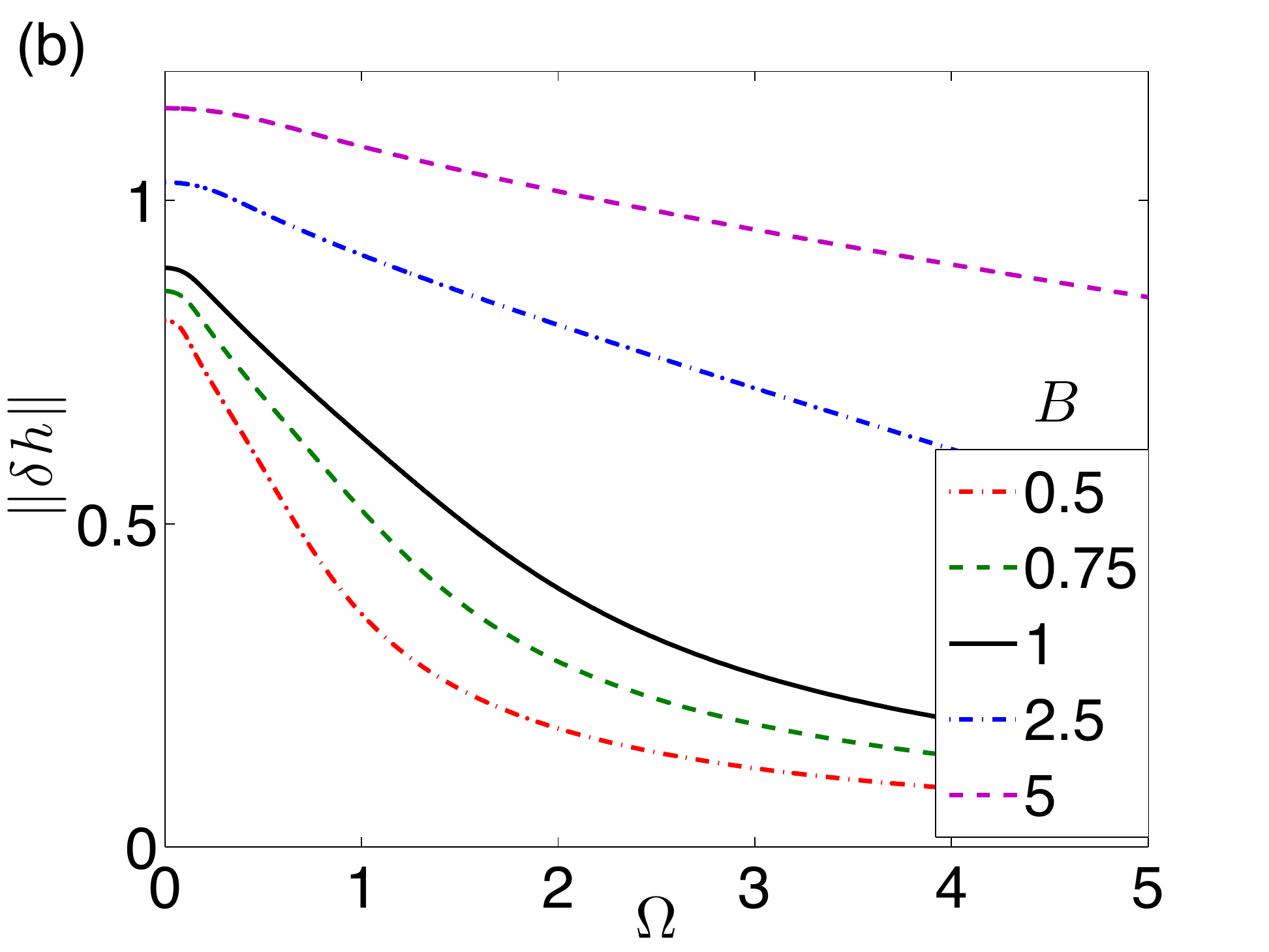}}\vspace{0.2cm}
\caption{(Color online) 
Shown are branches of steady states for different Bond numbers $B$ (as given in 
the legends) as functions of the rotation number $\Omega$ for equilibrium contact 
angles (a) $\beta_0=1$ and (b)  $\beta_0=0.1$.}\vspace{0.2cm}
\label{Fig10}
\end{figure}

Then for $\Omega>0$, continuous $2\pi$-periodic  steady solutions exist as the 
lateral driving creates a dynamic wetting layer. Increasing $\Omega$, the droplets 
change monotonically, the rotation shifts them towards the left and smears them 
out at the same time. Figure~\ref{Fig10}(a) gives for various $B$ the main branch 
structure as a function of $\Omega$ for a partially wetting liquid of a smaller contact 
angle $\beta_0=1$ while Fig.~\ref{Fig10}(b) gives this structure close to complete 
wetting at $\beta_0=0.1$.

From Fig.~\ref{Fig2}(b) \bfuwe{to Fig.~\ref{Fig10}(a), and further to Fig.~\ref{Fig10}(b)}, one 
observes a clear transition towards a much simpler branch structure \bfuwe{as the contact 
angle $\beta_0$ decreases.} Actually, all complex aspects of branch patterns discussed at Figs.~\ref{Fig2}(b) 
and~\ref{Fig3} disappear resulting, e.g., at $\beta_0=0.1$, in a rather simple branch of steady solutions with 
a monotonically decreasing $L^2$-norm. In particular, as the contact angle 
is decreased, all the SN bifurcations occurring for various values of the rotation and 
Bond numbers (cf.~Fig.~\ref{Fig4}) are eliminated in various codimension-two 
bifurcations. Thereby, the prominent SN bifurcation, that at $\beta_0=2$ is related 
to the SNIPER bifurcation, is \bfuwe{the bifurcation structure that exists for the largest range 
of $\beta_0$.} It is a particularly intriguing question what happens to the SNIPER bifurcation 
and the emerging branch of time-periodic solutions (stick-slipping drops) when the 
wettability increases, i.e., $\beta_0$ decreases. This is studied in 
Sec.~\ref{sec:complete}.

\renewcommand{\baselinestretch}{1}
\renewcommand{\baselinestretch}{1.5}

\renewcommand{\baselinestretch}{1}
\begin{figure}
\begin{center}
{\includegraphics[width=0.49\hsize]{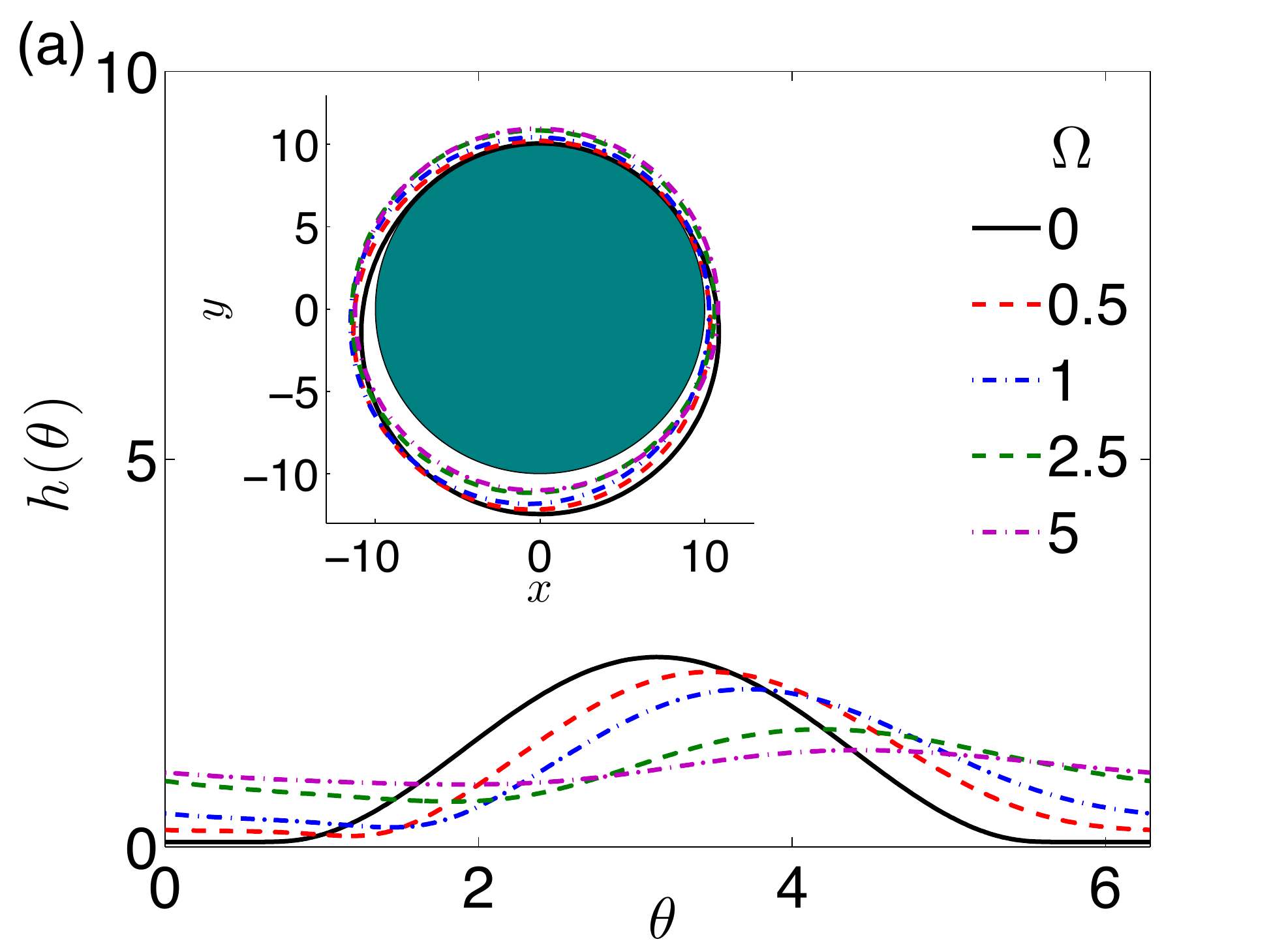}}
{\includegraphics[width=0.49\hsize]{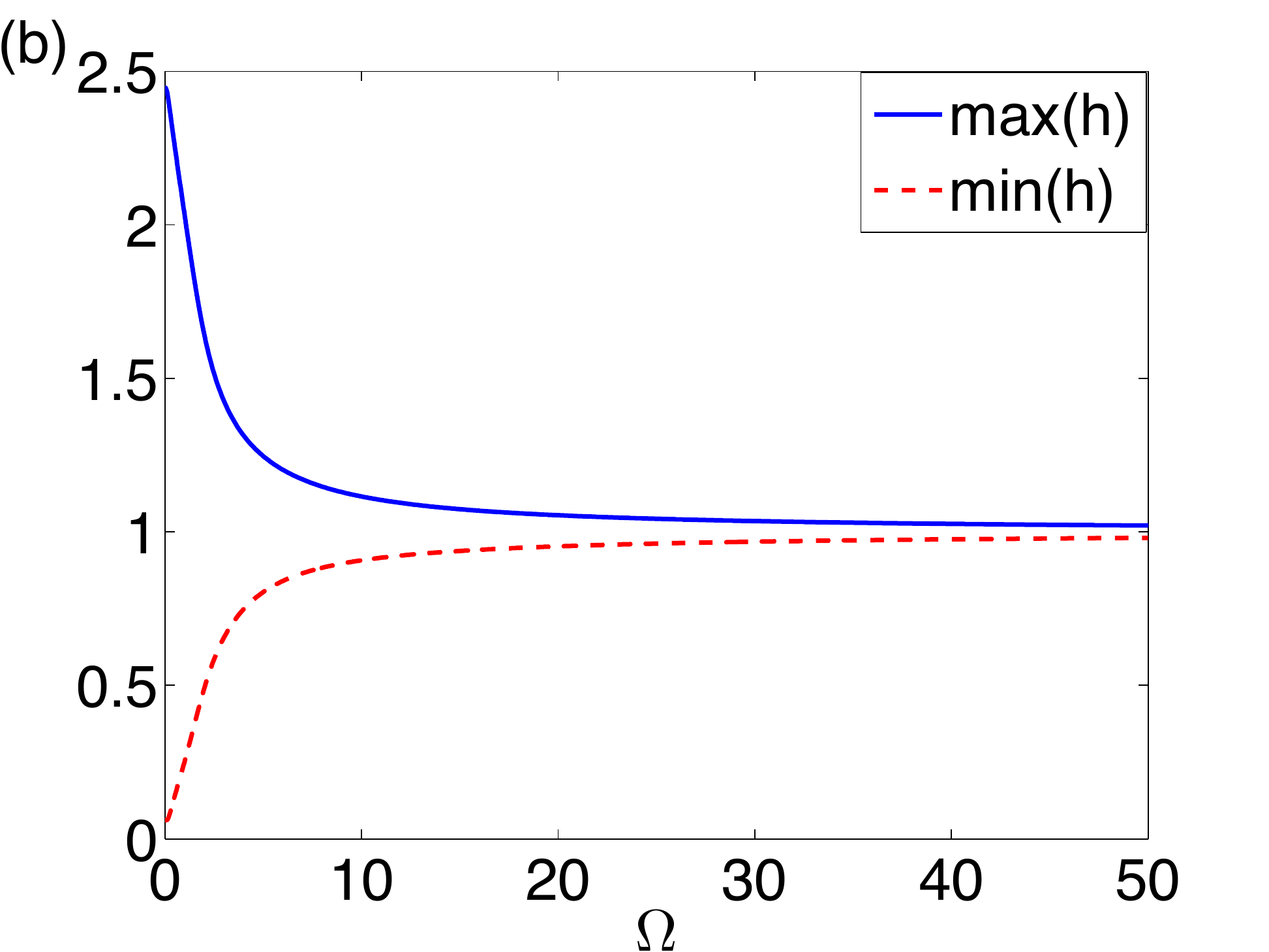}}
\caption{(Color online) 
(a) Steady film-thickness profiles for the approximate case of complete wetting 
($\beta_0=0.1$) with $B=1$ for several rotation numbers. 
(b) The maximum (blue solid line) and minimum (red dashed line) of the steady film 
thickness showing that the profile tends to a flat film with increasing $\Omega$.}
\label{Fig11}
\end{center}
\end{figure}
\renewcommand{\baselinestretch}{1.5}

The case of the smallest considered contact angle $\beta_0=0.1$ is already 
close to the complete wetting case. Selected steady drop profiles that correspond 
to the solution branch for $B=1$ in Fig.~\ref{Fig10}(b) are displayed in 
Fig.~\ref{Fig11}(a). For increasing $\Omega$, the driving drags the liquid along 
the surface of the cylinder, resulting in a steady drop whose center of mass moves 
towards a larger $\theta$. However, as wettability is large, the adhesion force that 
keeps the drop together is weak and the drop can now deform strongly. The drop 
becomes shallow as indicated by the monotonically decreasing/increasing 
maximum/minimum values of its profile, as shown in Fig.~\ref{Fig11}(b). 
Finally, at large $\Omega$, it approaches a flat-film solution. In other words, the 
drop does not survive as a ``coherent structure" due to the attenuated influence of 
partial wettability. We notice that the drops flatten and smear out well before their 
maxima reach $\theta = 3\pi/2$ (the left-hand-side of the cylinder), where gravity 
becomes tangential to the cylinder surface.

\vspace{-0.3cm}
\section{Branches of time-periodic flow states}
\mylab{sec:complete}

To better understand the behavior of the system, one also needs to obtain the 
branches of \bfuwe{ unsteady 
flows, i.e., time-periodic 
film and drop profiles as, e.g., co-rotating drops 
on the cylinder or surface waves. Another point to clarify is how the} 
time-periodic states emerge from the already 
discussed branches of steady profiles. Ref.~\onlinecite{Thiele2011a} only 
considered the large contact angle case $\beta_0=2$ and obtained a single 
time-periodic solution branch for $\Omega>\Omega_\mathrm{sn}$ using direct 
numerical time integration. This method allows one to establish that the main 
depinning transition for the partially wetting drop on a rotating cylinder is a SNIPER 
bifurcation similar to the case of a drop on a heterogeneous 
substrate~\cite{Thiele2006}. However, as the direct numerical simulation follows the evolution in time 
starting from an initial condition, it does not allow for a determination of unstable 
steady or time-periodic states. As these are needed to develop a full understanding of the involved 
qualitative transitions, in the current study, we obtain both, stable and unstable 
states, via numerical continuation as outlined above in Sec.~\ref{sec:model}. In 
what follows, we fix the Bond number, $B=1$, and investigate the changes of the 
bifurcation diagram when the equilibrium contact angle $\beta_0$ varies.

\vspace{-0.3cm}
\subsection{The case of a large contact angle ($\beta_0 = 2$)}
\mylab{sec:timeb2}

\begin{figure}
\begin{center}
{\includegraphics[width=0.49\hsize]{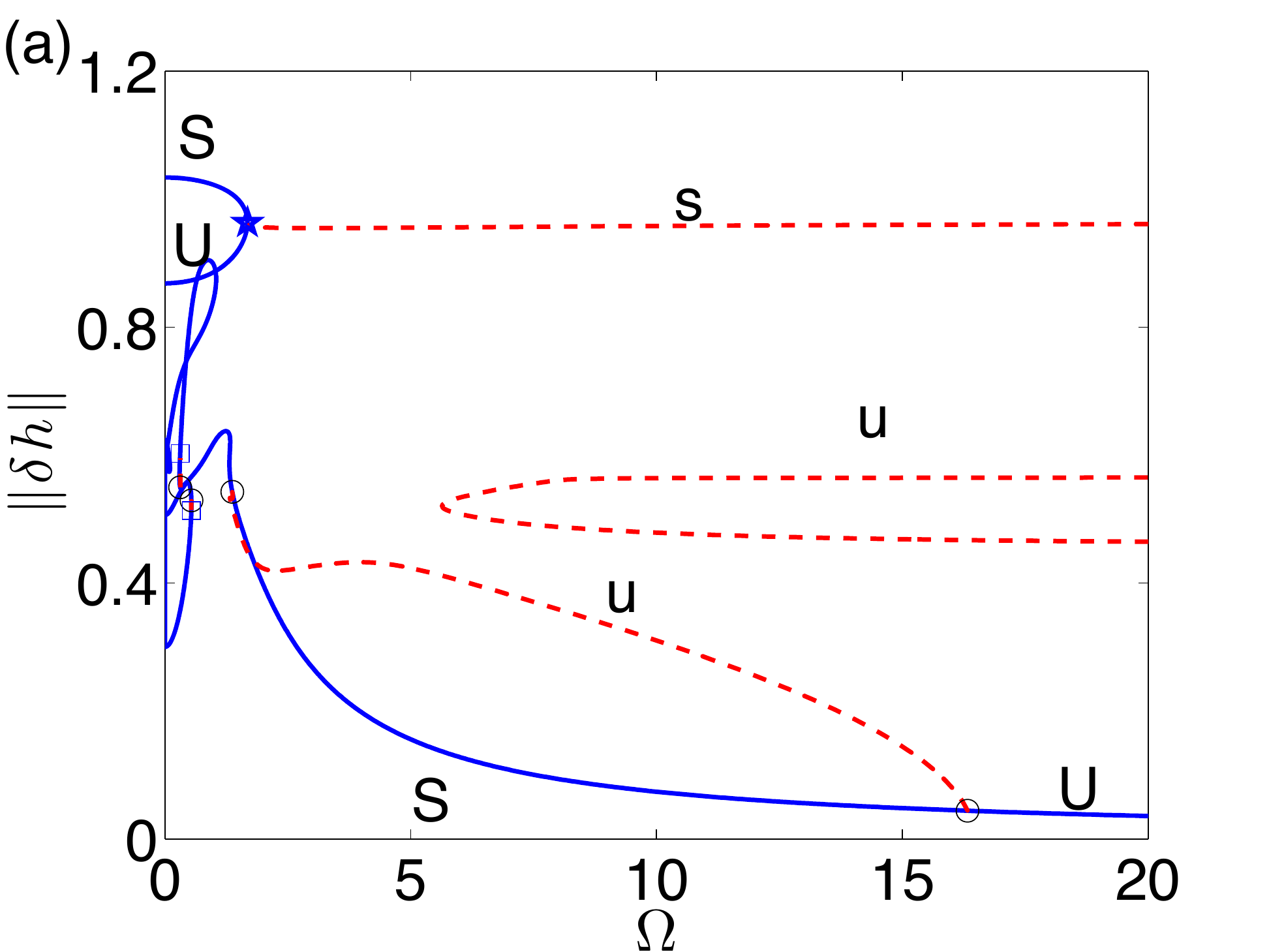}}
{\includegraphics[width=0.49\hsize]{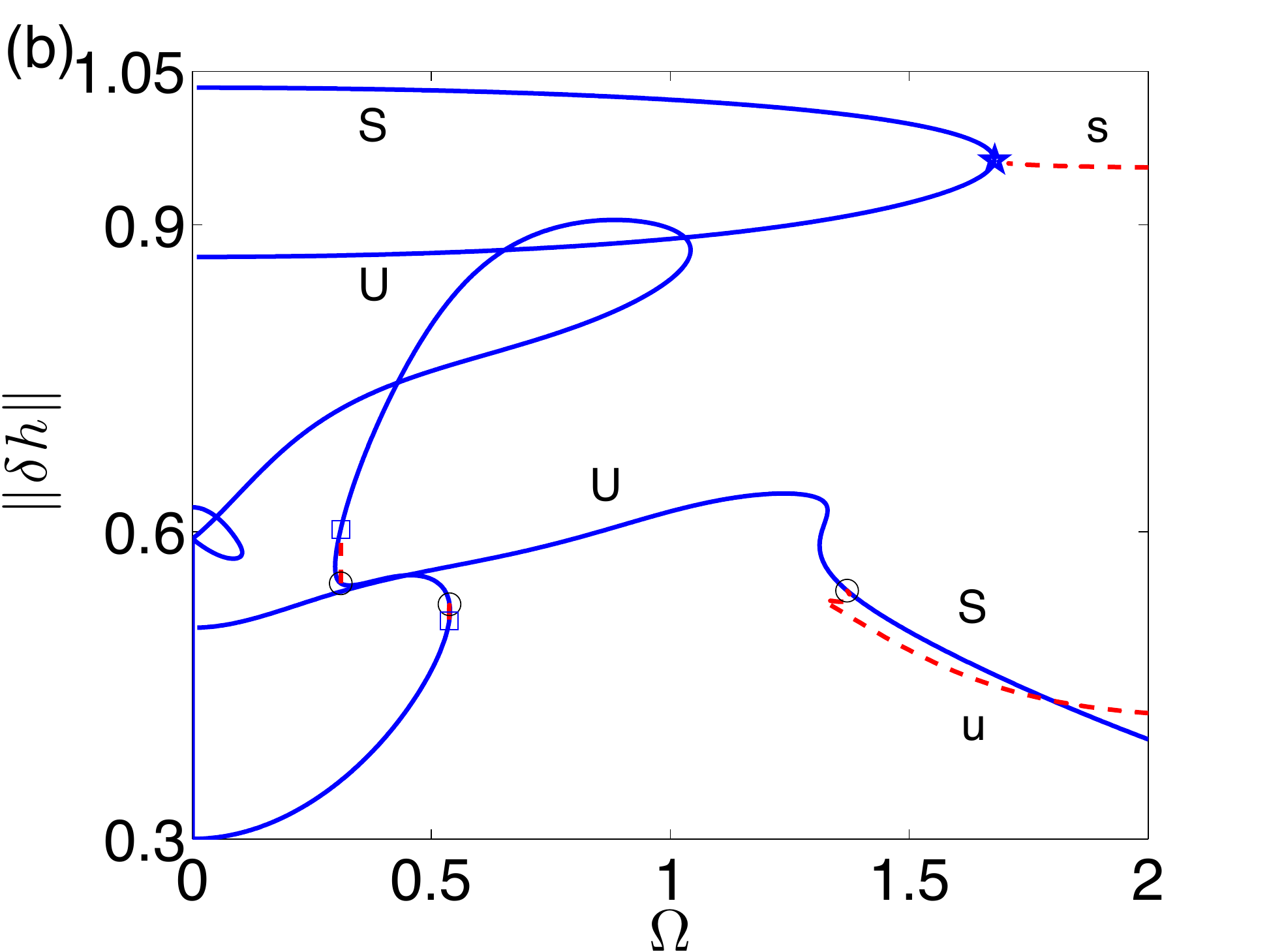}}
\caption{(Color online) 
The complete bifurcation diagrams for $\beta_0=2$ and $B=1$ giving the solution measure 
\bfuwe{$\|\delta h\|$} in dependence on the rotation number $\Omega$. It includes the steady 
profiles (blue solid lines) and time-periodic profiles (red dashed lines). The HBs are indicated 
by circles, the HCs are indicated by squares and the SNIPER is indicated by a star. 
(a) The complete bifurcation diagram. 
(b) A zoom of panel (a) into the small-$\Omega$ region where most of the bifurcation 
structure is located. \bfuwe{The stabilities of the states are discussed in the main text.}
}
\label{Figbeta2}
\end{center}
\end{figure}

We start with the case of a ``large" contact angle ($\beta_0=2$). The complete bifurcation diagram 
is presented in Fig.~\ref{Figbeta2} where panel (a) gives the solution measure $\| \delta h \|$, 
\bfuwe{defined by the $L^2$-norm in Eq.~(\ref{eq:normsteady}) for steady-state profiles and by the 
time-averaged $L^2$-norm in Eq.~(\ref{eq:normtime}) for time-periodic flow states,} as a function of 
the rotation number over the full range, while panel (b) presents a zoom into the region of small 
values of $\Omega$. Steady-state solution branches are shown as (blue) solid lines while the 
time-periodic solution branches are given as (red) dashed lines. \bfuwe{In this figure and in the figures that follow,} stable/unstable branches of 
steady-state solutions are indicated by ``S" and ``U", respectively, and  stable/unstable branches of 
time-periodic solutions are indicated by ``s" and ``u", respectively. Furthermore, the loci of Hopf 
bifurcations (HBs), homoclinic bifurcations (HCs) and SNIPER bifurcations are indicated by circles, 
squares and stars, respectively. In total, there exist four HBs, two HCs, one SNIPER and five 
time-periodic solution branches. 

\renewcommand{\baselinestretch}{1}
\renewcommand{\baselinestretch}{1.5}

It is found that steady-state solutions exist for all rotation numbers. In particular, 
for large enough $\Omega$, the solution measure of the steady-state 
solution decreases as $\Omega$ increases, and we find that the solution approaches 
the uniform coating state $h\equiv 1$. See Appendix~\ref{sec:B2a} for further details. 
Besides, below [above] the value $\Omega\approx 16.4$ at which the rightmost 
HB exists, steady-state solutions are linearly stable [unstable]. At this HB, a branch 
of unstable time-periodic solutions emerges subcritically. As can be seen in 
Fig.~\ref{Figbeta2}(a), the solution measure of the time-periodic solution increases as 
$\Omega$ decreases and the branch terminates at another HB at $\Omega\approx1.37$. 

The only branch of stable time-periodic solutions emerges in a SNIPER at the SN 
of the steady-state branch at $\Omega_\mathrm{sn}\approx 1.68$. This SNIPER was found using 
numerical time integration in Ref.~\onlinecite{Thiele2011a}. 
In the limit $\Omega\rightarrow\infty$ along this branch, the solution is equivalent 
to a drop solution at zero Bond number that co-rotates with the cylinder. A 
corresponding detailed asymptotic analysis is presented in Appendix~\ref{sec:B2b}. \bfuwe{We note that there is one more branch of time-periodic solutions that is not connected to any of the branches of steady-state solutions. It has an SN at $\Omega\approx 5.64$, and both the upper and lower parts of this branch extend to infinitely large values of $\Omega$. In the limit $\Omega\rightarrow\infty$ along the upper and lower parts of this branch, the solutions are equivalent 
to double-drop solutions at zero Bond number that co-rotate with the cylinder. We note that both the upper and lower parts of this branch are linearly unstable.}

\renewcommand{\baselinestretch}{1}
\begin{figure}
\begin{center}
{\includegraphics[width=0.49\hsize]{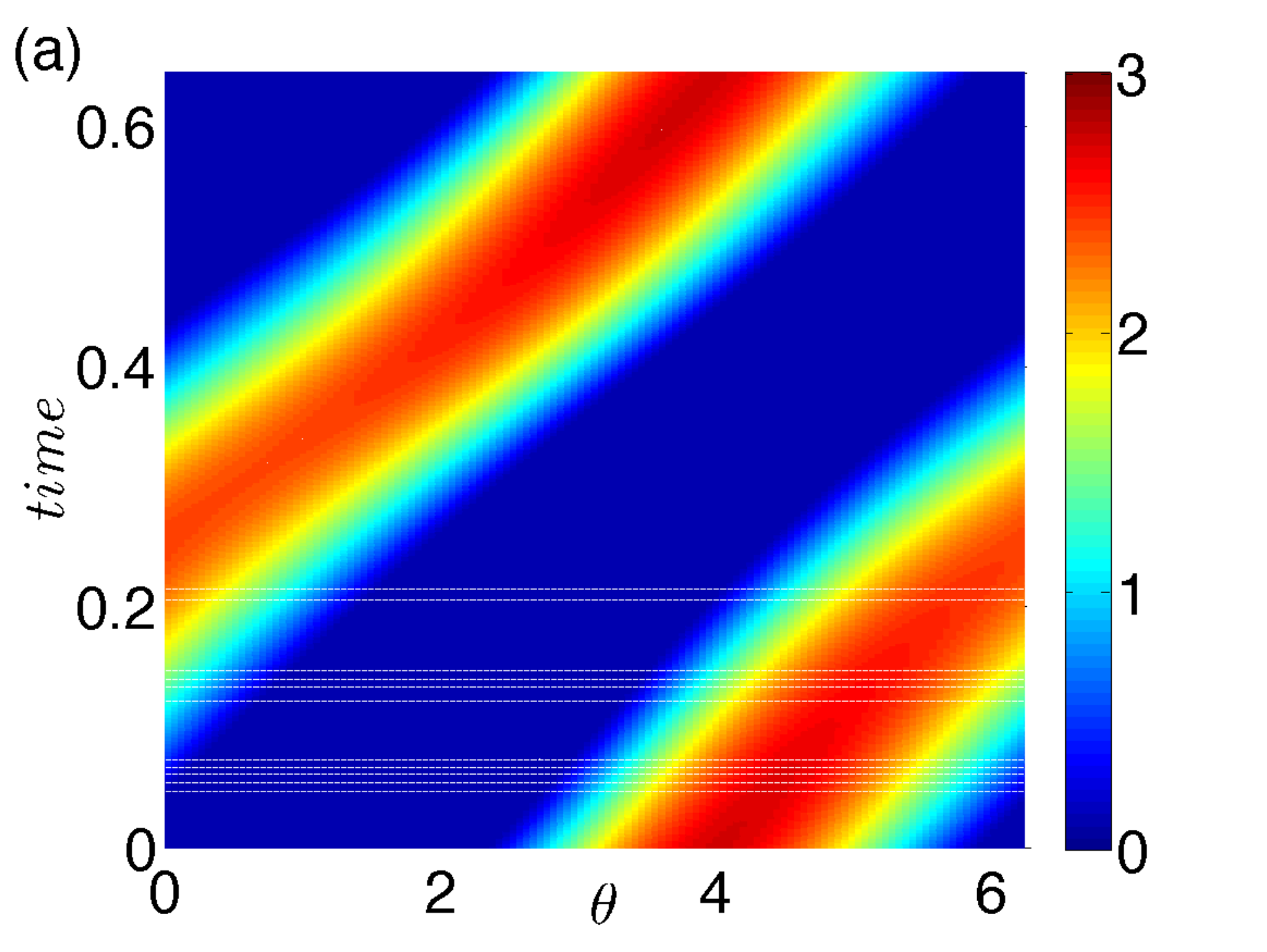}}
{\includegraphics[width=0.49\hsize]{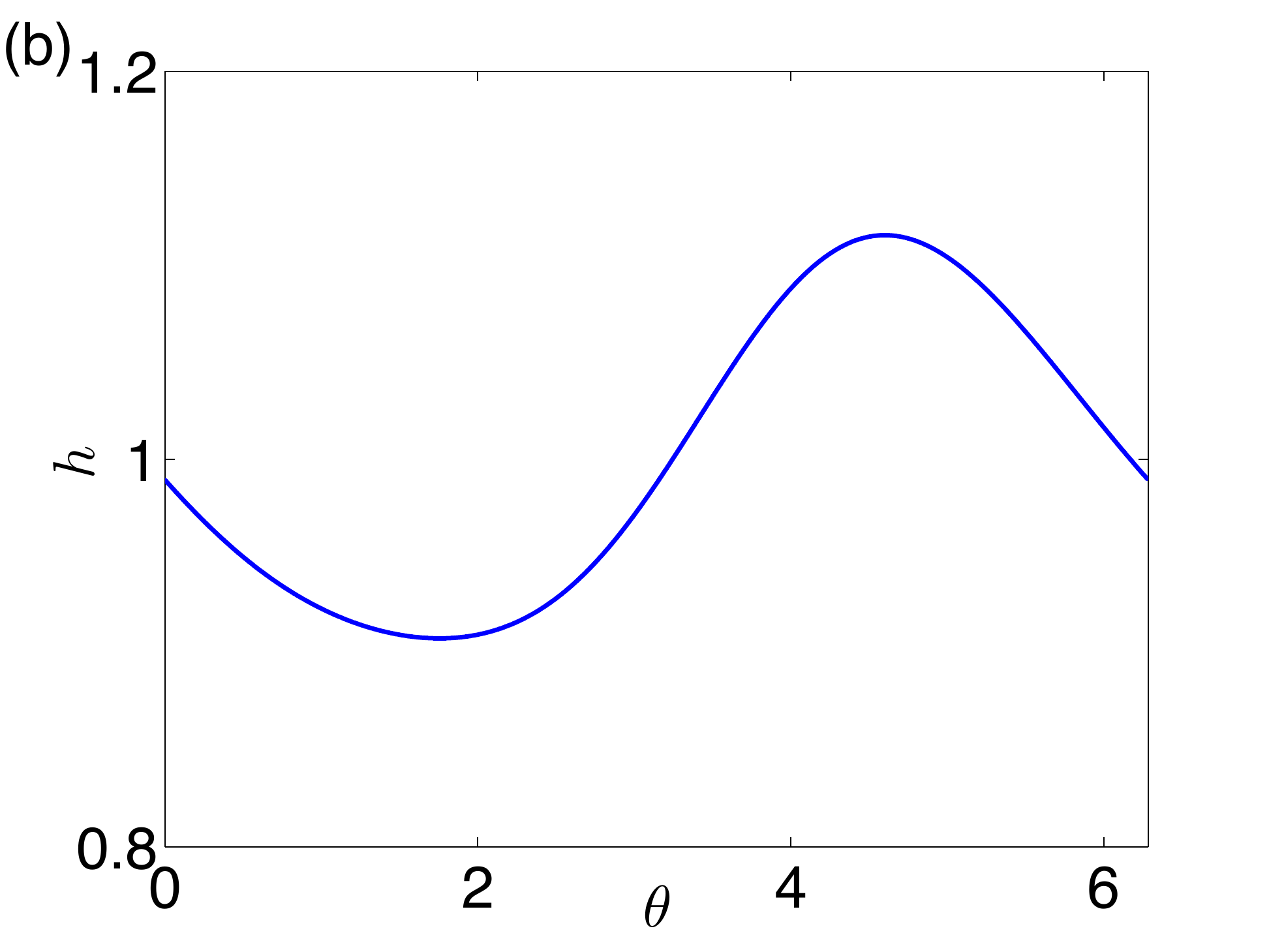}}
\caption{(Color online) An example of multistability: Two solutions that are stable 
at the identical parameter values, $\Omega=10$, $\beta_0=2$ and $B=1$. 
\bfuwe{Panel (a) shows the space-time contour plot of the stable time-periodic profile, 
while panel (b) shows the stable steady profile.}
}
\label{omega}
\end{center}
\end{figure}
\renewcommand{\baselinestretch}{1.5}

The bifurcation diagram in Fig.~\ref{Figbeta2}(a) suggest that extended ranges of 
multistability exist, i.e., depending on initial conditions, time simulations will show 
different long-time solutions at a given rotation number. For $1.68<\Omega<16.4$, 
the stable long-time solution may either be a small-amplitude steady profile or a 
large-amplitude time-periodic profile. For $\Omega>16.4$, the small-amplitude 
steady profile is unstable and only the large-amplitude time-periodic profile 
remains as an attractor. Figure~\ref{omega} shows as an example the two stable 
long-time states for $\Omega=10$. The large-amplitude time-periodic profile
is shown in the contour-plot in panel (a), while the small-amplitude
steady profile is shown in panel (b). 

We also note that at each of the HB points at $\Omega\approx 0.31$ and at 
$\Omega\approx 0.54$ \bfuwe{a short branch of unstable time-periodic states} emerges that
terminates in an HC on a branch of steady profiles. 

\subsection{The case of a medium contact angle ($\beta_0 = 1$)}
\mylab{sec:timeb1}

\begin{figure}
\begin{center}
{\includegraphics[width=0.49\hsize]{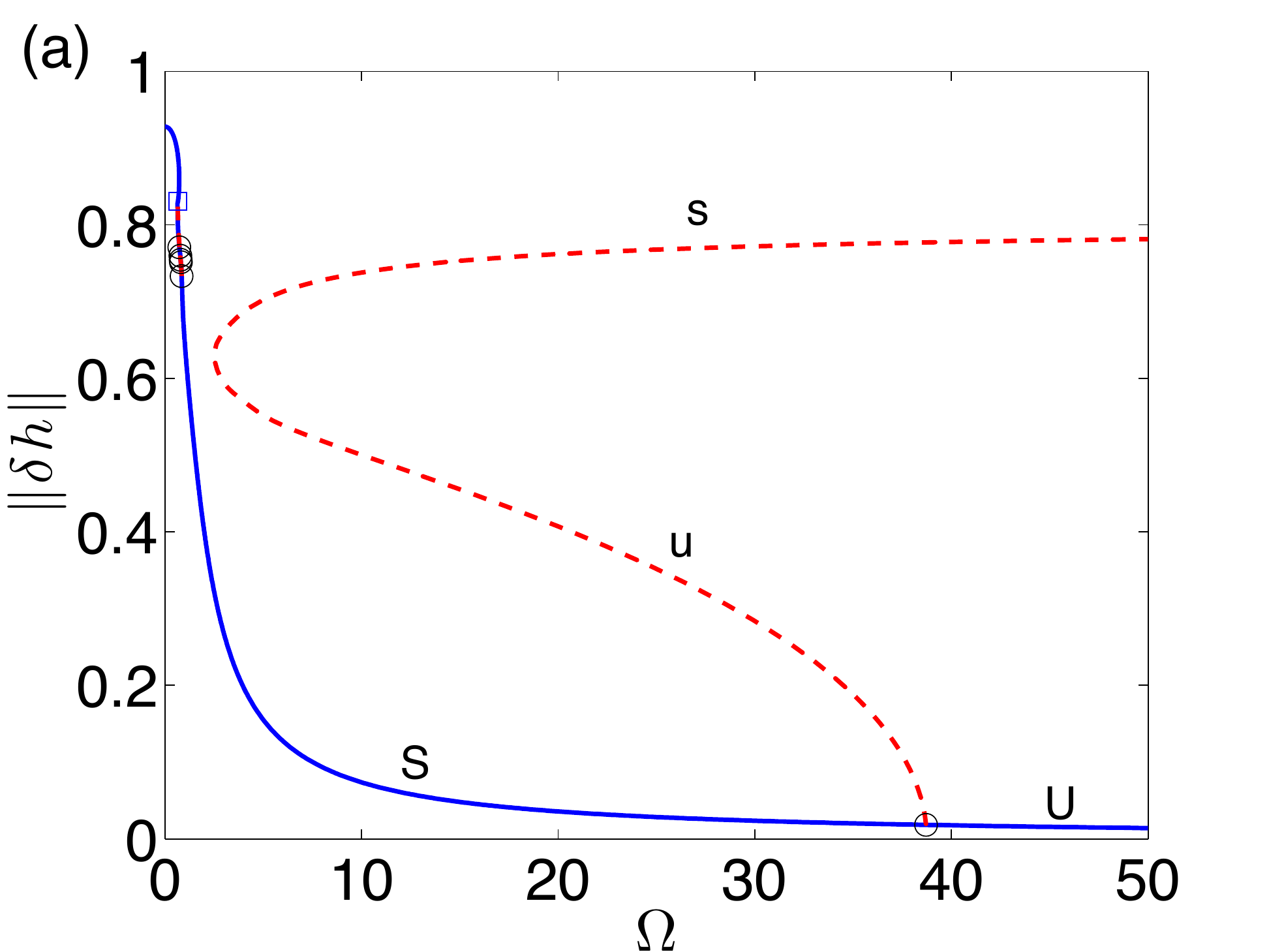}}
{\includegraphics[width=0.49\hsize]{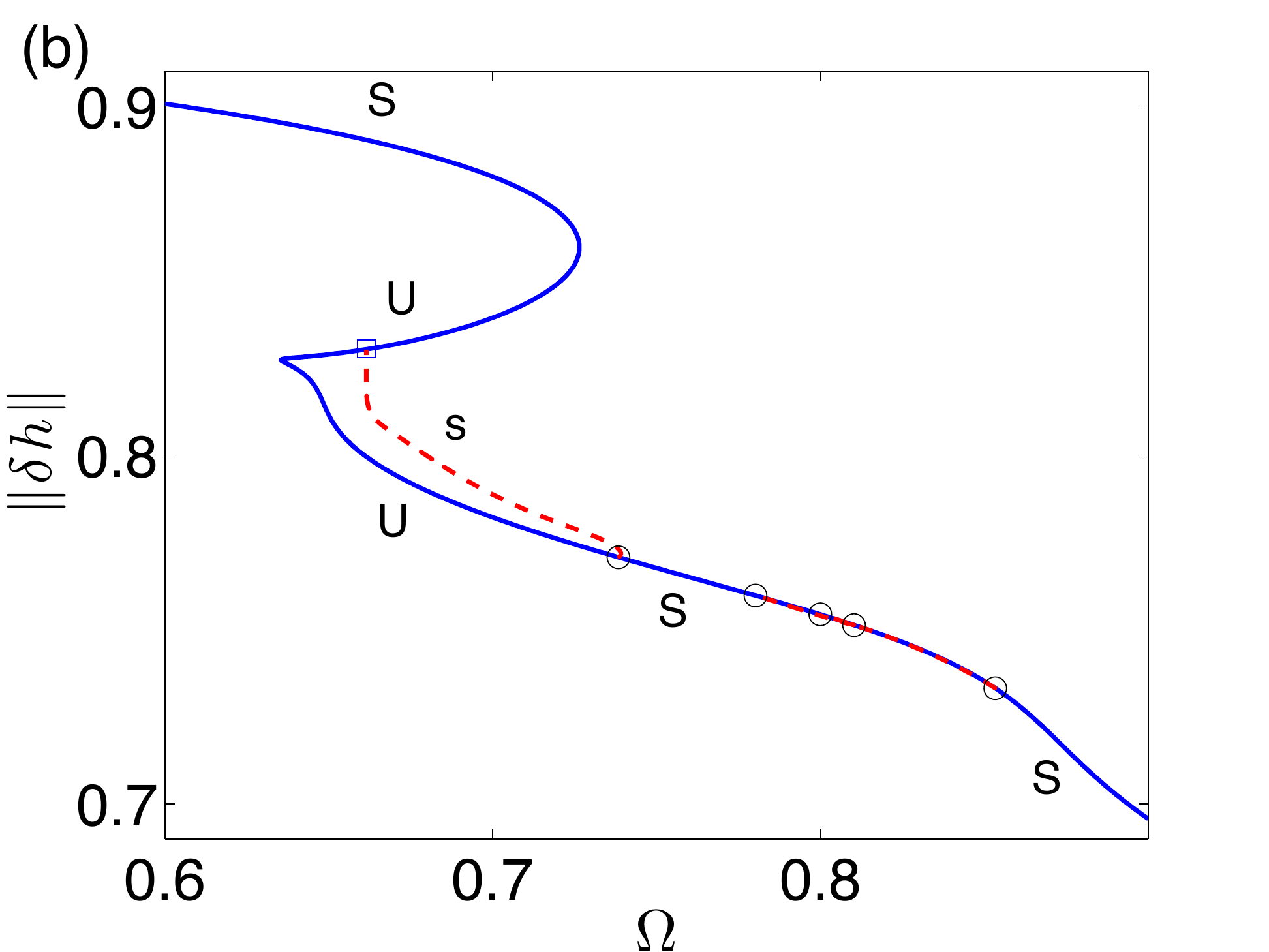}}
{\includegraphics[width=0.49\hsize]{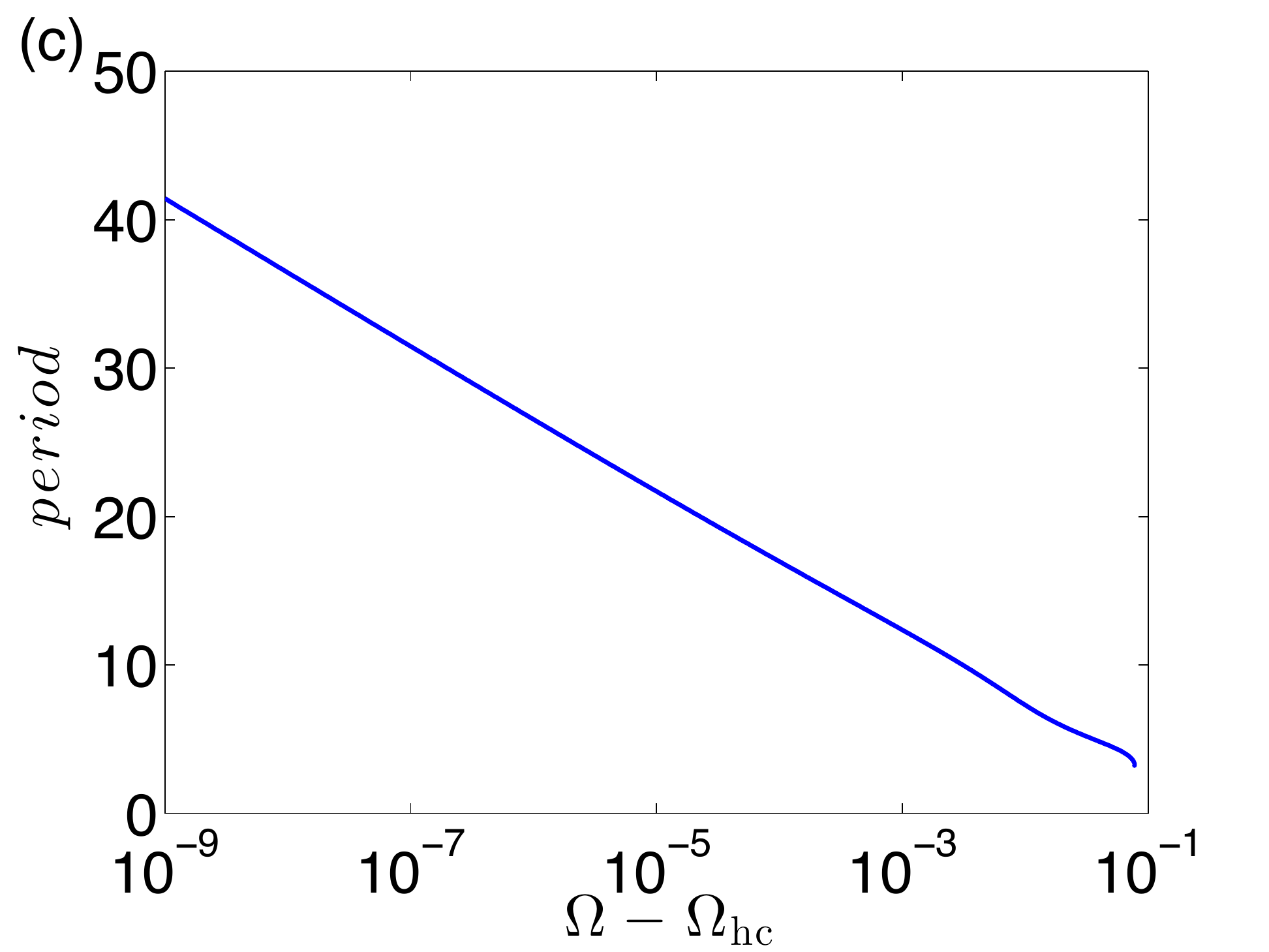}}
{\includegraphics[width=0.49\hsize]{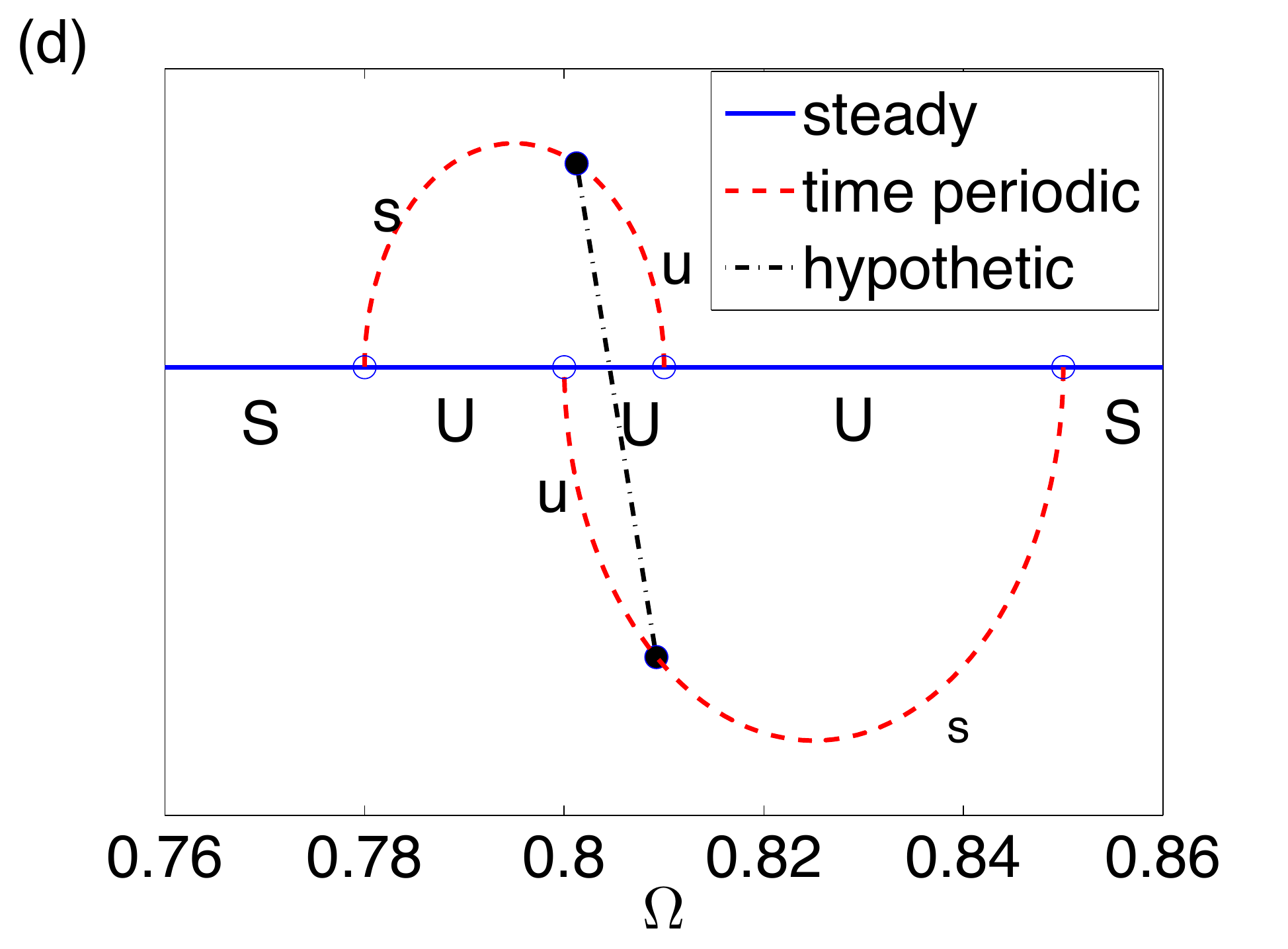}}\vspace{-0.5cm}
\caption{(Color online) 
(a) The complete bifurcation diagram for $\beta_0=1$ and $B=1$ giving the solution 
measure \bfuwe{$\|\delta h\|$} in dependence on the rotation number $\Omega$. It includes the 
steady and time-periodic states.  
(b) A zoom of panel (a) into the small-$\Omega$ region where most of the 
bifurcation structure is located. 
(c) A plot of the period over the logarithm of $\Omega-\Omega_\mathrm{hc}$ for the 
time-periodic branch originating \bfuwe{at $\Omega\approx 0.74$ in panel~(b).} 
(d) A schematic representation showing the connection of the four HBs at 
$\Omega\approx 0.78$, $0.80$, $0.81$ and $0.85$ in panel (b). The HBs are 
indicated by circles. There exist two time-periodic branches, one of them 
connects HBs at $0.78$ and $0.81$, and the other one connects HBs at $0.80$ 
and $0.85$. In addition, on each of the time-periodic branches there exist 
torus bifurcations that are indicated by filled circles. 
We expect that there exists a branch of quasi-periodic solutions that connects the 
two torus bifurcations. This \bfuwe{hypothetic branch is schematically represented by the black 
dot-dashed line, but has not been actually computed while all the other
branches and all the bifurcations have been computed.}
}
\label{Figbeta1}
\end{center}
\end{figure}

We next investigate the case of a ``medium" contact angle ($\beta_0=1$). 
The complete bifurcation diagram is presented in Fig.~\ref{Figbeta1}(a). It is found 
that there exist six HBs and one HC. The rightmost HB is at $\Omega\approx 38.7$ and 
the other five HBs are clustered around $\Omega\approx0.8$. 
Figure~\ref{Figbeta1}(b) shows a zoom of this region. 
Similar to the diagram for $\beta_0=2$, steady profiles exist for all rotation 
numbers. At the HB at $\Omega\approx38.7$, a branch of unstable time-periodic 
solutions emerges subcritically. As can be seen in Fig.~\ref{Figbeta1}(a), the solution 
measure of the time-periodic solution increases as $\Omega$ decreases and the 
branch turns back towards larger $\Omega$ at a saddle-node bifurcation (at 
$\Omega \approx 2.5$ ) where its stability changes. As $\Omega$ increases again, 
the solution measure of this stable solution increases as well. In the limit 
$\Omega\rightarrow\infty$, the solution is equivalent to a drop solution at zero Bond 
number that co-rotates with the cylinder.

\renewcommand{\baselinestretch}{1}
\renewcommand{\baselinestretch}{1.5}

Next, we analyze the solutions in the range of smaller $\Omega$. 
For $0\le \Omega< 0.64$, there exits 
only one solution for each rotation number that is a stable steady-state solution. 
As $\Omega$ increases, the solution measure of this solution decreases until it 
reaches an SN at $\Omega\approx 0.73$ where the branch becomes unstable and turns 
back towards smaller $\Omega$. See, e.g., Fig.~\ref{Figbeta1}(b). 
Following the branch with decreasing 
$\| \delta h\|$, one passes another SN at $\Omega\approx 0.64$, a 
second eigenmode becomes unstable and the branch turns again towards larger 
$\Omega$. Continuing towards larger $\Omega$, one crosses $5$ HBs (at 
$\Omega\approx 0.74$, $0.78$, $0.80$, $0.81$ and $0.85$). In this region, the 
steady-state solution is stable only for $0.74<\Omega< 0.78$ and for 
$0.85<\Omega< 38.4$. At each HB, a branch of time-periodic 
solutions emerges. The one that starts supercritically at the first HB at 
$\Omega\approx 0.74$ is stable and reconnects at 
$\Omega_\mathrm{hc}\approx 0.66$ in a homoclinic global bifurcation to the unstable part of 
the steady-state branch between the two SNs. This classification as a 
homoclinic bifurcation is supported by Fig.~\ref{Figbeta1}(c), where 
$\Omega-\Omega_\mathrm{hc}$ is plotted against the temporal period in a linear-log plot. The 
figure shows that the period diverges logarithmically as $\Omega-\Omega_\mathrm{hc}$ 
approaches zero. This is the signature of a homoclinic bifurcation~\cite{Strogatz1994} 
and is also a common scenario for systems where depinning 
occurs~\cite{Kopf2012,Kopf2014,Pailha2012,Herde2012}. 

The HBs at $\Omega\approx 0.78$ and $0.81$ are connected by one time-periodic 
branch and the HBs at $\Omega\approx 0.80$ and $0.85$ are connected by another 
time-periodic branch that in part overlaps with the first one. But since the solution 
measures $\|\delta h\|$ of these time-periodic solutions and the steady-state 
solutions are very similar to each other for the values of $\Omega$ in this range, 
the branches are almost indistinguishable in Fig.~\ref{Figbeta1}(b). To illustrate 
the connections, a schematic representation of this region is shown in 
Fig.~\ref{Figbeta1}(d) \bfuwe{where also the stabilities are indicated}. In addition, on 
the two time-periodic branches, we detect 
that the stability of the time-periodic solutions changes, which is an indication of 
torus bifurcations, at $\Omega\approx 0.802$ and $0.805$, respectively. The 
fact that the time-periodic solutions undergo torus bifurcations at these values of 
$\Omega$ is supported by the observation that at these points there exist pairs 
of complex-conjugate Floquet multipliers crossing the unit circle. Namely, at 
$\Omega\approx0.802$ and $0.805$, we find that the multipliers
$0.545 \pm 0.838\, i$ and $0.155 \pm 0.988 i$, respectively, cross the unit circle. 
We expect that there exists a branch of quasi-periodic solutions that connects the 
two torus bifurcations. However, with the current continuation tools we are not able 
to continue solutions on this solution branch. We, therefore, draw schematically a (black) dot-dashed 
 line in Fig.~\ref{Figbeta1}(d) to connect the two torus bifurcations to indicate 
the existence of such a stable solution branch, that has not actually
been computed \bfuwe{by continuation.}

\begin{figure}
\begin{center}
{\includegraphics[width=0.49\hsize]{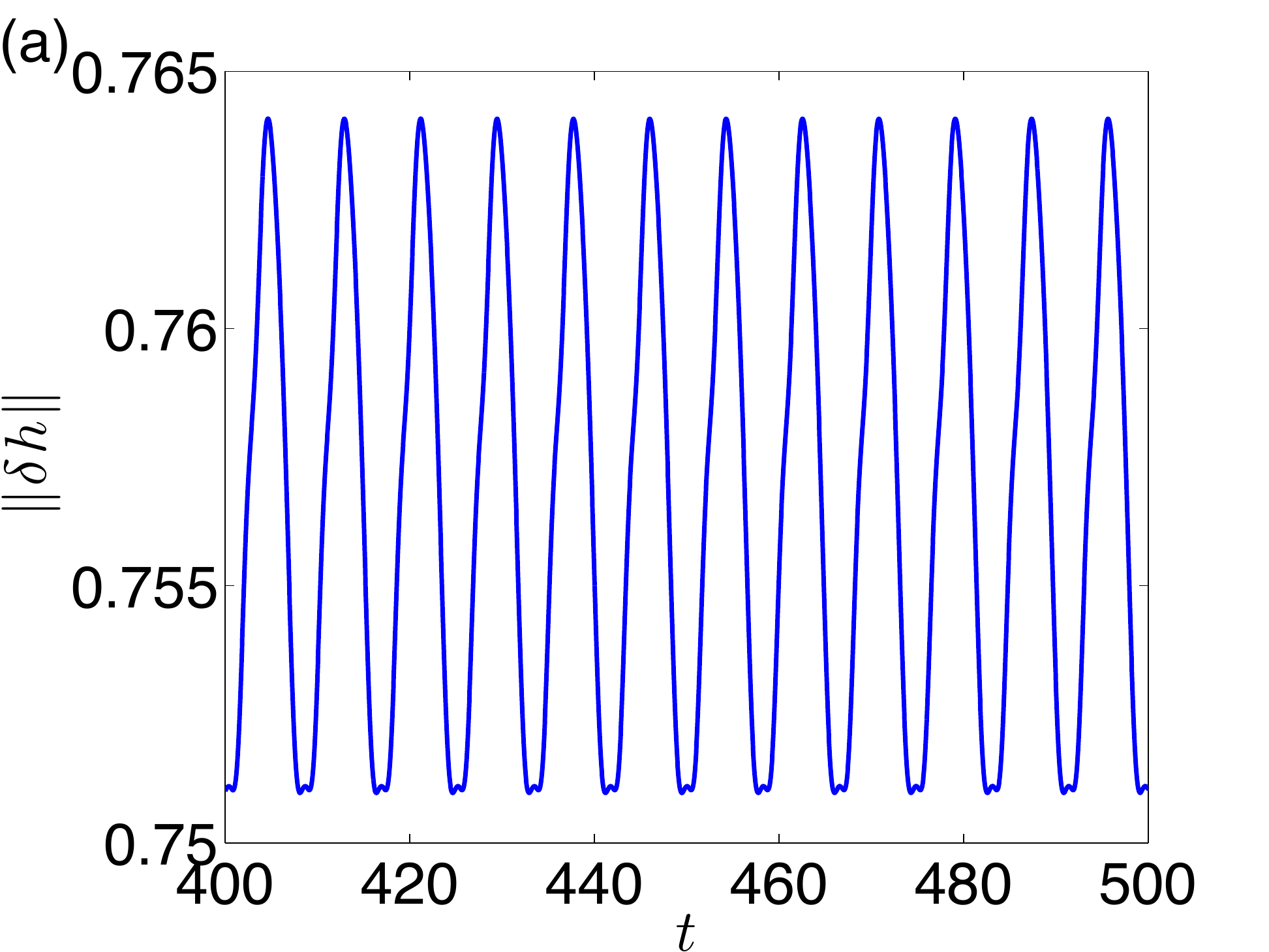}}
{\includegraphics[width=0.49\hsize]{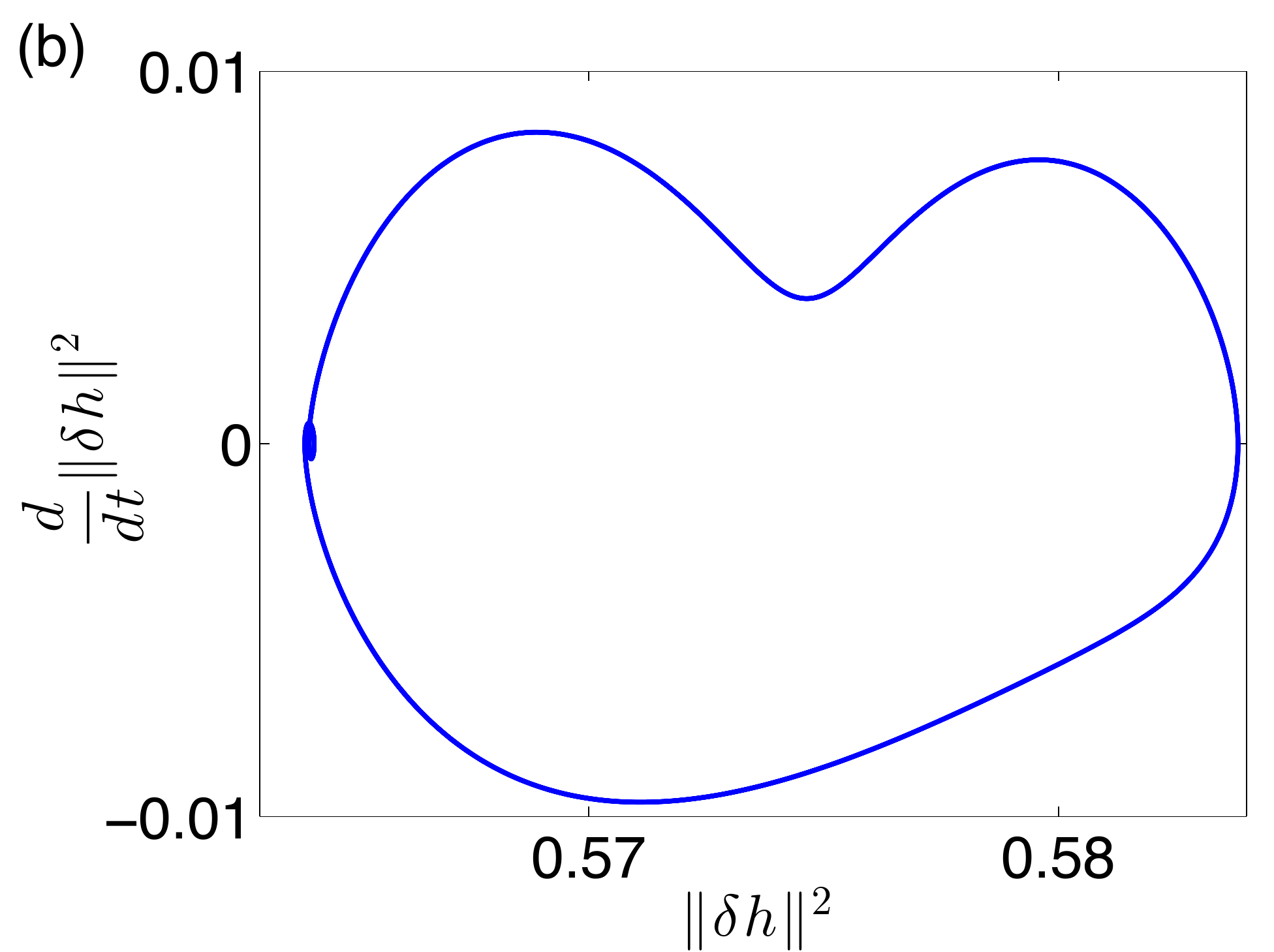}}
{\includegraphics[width=0.49\hsize]{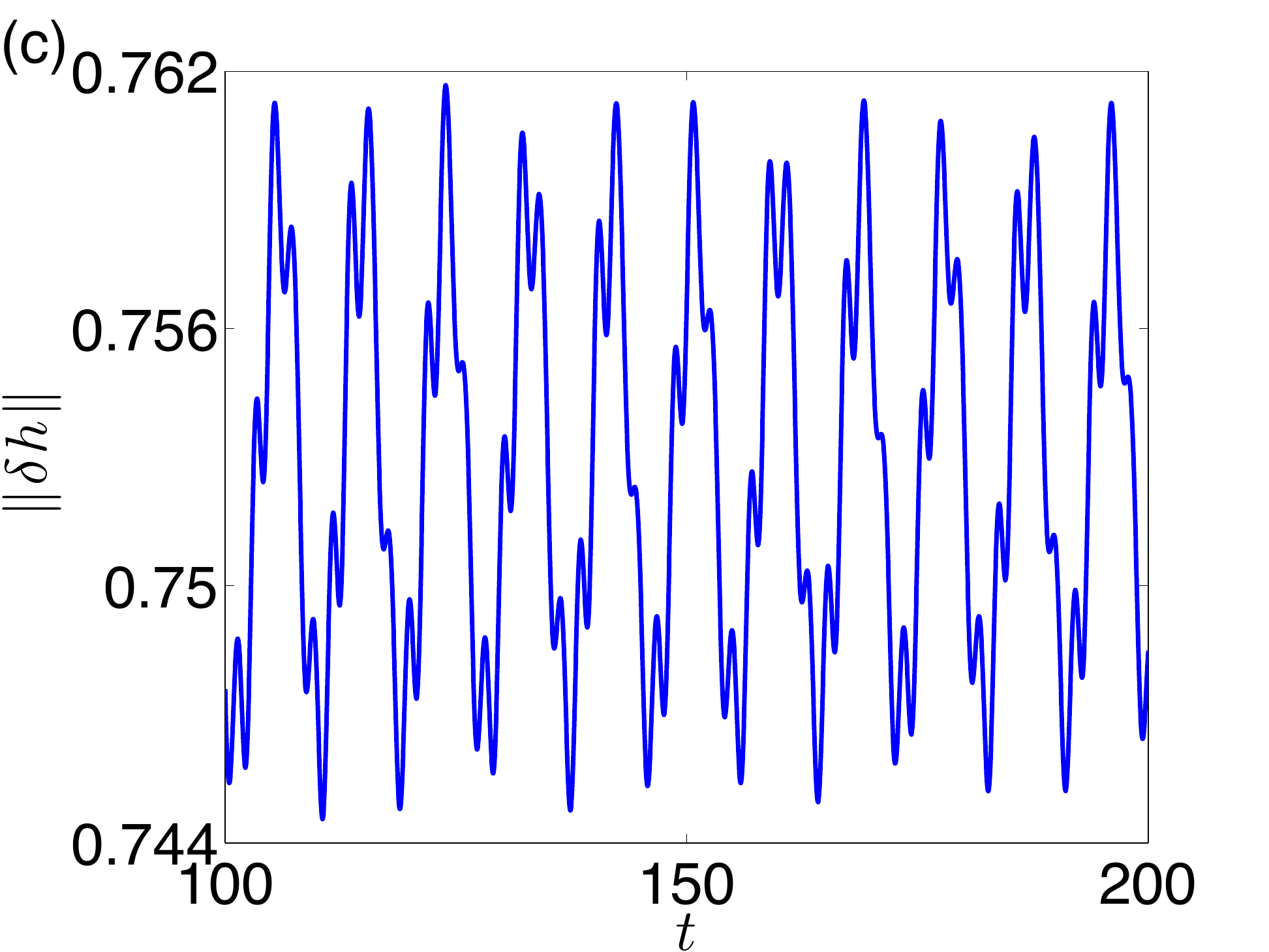}}
{\includegraphics[width=0.49\hsize]{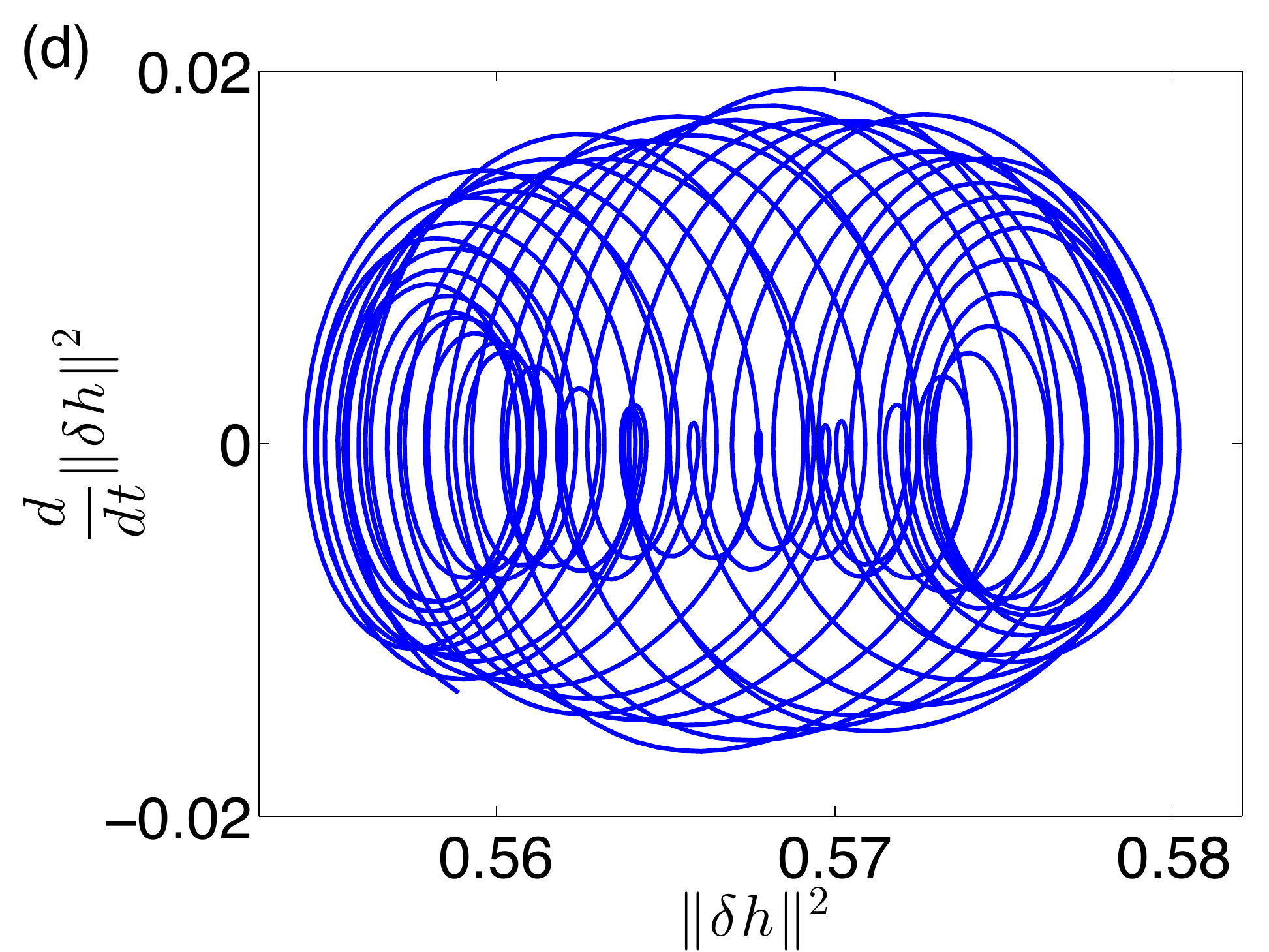}}
\caption{(Color online) 
Results of the numerical time integration of Eq.~(\ref{eq:timeevolution}) for 
$\beta_0=1$, $B=1$ and $\Omega=0.79$ (panels (a) and (b)) and 
$\Omega=0.804$ (panels (c) and (d)). Panels (a) and (c) show the time 
evolution of the $L^2$-norm,  $\|\delta h\|$, for $\Omega=0.79$ and $0.804$, respectively, 
and panels (b) and (d) show the corresponding curves in the phase plane 
$(\|\delta h\|^2,d(\|\delta h\|^2)/dt)$.
}
\label{Figbeta1more}
\end{center}
\end{figure}

With the information presented above, we conclude that the time-periodic solutions 
are stable in the ranges $0.78< \Omega< 0.802$ and $0.805< \Omega< 0.85$, 
which is also verified using numerical time integration of Eq.~(\ref{eq:timeevolution}). 
We choose $\Omega=0.79$ as an example. The initial condition is the state of 
homogeneous coating, $h(x,0)=1$, and the \bfuwe{$L^2$-norm of the solution, $\|\delta h\|$,}  is extracted in the 
range $400<t<500$, to study the long-time behavior. Figure~\ref{Figbeta1more}(a) 
shows \bfuwe{the $L^2$-norm of the solution,} $\|\delta h\|$, versus $t$ and Figure~\ref{Figbeta1more}(b) shows 
$\frac{d}{dt}(\|\delta h\|^2)$ versus $\|\delta h\|^2$. It clearly indicates that the solution 
evolves to a time-periodic solution.

However, in the range $0.802<\Omega<0.805$, \bfuwe{i.e., between the
  two torus bifurcations, no stable steady profiles or
  single-frequency time-periodic solutions exist. Therefore, it is
  likely that the attractor corresponds either to a double-frequency
  time-periodic state or a multi-frequency solution (if further
  bifurcations of the double-frequency time-periodic state occur in
  this very small $\Omega$ range). With identical initial conditions
  as before, at $\Omega=0.804$ a numerical time simulation gives the
  trajectory presented in Figs.~\ref{Figbeta1more}(c) and (d) (the
  $L^2$-norm of the solution is extracted in the range
  $100<t<200$). The results indicate a double-frequency time-periodic
  state consistent with the hypothesis formulated above.  Therefore,
  we expect that the hypothetical (black) dot-dashed branch in
  Fig.~\ref{Figbeta1}(d) joining the two torus bifurcations is a
  quasi-periodic solution branch that is at least in part of this
  $\Omega$-range linearly stable.}

\renewcommand{\baselinestretch}{1}
\renewcommand{\baselinestretch}{1.5}

\subsection{Transition from a large ($\beta_0 = 2$) to a medium ($\beta_0= 1$) 
contact angle}
\mylab{sec:timetransition}

After having analyzed the diagrams for large and medium contact angles, we next 
study the transitions in between. Comparing the diagrams for $\beta_0=2$ in 
Fig.~\ref{Figbeta2}(a) and $\beta_0=1$ in Fig.~\ref{Figbeta1}(a), one notices that 
the two time-periodic branches for $\beta_0=2$, where one is the unstable branch 
connecting the two HBs and the other is the stable branch that emerges in the 
SNIPER bifurcation, \bfuwe{have at  $\beta_0=1$ somehow merged into one branch that emanates from the 
HB at $\Omega\approx 38.7$.} One should also note that, 
as a consequence, when a \bfuwe{putative hydrodynamic experiment is
performed, the observed sequence of behaviors is}
very different when $\Omega$ is continuously increased for $\beta_0=2$ and 
for $\beta_0=1$. For liquids with the large contact angle ($\beta_0=2$), a pendent droplet 
directly evolves into a large-amplitude time-periodic state that represents a droplet co-rotating 
with the cylinder, while for liquids with the medium contact angle ($\beta_0=1$), a 
pendent droplet first changes into a steady uniform coating film before eventually 
changing into a large-amplitude time-periodic  wave for sufficiently large values of 
$\Omega$.

In what follows, we will see that the transitions of branches and bifurcations are rather 
complicated and involve several steps. We focus on the main transitions and show 
how the bifurcation diagram changes when $\beta_0$ is varied and, in particular, how 
the SNIPER bifurcation disappears. 

\begin{figure}
\begin{center}
{\includegraphics[width=0.49\hsize]{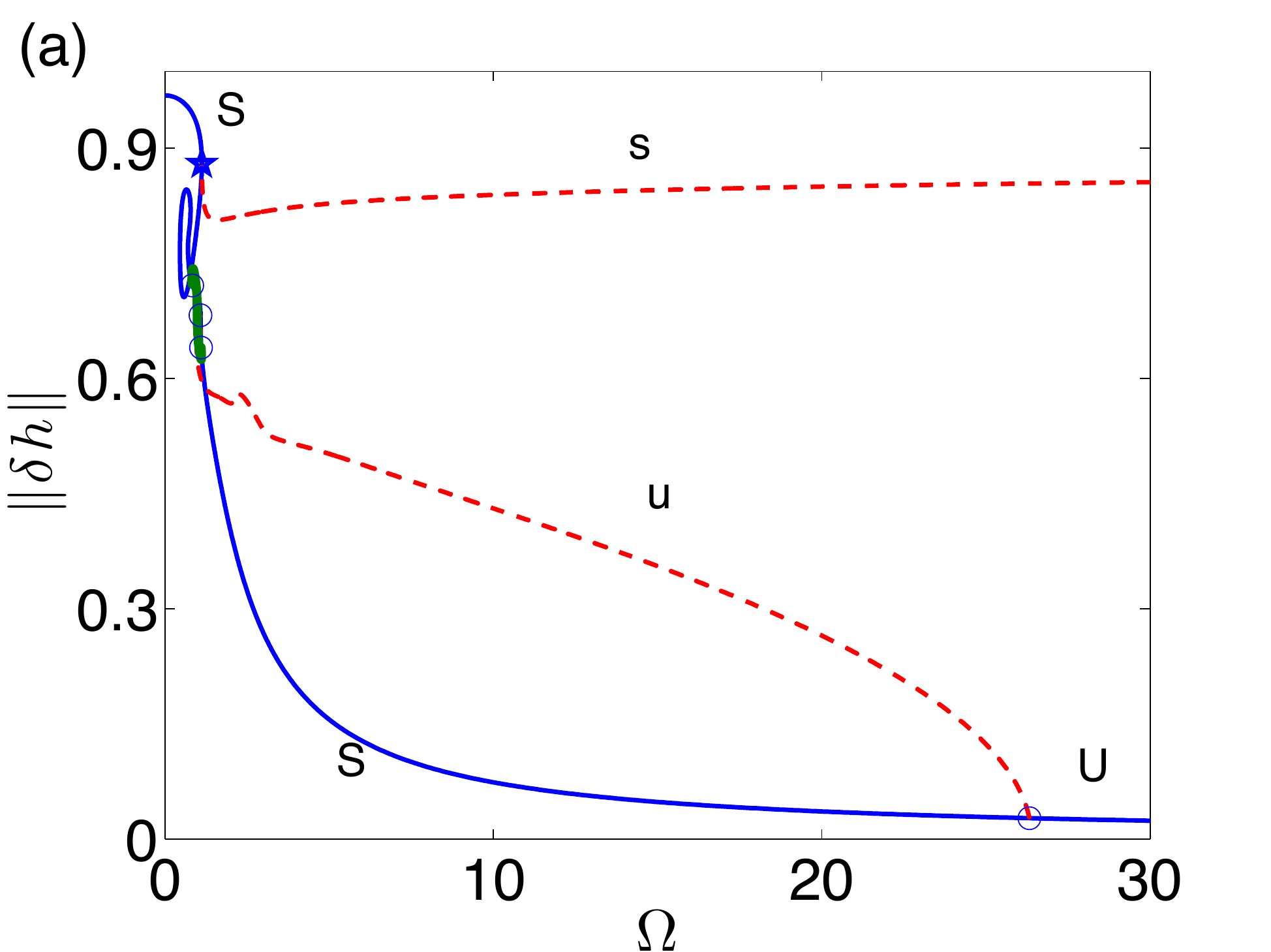}}
{\includegraphics[width=0.49\hsize]{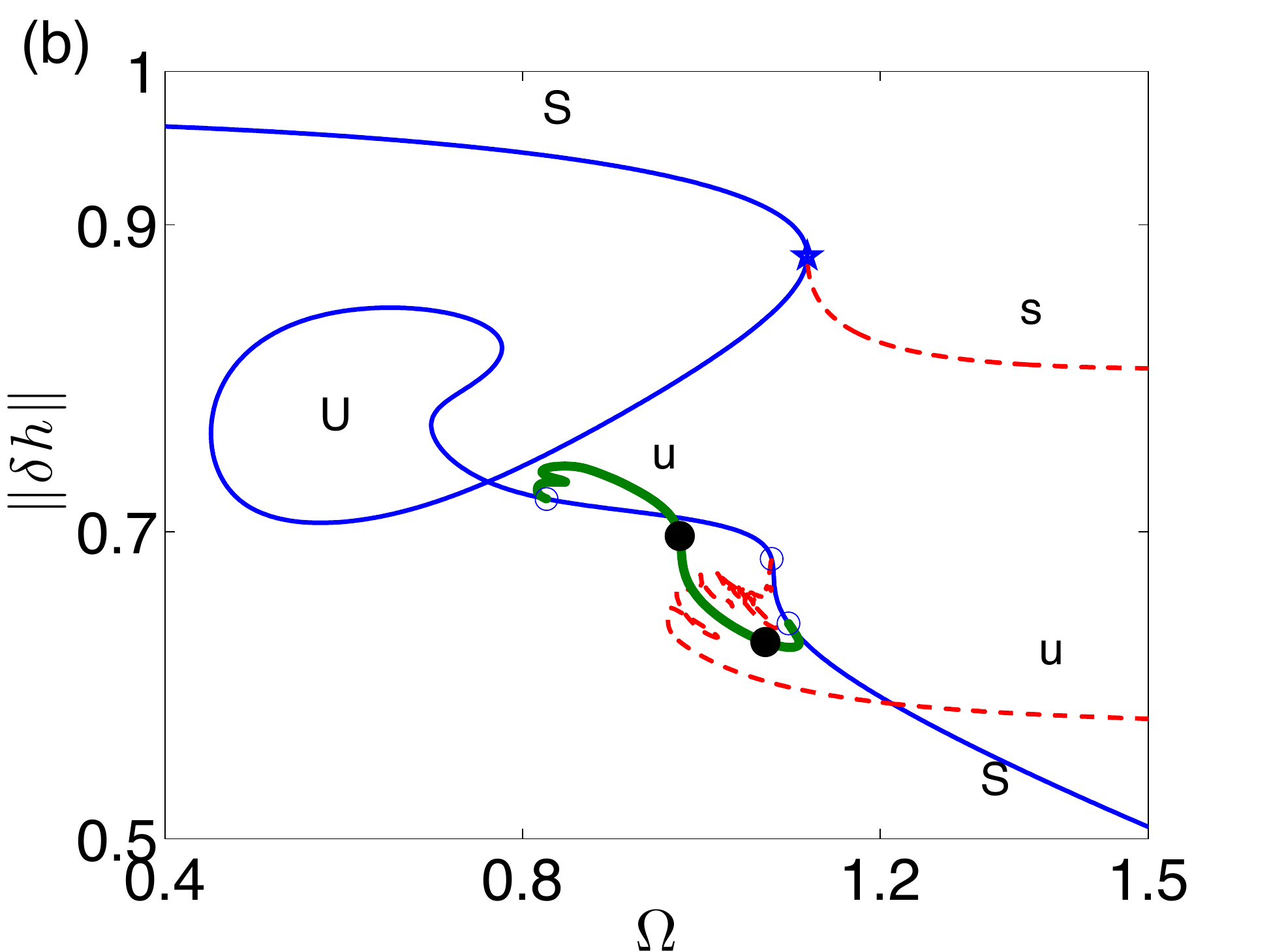}}
\caption{(Color online) 
(a) The complete bifurcation diagram for $\beta_0=1.4$ and $B=1$ giving the solution 
measure \bfuwe{$\|\delta h\|$} in dependence \bfuwe{on} the rotation number $\Omega$. The steady-state branch 
is shown by the (blue) solid line. \bfuwe{The various time-periodic branches
  are shown by the (red) dashed and (green) thick solid  lines.} 
Panel (b) shows a zoom of panel (a) into the small-$\Omega$ region \bfuwe{($0.4\leq \Omega\leq 1.5$)}
where most of the bifurcation structure is located. \bfuwe{The filled black circles on the (green) thick solid  line of time-periodic solutions indicate torus bifurcations, and the solutions on the part of the branch between these points are stable.}
}
\label{Figbeta14}
\end{center}
\end{figure}

\begin{figure}
\begin{center}
{\includegraphics[width=0.49\hsize]{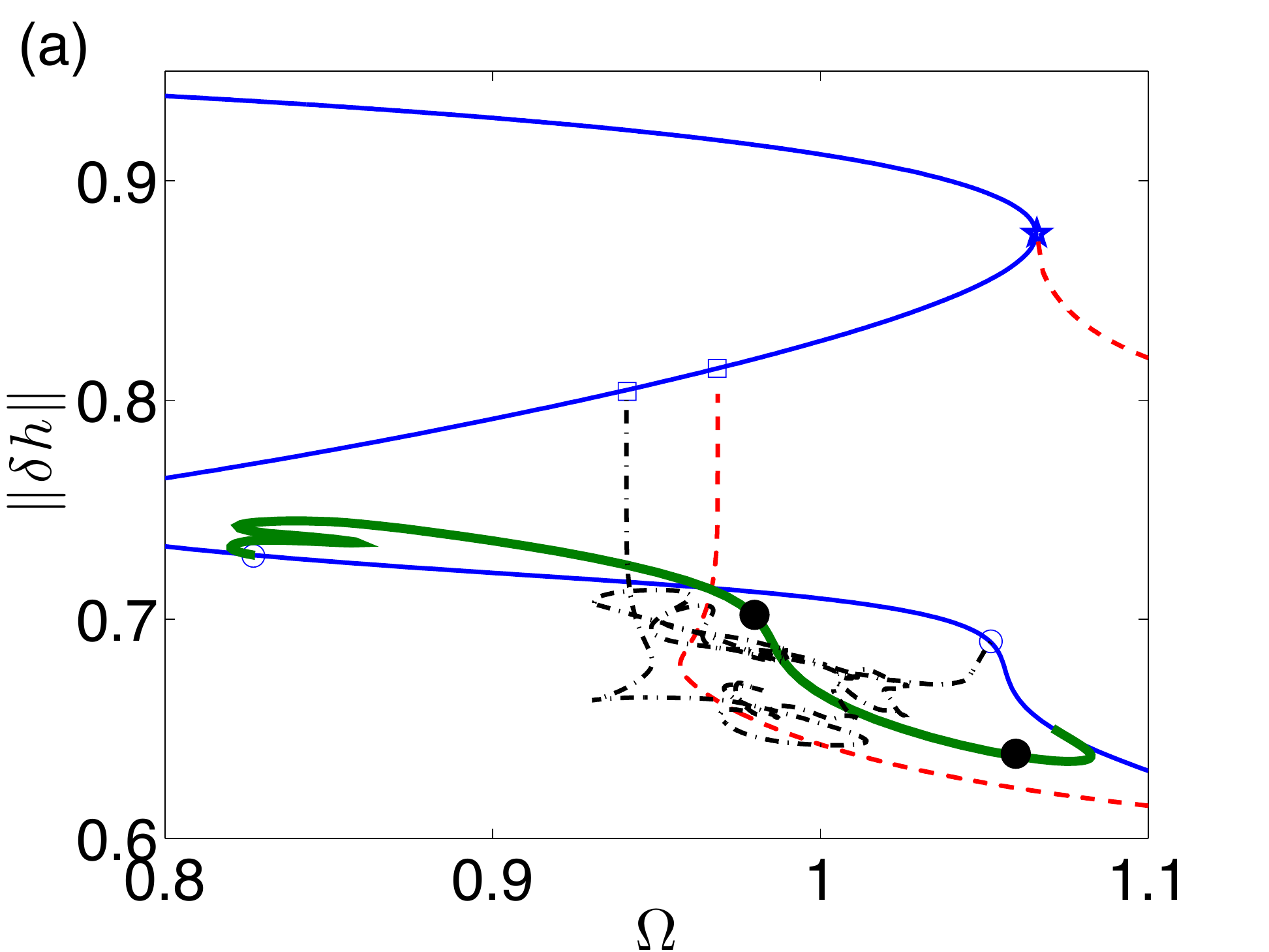}}
{\includegraphics[width=0.49\hsize]{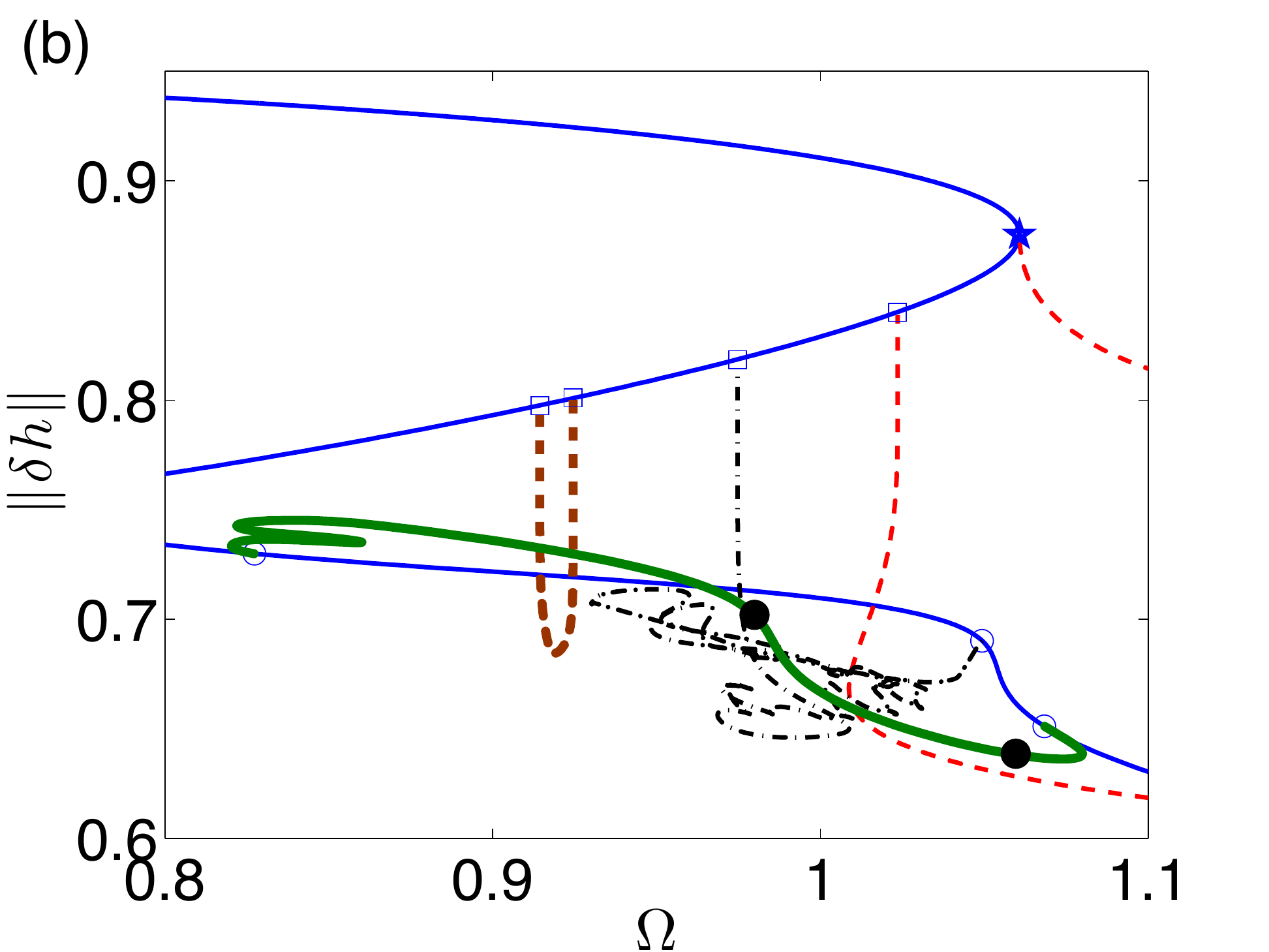}}
{\includegraphics[width=0.49\hsize]{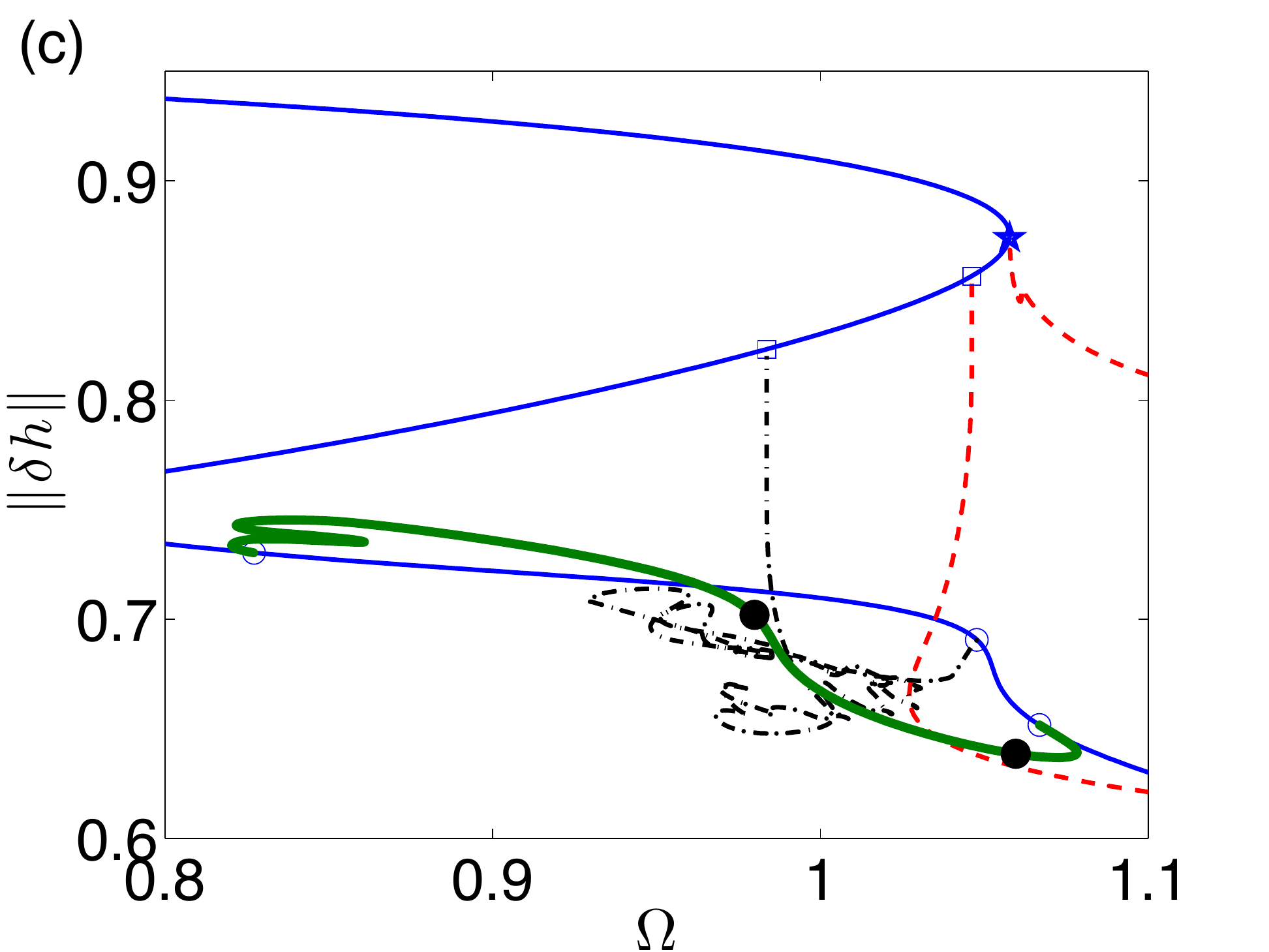}}
{\includegraphics[width=0.49\hsize]{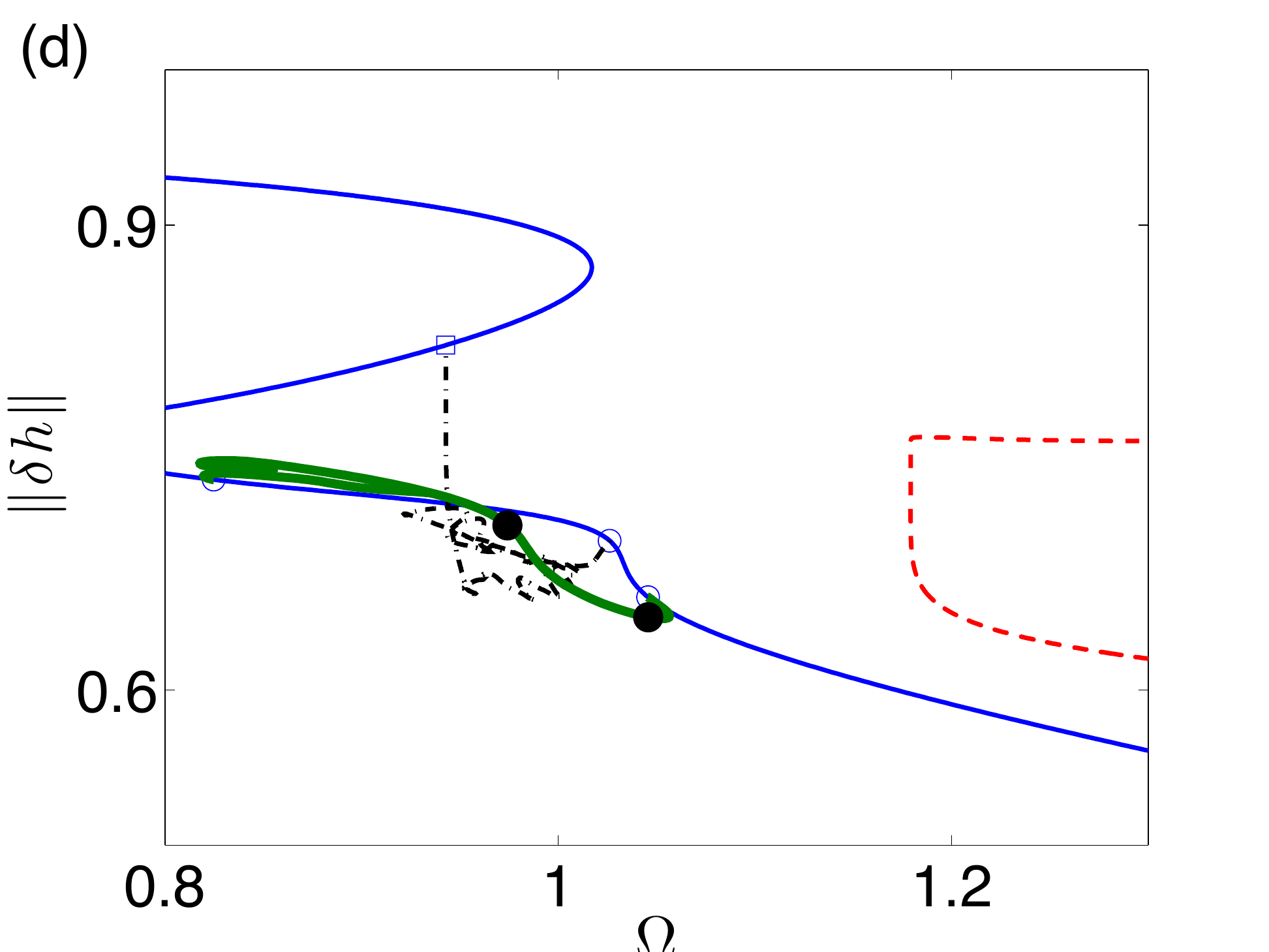}}
\caption{(Color online) 
The zoom-ins of the bifurcation diagrams in the small-$\Omega$ regions 
showing how the branches/bifurcations evolve with decreasing $\beta_0$. 
(a) $\beta_0=1.348$; (b) $\beta_0=1.343$; (c) $\beta_0=1.34$; (d) $\beta_0=1.3$. 
\bfuwe{The steady-state branch 
is shown by the (blue) solid line. The various time-periodic branches are
shown by the (red) dashed, (green) thick solid, (brown) thick dashed
and (black) dot-dashed lines. The filled black circles on the (green) thick solid  lines of time-periodic solutions indicate torus bifurcations. As the figure aims at illustrating the various reconnections, the branch stabilities are not indicated. They can in part be deduced from Figs.~\ref{Figbeta14} and \ref{figbeta150}. 
}
}
\label{Figbeta13}
\end{center}
\end{figure}

The bifurcation diagram for $\beta_0=1.4$ is shown in Fig.~\ref{Figbeta14}, and
Fig.~\ref{Figbeta13} shows zooms into the small-$\Omega$ regions of the 
bifurcation diagrams for $\beta_0=1.348$, $1.343$, $1.34$ and $1.3$ illustrating how 
the SNIPER bifurcation disappears as $\beta_0$ decreases. The steady-state branch 
is shown as (blue) solid line. \bfuwe{The various time-periodic branches are shown as
(red) dashed, (green) thick solid, (brown) thick dashed and (black) dot-dashed 
lines.}

\renewcommand{\baselinestretch}{1}
\renewcommand{\baselinestretch}{1.5}

Comparing the cases of $\beta_0=2$ in Fig.~\ref{Figbeta2} and of $\beta_0=1.4$ in 
Fig.~\ref{Figbeta14}, we notice that in both cases the time-periodic branch emerges from the 
rightmost HB (at $\Omega\approx16.4$ and $26.3$ for $\beta_0=2$ and $1.4$, respectively) 
and terminates at another HB. However, this structure changes at smaller $\beta_0$. As shown 
in Fig.~\ref{Figbeta13}(a) for $\beta_0=1.348$, there appear two additional HCs at 
$\Omega \approx 0.94$ and $0.97$. The time-periodic branch emerging from the rightmost HB 
now terminates at the HC at $\Omega\approx0.97$ and, correspondingly, the HB at $\Omega=1.05$ 
terminates at the HC at $\Omega\approx0.94$. We note that all the remaining structures are the same 
between \bfuwe{$\beta_0=1.4$} and $1.348$. There is one steady-state branch and two more 
time-periodic branches, where one emerges in a SNIPER bifurcation and the other one connects two HBs.

Going further down in $\beta_0$, at $\beta_0=1.343$ there appear two more HCs 
(at $\Omega\approx 0.915$ and $0.925$) that are connected by a new time-periodic branch, 
\bfuwe{(brown) thick dashed line} in Fig.~\ref{Figbeta13}(b). This new branch has again disappeared at 
$\beta=1.34$ (Fig.~\ref{Figbeta13}(c)). Then, at $\beta_0=1.3$ as shown in Fig.~\ref{Figbeta13}(d), 
the time-periodic branch emerging from the rightmost HB 
and the time-periodic branch emerging in a SNIPER bifurcation now merge into one single 
branch. As a consequence, the SNIPER bifurcation and one HC bifurcation disappear.

\renewcommand{\baselinestretch}{1}
\renewcommand{\baselinestretch}{1.5}

\renewcommand{\baselinestretch}{1}
\begin{figure}
\begin{center}
\includegraphics[width=0.7\hsize]{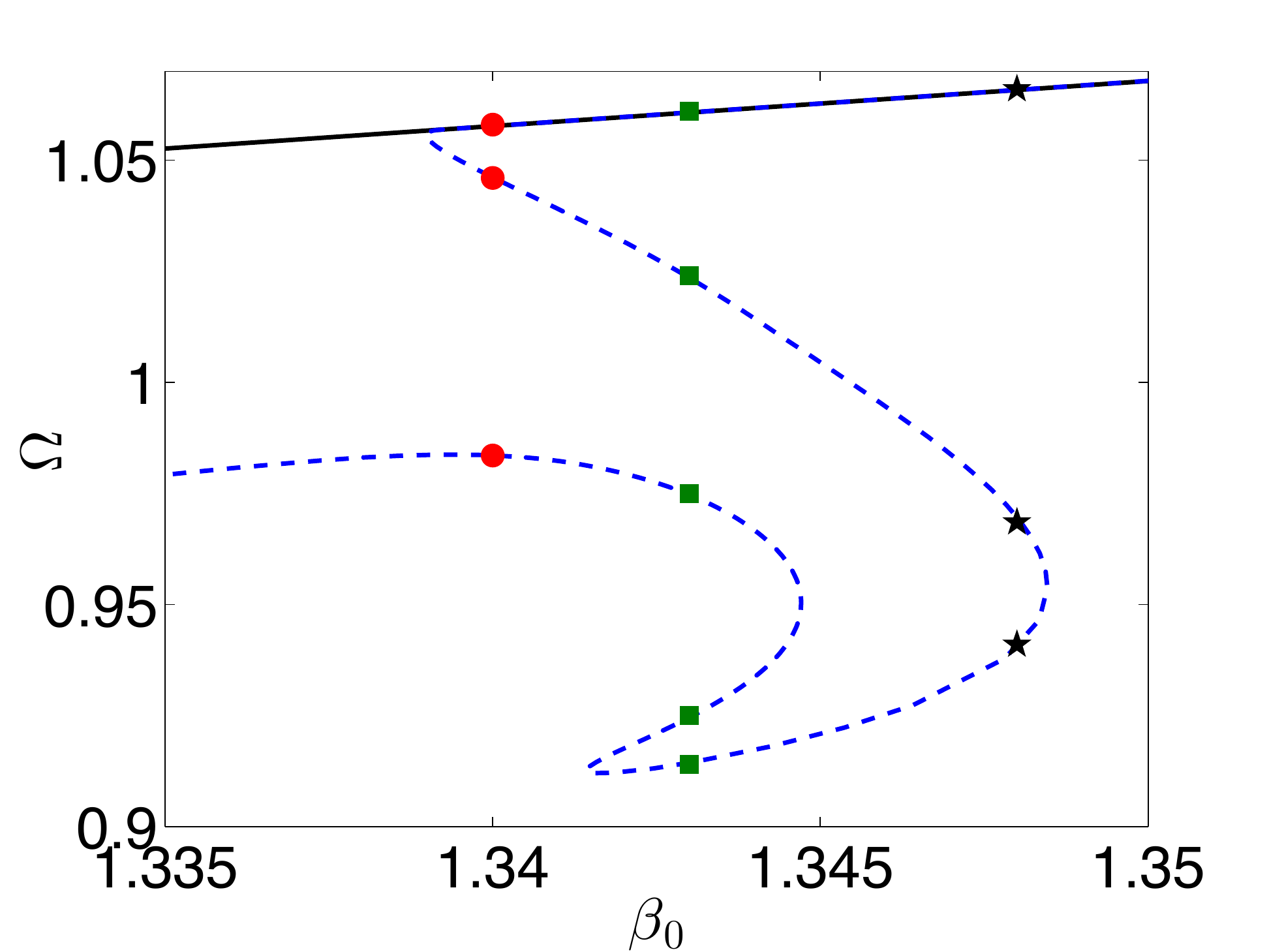}
\caption{(Color online) Numerical two-parameter continuation in the
  plane spanned by the contact angle, $\beta_0$, and the rotation number,
  $\Omega$.  Shown are (i) the loci of the saddle-node bifurcation that at
  $\beta_0=2$ `hosts' the SNIPER bifurcation by the (black) solid line and (ii)
  the loci of time-periodic solutions very close to the global
  bifurcation by the (blue) dashed  line.
  Here, the temporal period is fixed to equal $1000$.  
   (Red) circles, (green) squares and (black) stars correspond to period-$1000$
  time-periodic solutions at $\beta_0=1.34$, $1.343$ and $1.348$,
  respectively.}
\label{Figsaddlenode}
\end{center}
\end{figure}
\renewcommand{\baselinestretch}{1.5}

To analyze in more detail how the branches/bifurcations appear and disappear, we 
perform continuations on the SNIPER and HC bifurcations to see how they evolve 
as $\beta_0$ varies. First, we focus on the SN that for $\beta_0=2$ is located 
at $\Omega\approx 1.68$ and perform a two-parameter fold-continuation in $\Omega$ 
and $\beta_0$ for decreasing $\beta_0$. The results is the (black) solid line in 
Fig.~\ref{Figsaddlenode}. Each of its points represents the location of an SN at the 
corresponding $\beta_0$. Next, we take a time-periodic solution for $\beta_0=2$ 
that is very close to the SNIPER bifurcation and perform a similar two-parameter 
continuation at a fixed temporal period (here at $1000$). The result is the (blue) dashed 
 line in Fig.~\ref{Figsaddlenode} that itself folds back and forth several times. 
We note that the upper part of the line for $\beta_0 \ge 1.338$ (about 
$\Omega\approx 1.06$) actually coincides with the SN branch indicating that we are 
indeed tracking the SNIPER bifurcation. At $\beta_0 = 1.338$ the dashed line folds 
back towards larger $\beta_0$ indicating that 
(i) for $\beta_0 >1.338$ there exist a SNIPER and a homoclinic bifurcation, 
(ii) the two global bifurcations approach each other as $\beta_0$ decreases and 
(iii) annihilate each other at (or very near to) the SN about $\beta_0\approx 1.338$ 
thereby joining the two branches of time-periodic solutions. Or seeing the process 
as occurring for increasing $\beta_0$: The time-periodic solution branch that emerges 
from the HB at $\Omega\approx38.7$ in Fig.~\ref{Figbeta1}(a) approaches the branch of 
unstable steady solutions when increasing $\beta_0$, hits it finally at 
$\beta_0 \approx1.338$ at (or very close to) the SN. The time-periodic branch splits 
and the two ends are glued to the steady branch as a SNIPER and an HC. Similar 
transitions have been analyzed in the context of line formation in Langmuir-Blodgett 
transfer described via an amended Cahn-Hilliard equation~\cite{Kopf2014}. 
Another glimpse at Fig.~\ref{Figsaddlenode} indicates that this is not the full picture 
as the dashed line that represents the homoclinic bifurcation undergoes another three 
folds indicating several annihilation and creation events of pairs of homoclinic bifurcations 
and even the emergence of entire new branches. Following the dashed line further 
down to $\beta_0=1$, we find that it then corresponds to the single HC described above 
in Sec.~\ref{sec:timeb1}.

With the obtained information, we are now able to describe how the branches 
change. For $\beta_0> 1.3485$, there exists a SNIPER bifurcation and there are 
no HCs. With decreasing $\beta_0$, at $\beta_0\approx1.3485$ one of the time-periodic 
branches touches the unstable part of the steady-state branch and splits into two 
branches. Each of the branches then terminates at the steady-state branch in an HC, 
see, e.g., the case of $\beta_0=1.348$ in Fig.~\ref{Figbeta13}(a). At $\beta_0\approx 1.3447$ one of the 
time-periodic branches touches again the unstable part of the steady-state branch 
that generates two more HCs that are connected by a new time-periodic branch, 
see, e.g., the case of $\beta_0=1.343$ in Fig.~\ref{Figbeta13}(b). This new branch disappears 
at $\beta_0\approx1.3415$ when the two HCs annihilate each other. Finally, at 
$\beta_0\approx 1.338$ one of the HCs and the SNIPER annihilate each other and the two 
corresponding time-periodic branches join together, see, e.g., the case of $\beta_0=1.3$ in 
Fig.~\ref{Figbeta13}(d).

\subsection{Transition from a medium ($\beta_0= 1$) to a small ($\beta_0\ll 1$) contact angle}
\mylab{sec:timetransitionsmall}

\begin{figure}
\begin{center}
{\includegraphics[width=0.49\hsize]{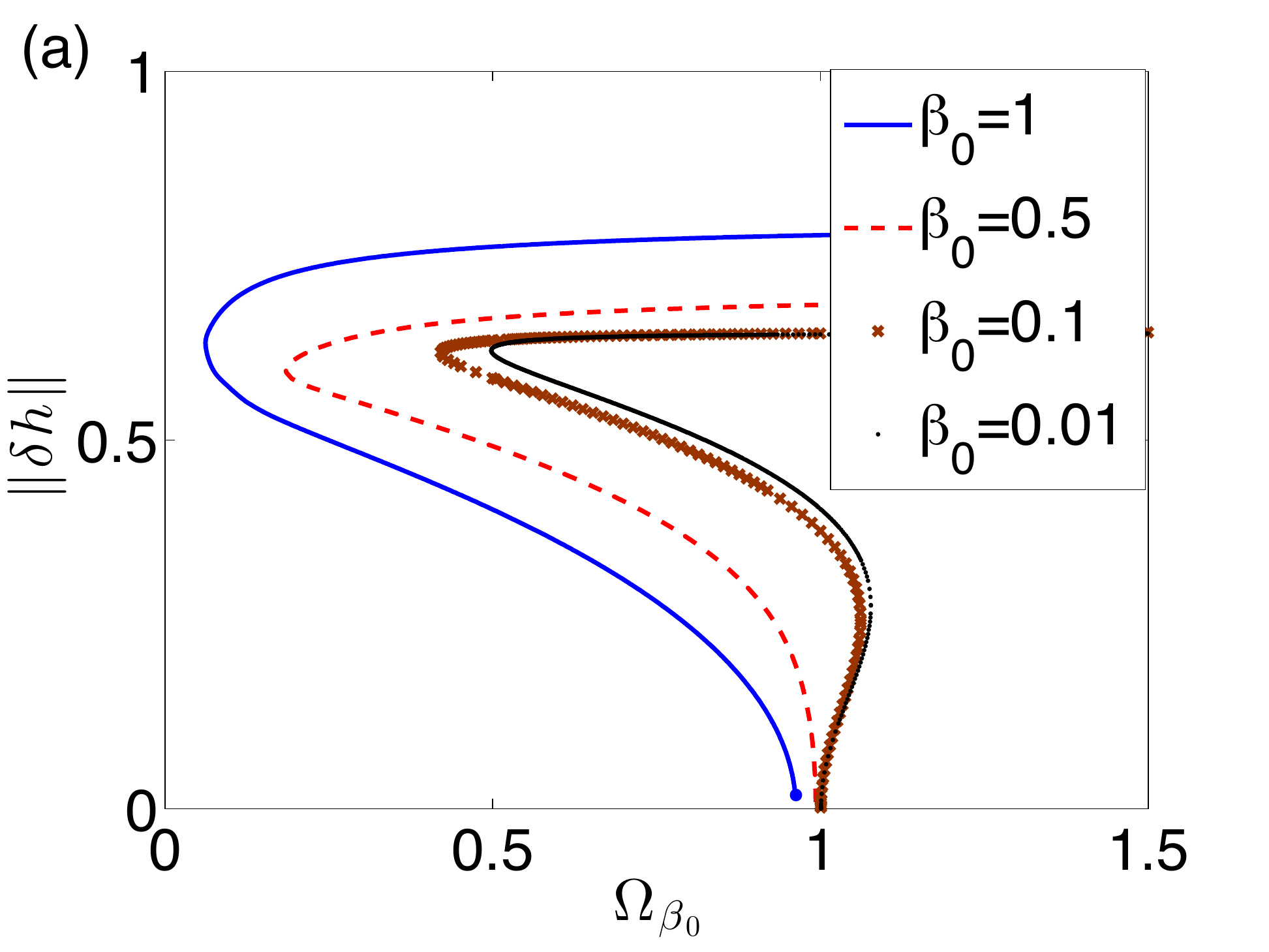}}
{\includegraphics[width=0.49\hsize]{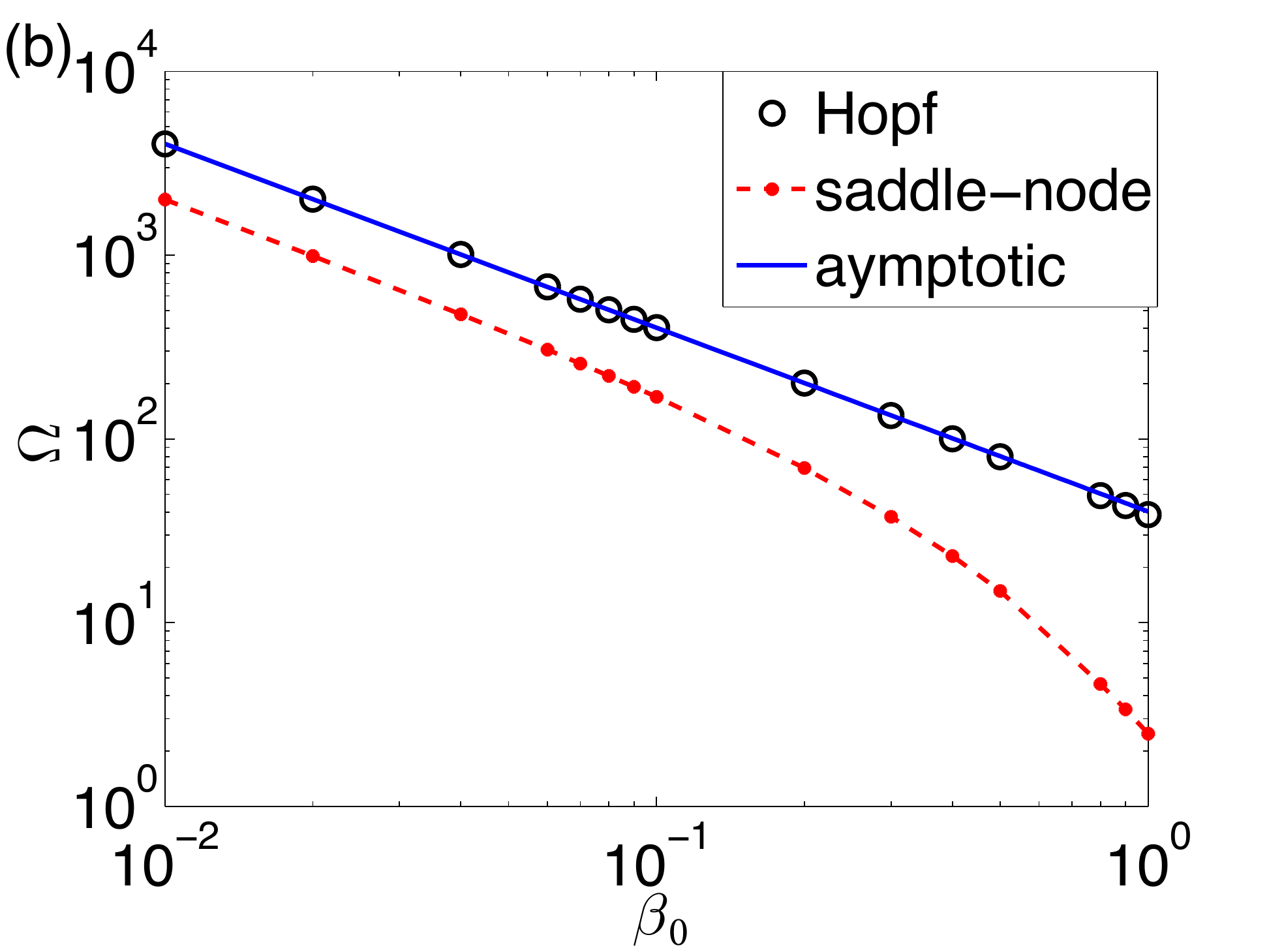}}
\caption{(Color online) 
(a) Time-periodic branches \bfuwe{for} different  values of $\beta_0$. The horizontal axis is the modified 
rotation number $\Omega_{\beta_0}$ given by Eq.~(\ref{eq:omegabeta}). 
(b) Locations of the numerically computed HBs (circles) and the corresponding asymptotic 
prediction (blue solid line) and locations of the numerically found SNs 
(red dot-dashed line) on the time-periodic branch in dependence on
$\beta_0$.
}
\label{largerotation}
\end{center}
\end{figure}

\bfuwe{As the rich structure of time-periodic states is also related to partial wettability, it has to 
simplify and ultimately disappear with decreasing the equilibrium contact angle $\beta_0$, similar 
to the changes observed for branches of steady states discussed in Sec.~\ref{sec:completewetting}.} 
We have found that, as $\beta_0$ decreases, only the rightmost HB and the corresponding 
time-periodic branch survive. The location of the HB and the whole time-periodic branch move 
towards larger $\Omega$ as $\beta_0$ decreases. In fact, as explained in Appendix~\ref{sec:B2}, 
asymptotically one can show that there is always a HB located at large enough 
$\Omega$ for all non-zero $\beta_0$. Furthermore, we find that the location of the 
HB approaches infinity as $\beta_0$ approaches zero and the scaling is given by 
Eq.~(\ref{eq:hopf_cond2}). As a result, we define the modified rotation number as 
\begin{equation}
\Omega_{\beta_0} = \frac{\sqrt{5} h_0 \beta_0 \Omega}{9}.
\mylab{eq:omegabeta}
\end{equation}
Figure~\ref{largerotation}(a) shows the time-periodic branch for different values of 
$\beta_0$ with $\Omega_{\beta_0}$ as the horizontal axis. For each $\beta_0$, 
the HB is located where the respective branch starts. One can see that all the HBs 
are located close to $\Omega_{\beta_0}= 1$, meaning that the scaling \bfuwe{result obtained by the asymptotic analysis is in an excellent agreement with the numerical results.} 
Besides, one can observe that the time-periodic branches are subcritical for 
$\beta_0=1$ and $0.5$, i.e., \bfuwe{initially the branch turns towards smaller values of 
$\Omega_{\beta_0}$ before turning back towards larger values of $\Omega_{\beta_0}$ 
in an SN. However, as $\beta_0$ gets smaller, the HB where the time-periodic branch emerges becomes 
supercritical, see, e.g., the curves for $\beta_0 =0.1$ and $0.01$. Then, the branch turns at first towards 
larger values of $\Omega_{\beta_0}$, before it undergoes two SNs. This implies that another range of
multistability exists. 
For example, for $\beta_0=0.1$ and $\beta_0=0.01$ in Fig.~\ref{largerotation}(a) in the range 
$1\lesssim\Omega_{\beta_0}\lesssim 1.1$, there exist two stable and one 
unstable time-periodic states.} Depending on the initial condition, the long-time 
solution may be either a large-amplitude droplet co-rotating with the cylinder 
or a small-amplitude surface wave co-rotating with the cylinder. 

\renewcommand{\baselinestretch}{1}
\renewcommand{\baselinestretch}{1.5}

In Fig.~\ref{largerotation}(b), we show the location of the numerically found HBs (circles) for various 
values of $\beta_0$ and the corresponding asymptotic predictions ((blue) solid line), Eq.~(\ref{eq:hopf_cond2}). 
One can see that the two match very well indicating that \bfuwe{the approximation provided by the asymptotic 
analysis} is excellent and the location of the HB indeed scales as $\beta_0^{-1}$ as $\beta_0$ 
approaches zero. On the other hand, we show as a (red) dot-dashed line 
the location of the numerically found SN on the time-periodic 
branch. It is found that the scaling of the SN is also 
$\beta_0^{-1}$. As a result, we conclude that the whole time-periodic 
branch moves towards $\Omega=\infty$ as $\beta_0$ decreases to $0$, 
and the scaling is $\beta_0^{-1}$. That is, in the limiting case 
$\beta_0=0$, which \bfuwe{corresponds to complete wetting}, the 
time-periodic branch should disappear and there exists only the 
steady-state branch. See also the discussion in Appendix~\ref{sec:B2b}.

\subsection{Other interesting phenomena}

As shown in the previous sections, the transitions between the bifurcation diagrams 
involve several reconnections of branches and bifurcations. What we have 
explained so far is only part of the story. In fact, there exist some more complicated 
structures including, e.g., the appearance of the torus bifurcations mentioned in 
Sec.~\ref{sec:timeb1} and several \bfuwe{codimension-two bifurcations involving two HBs} 
that occur when varying 
$\beta_0$. We are not able to provide all the details of these, but we would like to 
point out a very interesting case \bfuwe{occuring, e.g., for $\beta_0=1.5$.}

\renewcommand{\baselinestretch}{1}
\begin{figure}
  \begin{center}
    {\includegraphics[width=0.49\hsize]{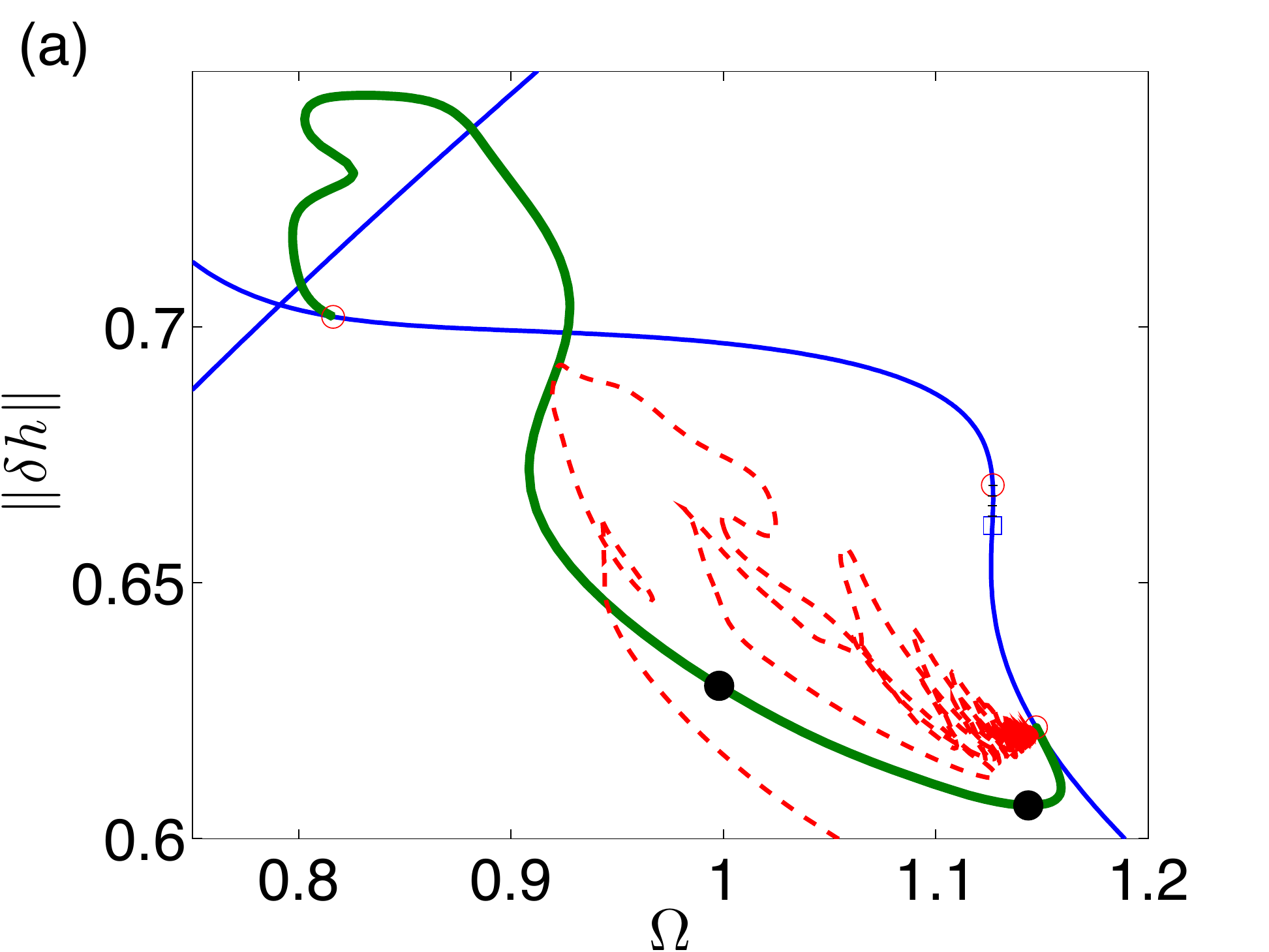}}
    {\includegraphics[width=0.49\hsize]{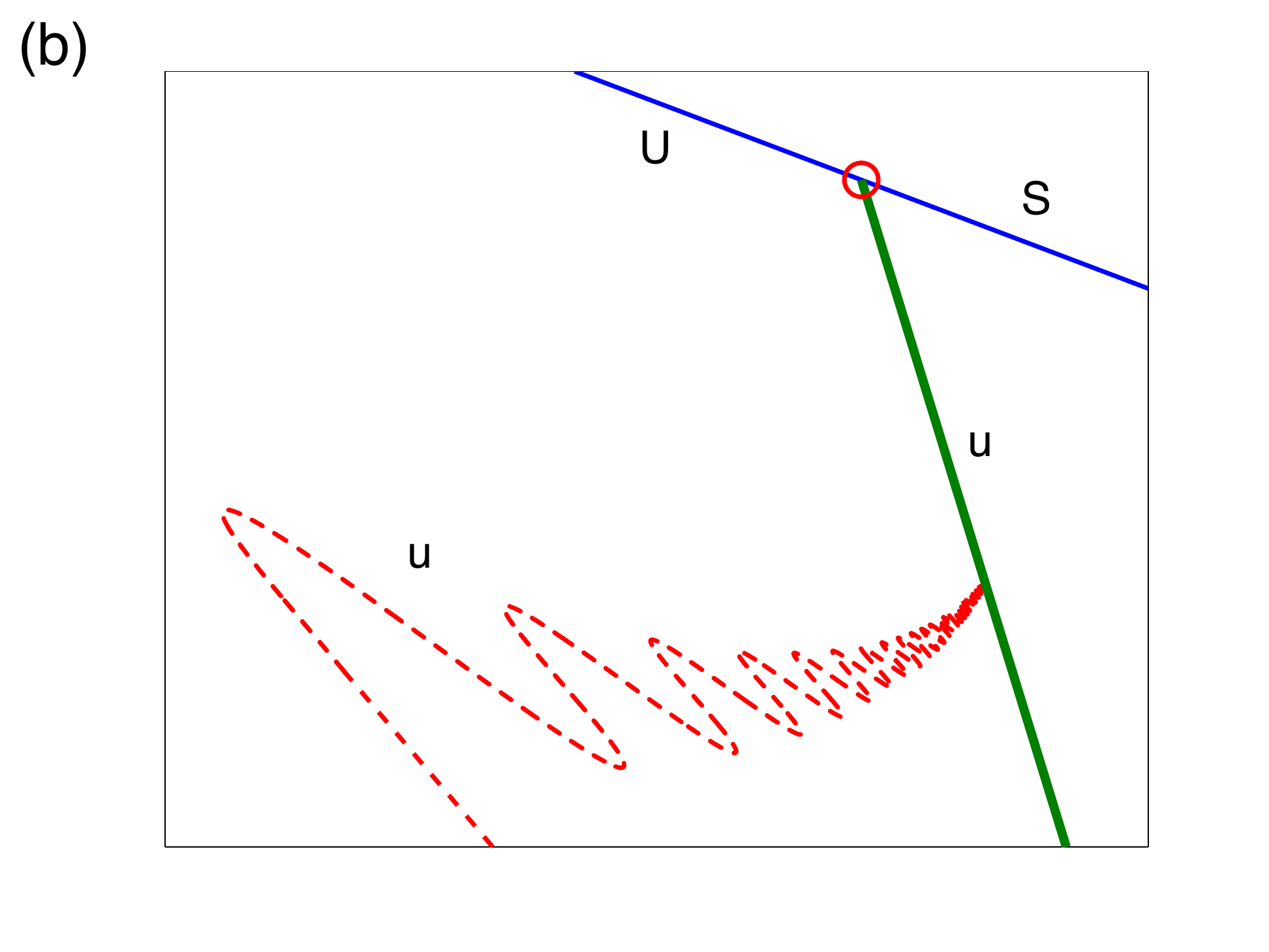}}
    {\includegraphics[width=0.49\hsize]{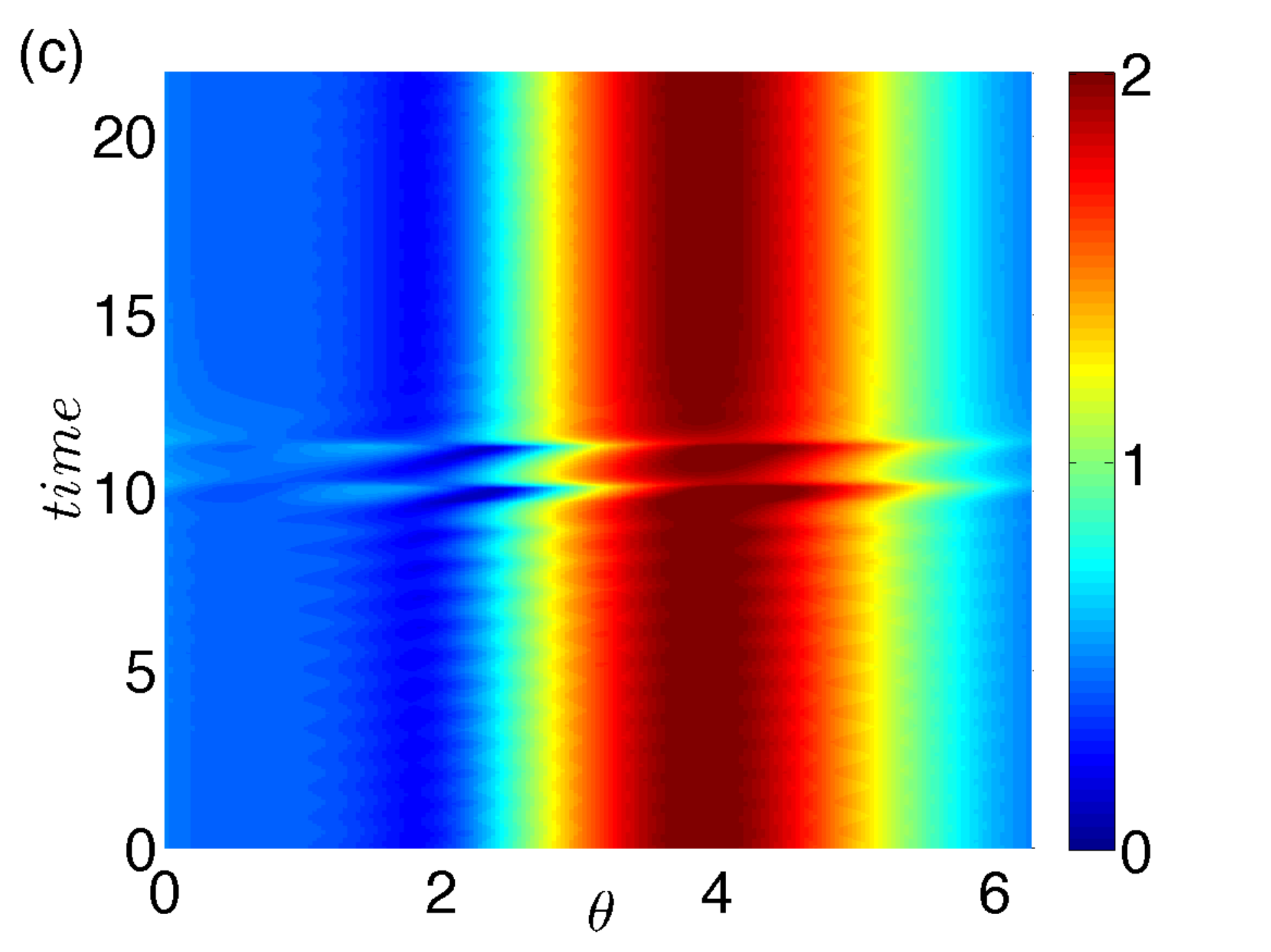}}
    {\includegraphics[width=0.49\hsize]{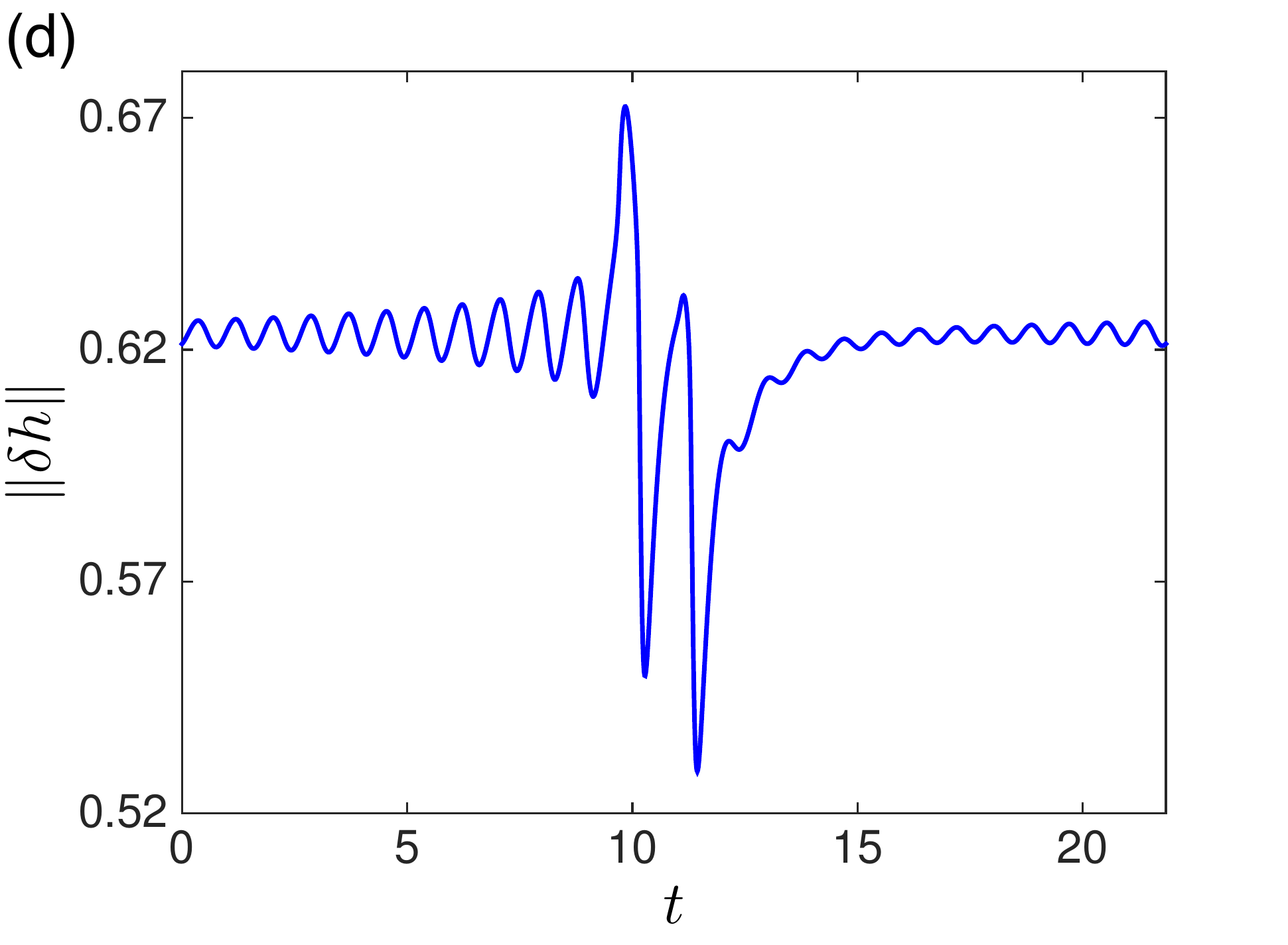}}
  \end{center}
  \caption{(Color online) (a) The zoom-in of the bifurcation diagram for $\beta_0=1.5$ and $B=1$. 
  The graph style is the same as in Fig.~\ref{Figbeta14}. \bfuwe{The filled black circles on the (green) thick solid  line of time-periodic solutions indicate torus bifurcations, and the solutions on the part of the branch between these points are stable.}
  (b) A schematic representation showing the connection of the \bfuwe{time-periodic 
  solution branch that shows tilted snaking and the time-periodic solution branch emerging} from the HB. 
  (c, d) Space-time contour plot and time evolution of \bfuwe{the $L^2$-norm, $\|\delta h\|$,} of the solution at the 
  tilted snaking time-periodic solution branch at $\Omega=1.145$. 
  }
  \mylab{figbeta150}
\end{figure}
\renewcommand{\baselinestretch}{1.5}

Figure~\ref{figbeta150}(a) shows a zoom of the bifurcation diagram for $\beta_0 = 1.5$ 
and $B=1$. The steady-state branch is shown as the (blue) solid line and the time-periodic 
branches are shown as the (green) thick and (red) dashed lines to distinguish two different 
branches. In this figure, there exist three HBs and one HC. The time-periodic branch shown 
as the (red) dashed line is the one that emerges at the rightmost HB \bfuwe{at large $\Omega$ 
outside the figure range}. \bfuwe{An interesting observation is that it actually terminates on another unstable 
time-periodic branch shown by the (green) thick line that itself ends nearby subcritically 
at an HB on the (blue) solid line representing the branch of steady states.} 
A schematic representation of the connections is shown in panel (b) of the figure. We find 
that the (red) dashed time-periodic branch approaches the (green) thick branch along a tilted 
snaking path with its time-period approaching infinity. 
One of the solutions on the (red) dashed branch at $\Omega=1.145$ is presented in panels (c) and (d) 
showing the space-time contour plot and the time evolution of its $L^2$-norm, respectively. 
\bfuwe{The time period for this solution is approximately $20.8468$. The solution profile corresponds to 
a pendent drop that continuously oscillates in time (small wiggles). This small-amplitude `background oscillation' 
is interrupted by a temporally localized large-amplitude oscillation about $t=10$. 
The branch seems to approach the global bifurcation 
in a snaking manner with an ever decreasing amplitude of `wiggling' in the bifurcation diagram.} 
Each time the branch has completed one wiggle of the snake in Fig.~\ref{figbeta150}(a), one more of 
the `background oscillations' is added and the period 
increases by a fixed value (not shown). We expect that as the (red) dashed branch advances towards the 
bifurcation, the number of the wiggles approaches infinity and the period of the solution diverges.
\bfuwe{We also note that the solution profiles on the approached (green) thick branch correspond to 
single-frequency oscillations similar to the small-amplitude `background oscillations'. }

\section{Conclusion} \mylab{sec:conc}

We have analyzed the behavior of a partially wetting liquid on a
rotating cylinder using the long-wave model introduced in
Ref.~\onlinecite{Thiele2011a} as the governing equation.  This model,
in addition to accounting for the effects of gravity, viscosity,
rotation and surface tension, includes the effect of wettability via
the introduction of a Derjaguin (or disjoining) pressure
term. \bfuwe{Here, we have extended the preliminary results of
  Ref.~\onlinecite{Thiele2011a} that was mainly concerned with  the main depinning
  transition in two ways. We have shown that
  the complete bifurcation diagrams that describe continuous and
  discontinuous transitions between different steady and time-periodic
  thickness profiles and accompanying flow states
are much richer than anticipated. This applies even to the case
  without rotation, not to speak of the case with driving where we
  have now discussed several branches of time-periodic solutions.}

\bfuwe{First, this has resulted in a deeper understanding of the
specific case at large contact angle studied in Ref.~\onlinecite{Thiele2011a}.
Second, and more importantly, we have analyzed how the
bifurcation behavior (the depinning transition in particular) changes as the
wettability is changed. In other words, we have investigated how the
complex bifurcation structure found at large contact angles transforms
into the much simpler behavior at zero contact angle. The calculations
have been performed employing three approaches, namely, on one
hand, different numerical continuation techniques 
that have allowed us to track stable and unstable steady \textit{and} time-periodic
states as well as, on the other hand, direct numerical simulations.}

At equilibrium, the transition between complete and partial wetting
corresponds to a phase transition called the wetting transition
\cite{BEIM2009rmp}. A lateral driving force, like the one related to
the rotation of the cylinder, brings the system permanently out of
equilibrium and may cause a dynamic wetting transition. For instance,
a thick film may be drawn out of a finite contact angle meniscus above
a critical lateral driving~\cite{SZAF2008prl,GTLT2014prl} or an array
of drops sliding down an incline may transform into a
film~\cite{TVNB2001pre}.  Placed within this wider context, the main
purpose of the present study has been to investigate the interplay of
depinning and wetting transitions for a specific well-defined
system. In particular, we have analyzed in detail how the depinning
behavior changes when the wettability of the liquid is increased
[decreased], i.e., when the equilibrium contact angle is decreased
[increased]. With this aim, we have determined bifurcation diagrams
with the rotation number as the main control parameter for various
values of the equilibrium contact angle.

\bfuwe{As only a picture that contains all stable and unstable steady film and 
drop states and their relations allows one to understand the emerging complex 
dynamics, we have first completed the bifurcation diagrams of steady profiles for 
$\beta_0=2$ that before\cite{Thiele2011a} were limited to a subset of the branches 
of steady-state solutions. In particular, we have discussed additional branches that 
represent unstable symmetric and asymmetric double-drop and double-hole 
(nucleation) solutions.} Furthermore, we have shown that although triple-drop 
solutions do not exist on a  resting cylinder (i.e., at equilibrium) at the chosen 
parameters, the non-equilibrium driving can bring them into existence when the 
cylinder is rotated.

In the second part, we have employed numerical continuation techniques to also 
determine all the branches of time-periodic solutions for various selected values of 
the contact angle. The completed bifurcation diagrams have allowed us to discuss 
how the global SNIPER bifurcation related to depinning of partially wetting drops 
\bfuwe{disappears through a number of transitions when the contact angle is decreased 
towards small values. However, instead of the hoped-for simple scenario, where one 
branch of time-periodic solutions emerges from two separate ones when two global 
bifurcations collide on a branch of unstable steady solutions,} we have encountered a 
rather intricate sequence of codimension-two bifurcations involving a dance of homoclinic 
bifurcations in the parameter plane, and a \bfuwe{branch of tilted snaking of time-periodic 
solutions.} This implies that the behavior is even more complicated than the behavior of 
the line deposition process described in Ref.~\onlinecite{Kopf2014} that may also be seen 
as a depinning process, as discussed in Refs.~\onlinecite{Kopf2012,FrAT2012sm}. It 
remains an interesting question how the described behavior is amended when further 
physical effects such as inertia and/or higher-order terms are incorporated, e.g., along the 
lines of Refs.~\onlinecite{NKR06,Kelm2009jfm}, or if one goes beyond the long-wave 
(small physical contact angle) case and uses a full Stokes~\cite{HiKe2003prslsapes} 
or Navier-Stokes description incorporating wettability.

In addition to the numerical investigation, we have performed asymptotic analyses of 
steady-state solutions for zero rotation number and small contact angle, and of steady-state 
and time-periodic solutions for large rotation speeds, and found excellent agreement with 
the numerical results. We have also corroborated our findings by full time-dependent 
simulations of the underlying model.

\bfuwe{The presented hydrodynamic system of a droplet on a rotating
  cylinder has been considered as a prototype for other depinning
  transitions in hydrodynamic systems, more general soft matter
  systems and beyond. These systems are characterized by an interplay
  of imposed spatial heterogeneity, a lateral driving force and a
  cohesive force resulting in coherent structures.  As a weakening of
  the cohesive force will result in a first- or second-order phase
  transition (e.g., the wetting transition for the present system), the
  depinning transition will dramatically change its character in the
  proximity of the phase transition. However, our study has shown that
  the bifurcation behavior in the specific system studied is actually
  richer (more complicated) than expected. On one hand, this
  implies that the particular system of the rotating cylinder merits
  further studies and, on the other hand, it indicates that the quest
  for a simple though physically realistic model system is still
  open.}\clearpage

\begin{acknowledgements}
  The authors would like to thank Andrey Pototsky for his insightful
  input in the development of the Auto07p+FFTW continuation code. He
  had showed to UT and DT how FFTs can be combined with Auto07p being
  himself inspired by codes developed by Grigory Bordyugov for the
  continuation of spiral waves. The rotating cylinder code was
  first developed by SR and UT as an Auto07p+FD (finite difference)
  code that was able to continue both on steady-state branches as well
  as time-periodic branches. However, the possible spatial resolution
  was not high enough and numerical instabilities always occured at
  high $\beta_0$. DT developed the first version of a new Auto07p+FFTW
  code that was able to continue the steady-state branches and to
  detect Hopf bifurcations. Later, with the advice of AP, TS compiled
  the Auto07p+FFTW code without OpenMp, which allowed us to also
  continue the branches of time-periodic solutions. Afterwards, the code
  was further improved by TS.

  The work of TS was partly supported by the EPSRC under grant
  EP/J001740/1. The work of DT was partly supported by the EPSRC under
  grants EP/J001740/1 and EP/K041134/1.  TS and DT thank the Center of
  Nonlinear Science (CeNoS) of the Westf{\"a}lische Wilhelms
  Universit\"at M\"unster (CeNoS) for its support of our collaboration
  meetings and for including the rotating cylinder as one of their
  ``M\"unsteranian Torturials'' on continuation
  (http://www.uni-muenster.de/CeNoS/Lehre/Tutorials/auto.html).
\end{acknowledgements}

\appendix

\section{Analysis of steady drops at $\Omega=0$}
\mylab{app:steady_drops}

\subsection{Drops at $B=0$ on an extended domain}
\mylab{sec:steady_drops_1}

\renewcommand{\baselinestretch}{1}
\begin{figure}
\begin{center}
\includegraphics[width=0.6\hsize]{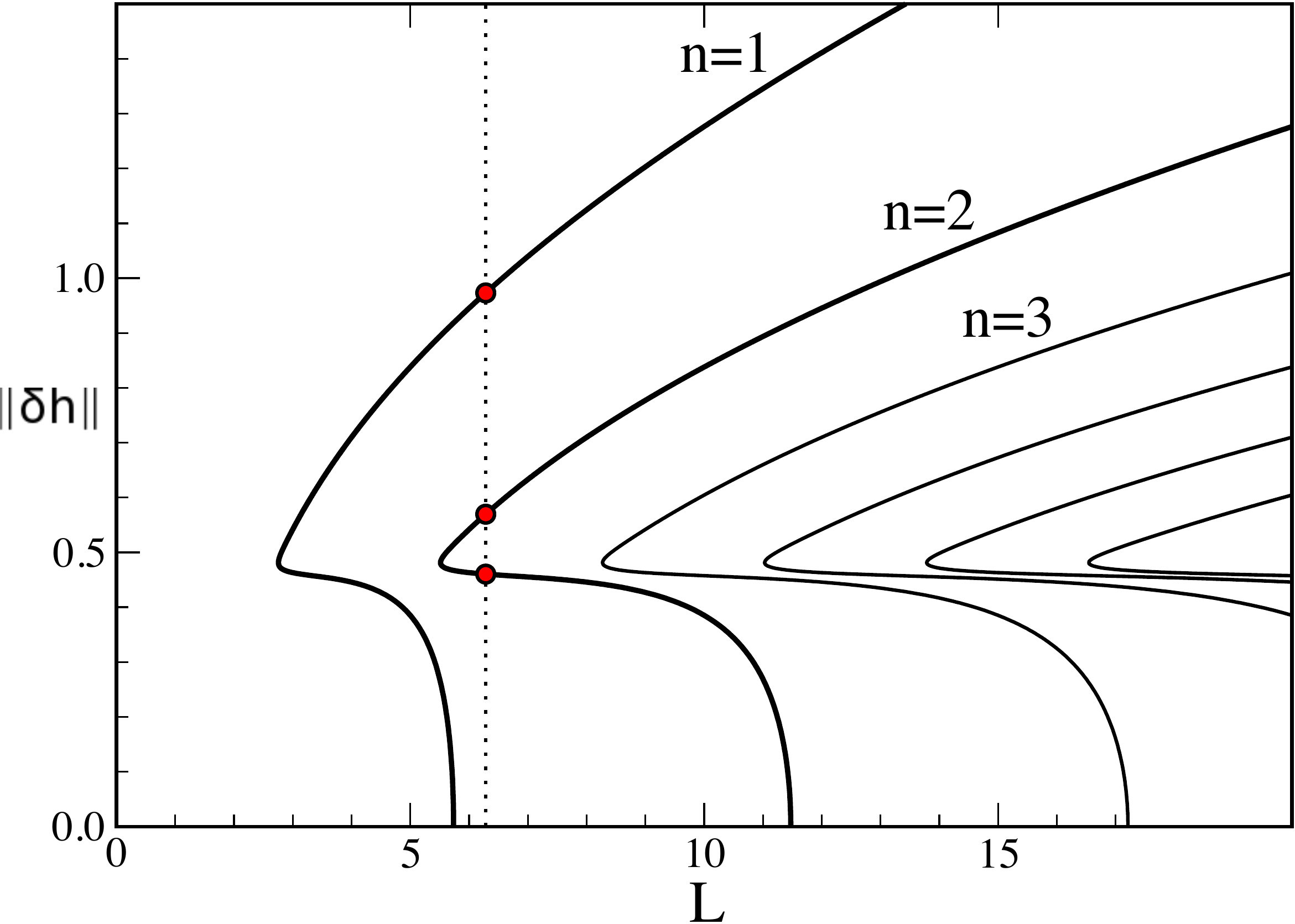}
\caption{(Color online) 
\bfuwe{The bifurcation diagram for contact angle $\beta_0=2$ for the simplified 
system without gravity ($B=0$) and without rotation ($\Omega=0$) but for an 
extended domain size $L$ that is employed as the control parameter. The dotted 
vertical line indicates the value $L=2\pi$ that corresponds to our system in the chosen 
scaling. The various curves represent $n$-drop solutions upon the cylinder. The three 
crossings of branches and the vertical line represent one stable single-drop solution 
and two unstable double-drop solutions, which is consistent with the number of 
solution seen in Fig.~\ref{Fig6} for $B=0$.}}
\label{Fig8}
\end{center}
\end{figure}
\renewcommand{\baselinestretch}{1.5}

In this Appendix, we briefly consider the case $\Omega=0$, $B=0$ in an extended 
parameter space, to understand why Fig.~\ref{Fig6} has to have at $\Omega=0$ and 
$B=0$ the two unstable double-drop solutions that give rise to the entire 
branch structure at $B>0$ even without driving. This will also allow us to understand 
why unstable triple-drop solutions may appear at finite $\Omega$. In 
particular, we continue $n$-drop solutions for $n=1,2,3,\dots$ in the domain size $L$. 
The result for $\beta_0=2$ is given in Fig.~\ref{Fig8} where the vertical dotted line 
indicates $L=2\pi$ that is the normally fixed value for our system in the chosen scaling. 
The three intersections correspond to the three non-trivial steady thickness profiles at 
$B=0$ in Fig.~\ref{Fig6}, one single-drop and two double-drop profiles. This is a result 
of the subcritical character of the primary bifurcation in Fig.~\ref{Fig8} that also 
explains why part of the solutions correspond to unstable nucleation (or hole) solutions 
(see discussions in Refs.~\onlinecite{Thiele2004} and~\onlinecite{TNPV2002csa}).

There are no intersections of the vertical line with the unstable solution branches for 
$n\ge3$ (three drops and above) and, therefore, there are no `equilibrium' solutions at 
$\Omega=0$ with more than two drops. However, as triple-drop solutions exist 
at equilibrium at parameters not very far away from the considered system (at 
$L\approx8$ in Fig.~\ref{Fig8}), the non-equilibrium driving can bring them into 
existence \bfuwe{also at $L=2\pi$, i.e., the circumference of the cylinder in our 
scaling}. Here, this occurs for $B=1$ between $\Omega\approx 0.68$ and $0.95$ 
(Figs.~\ref{Fig3}(b) and~\ref{Fig5}). This observation might be relevant for a number of 
other systems. 

\subsection{Changes in steady double-drop profiles with increasing Bond number}
\mylab{sec:steady_drops_2}

\renewcommand{\baselinestretch}{1}
\begin{figure}
\begin{center}
{\includegraphics[width=0.45\hsize]{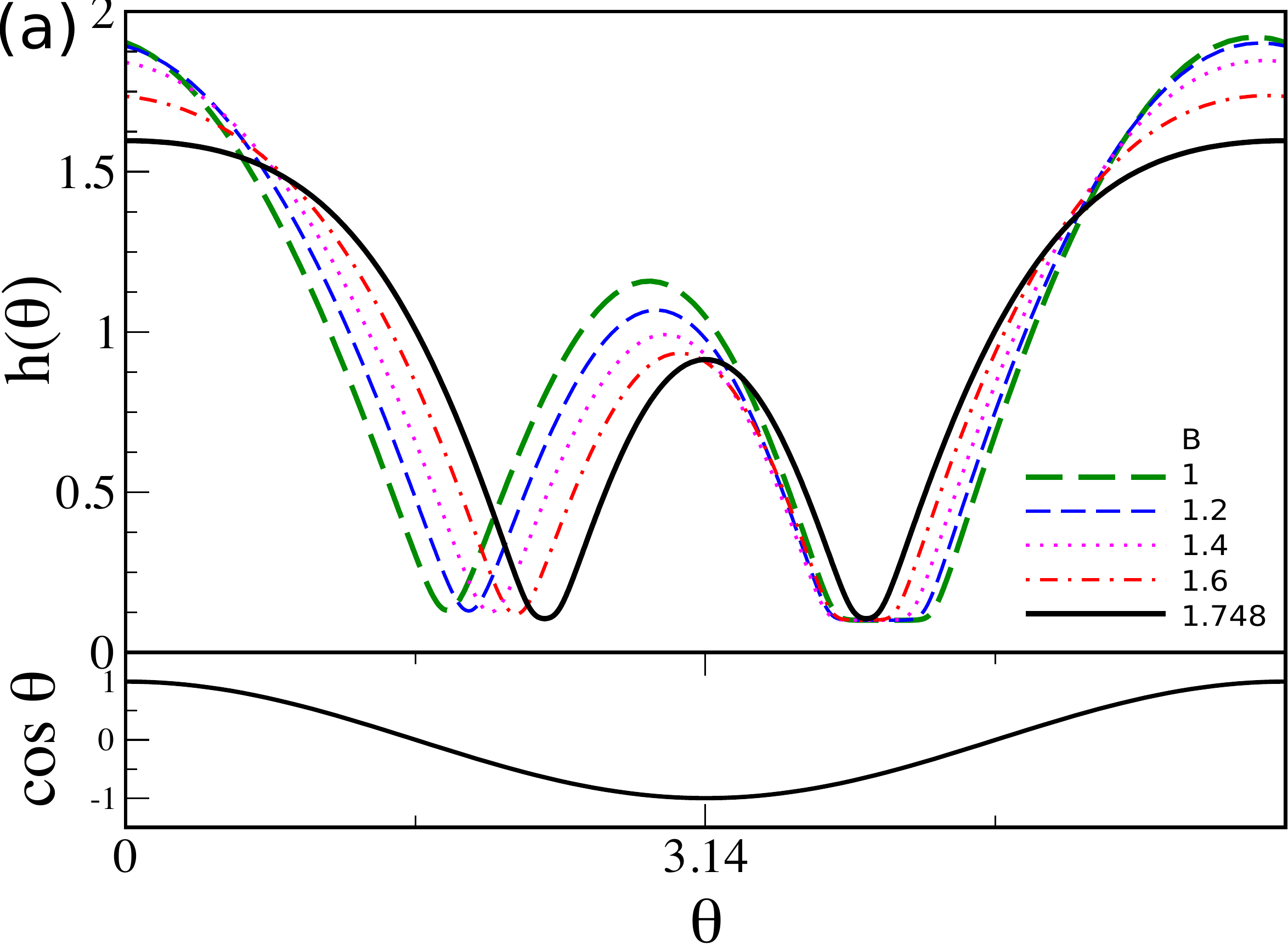}}\hspace{0.5cm}
{\includegraphics[width=0.45\hsize]{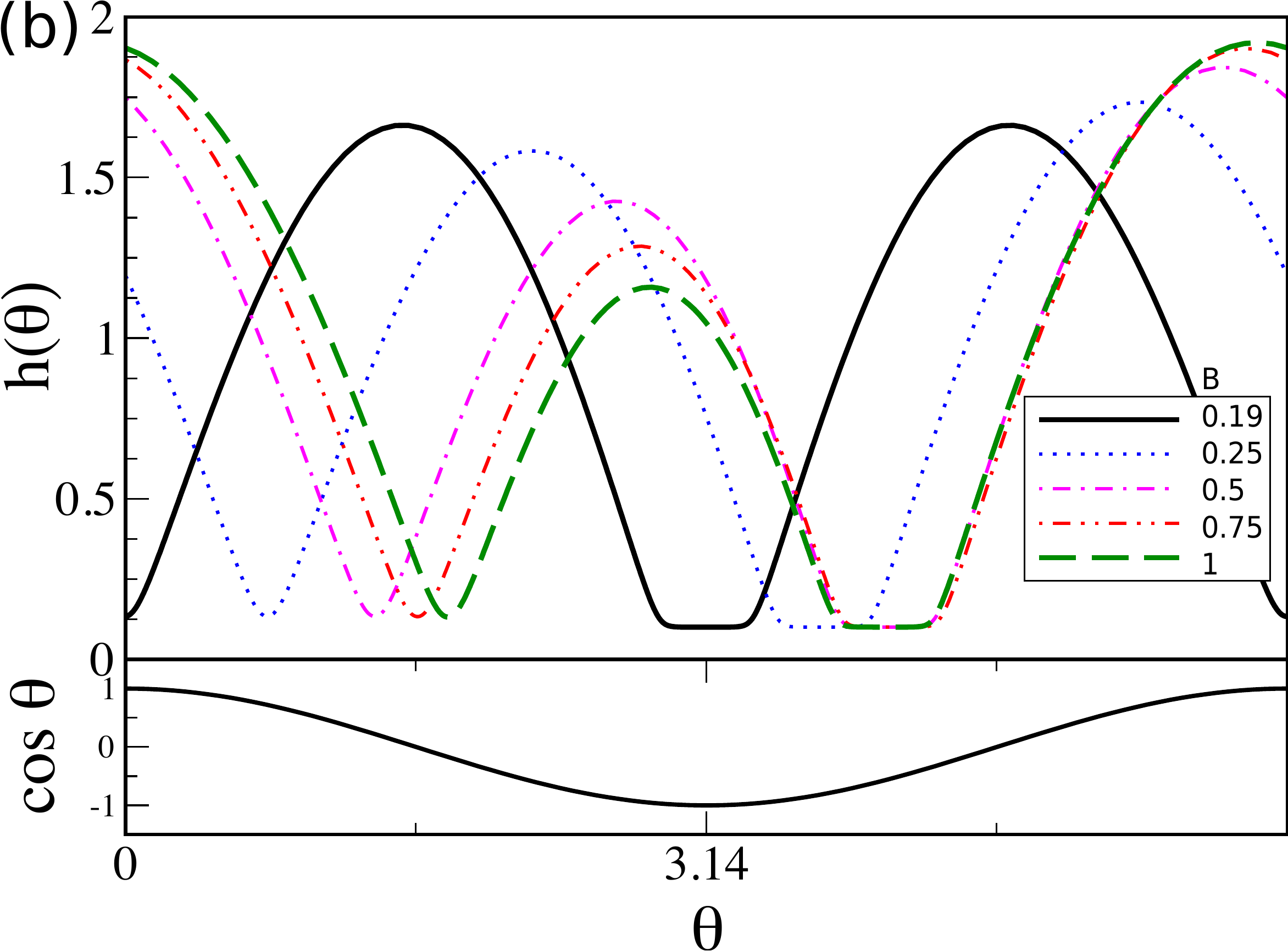}}
{\includegraphics[width=0.45\hsize]{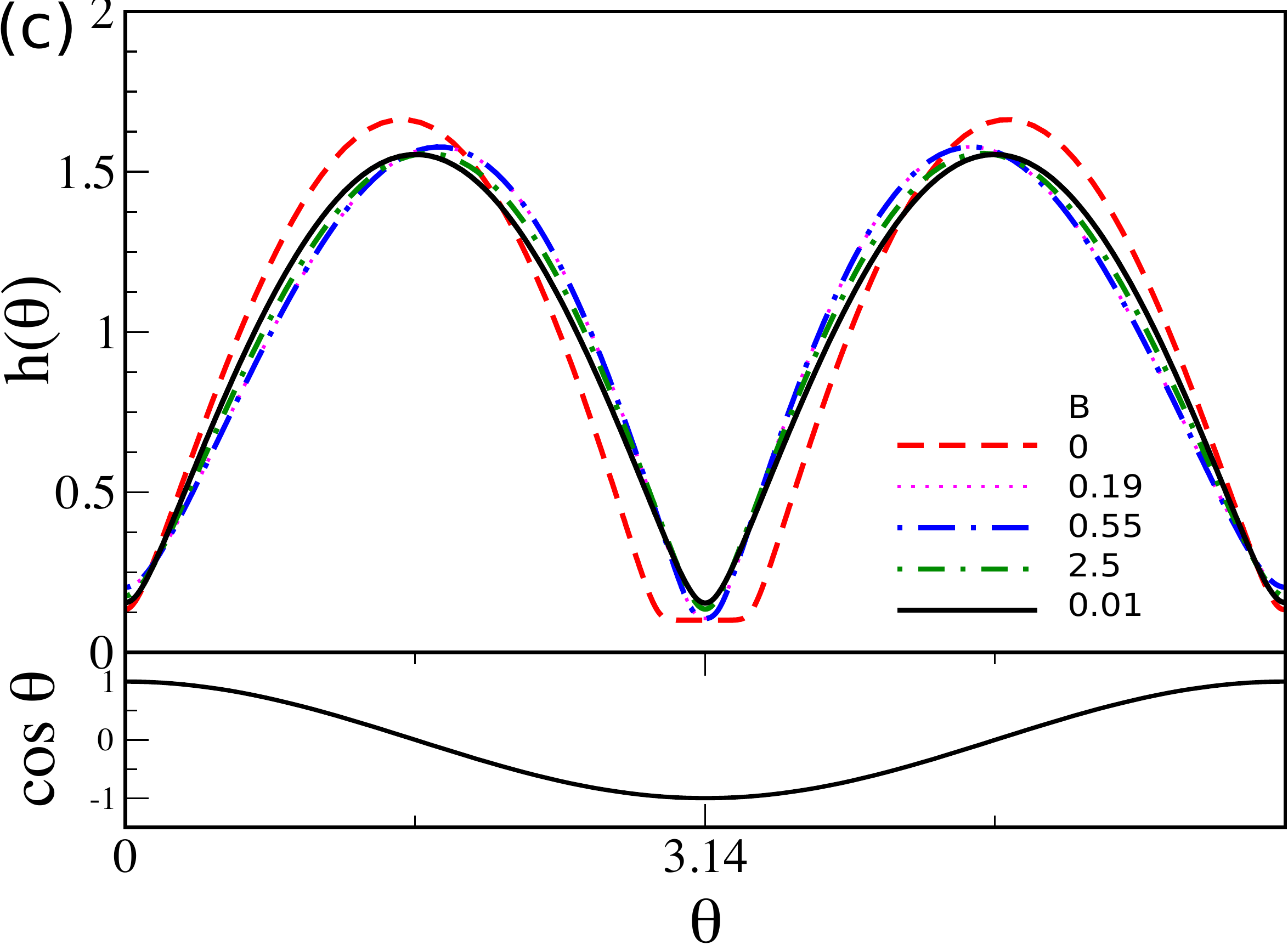}}\hspace{0.5cm}
{\includegraphics[width=0.45\hsize]{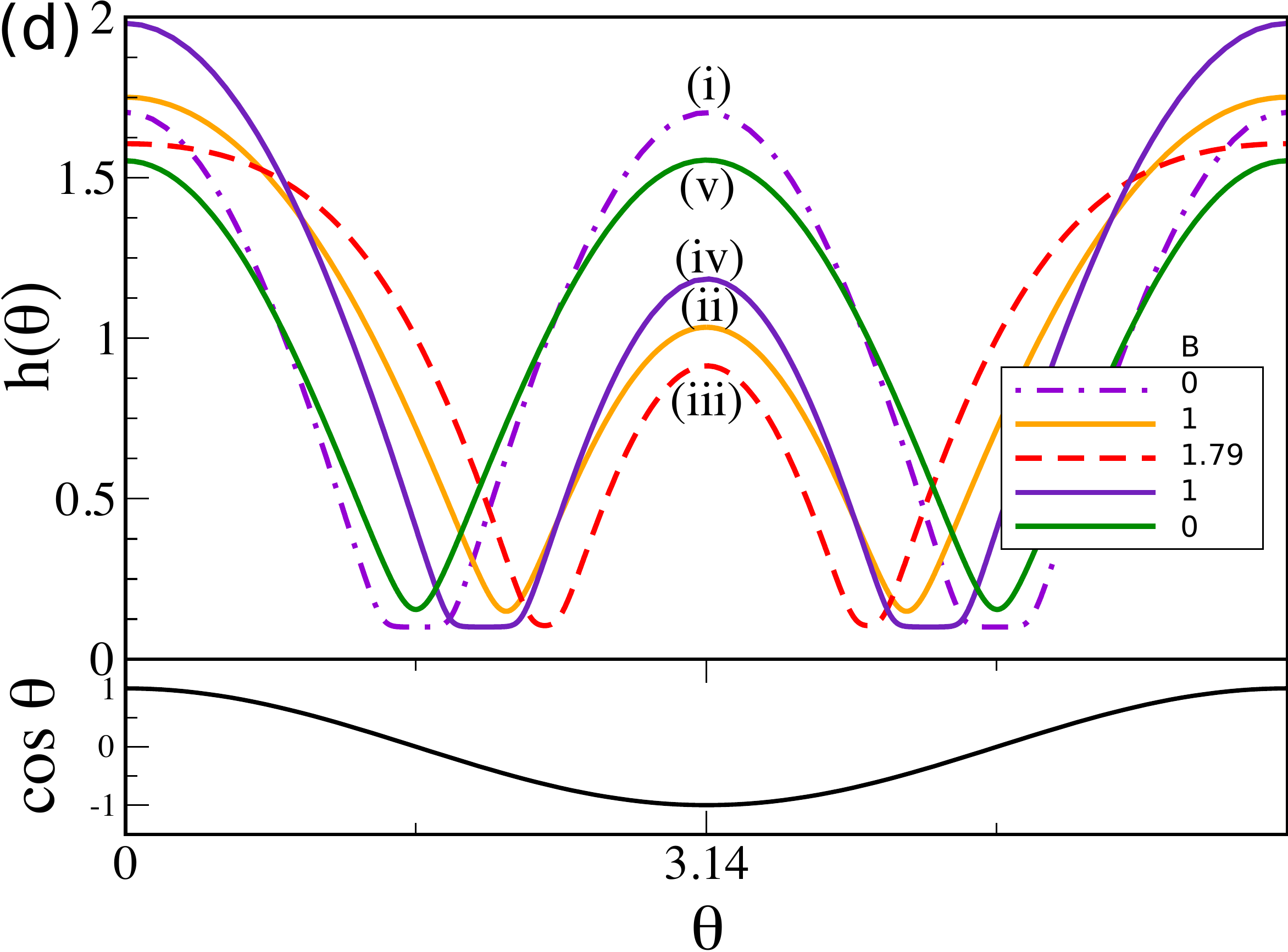}}
\caption{(Color online) 
Shown are selected double-drop profiles at various Bond numbers on the 
different branches in Fig.~\ref{Fig6} for $\Omega=0$ and $\beta_0=2$. Panel (a) gives 
profiles along the asymmetric branch a$\mathrm{1}$ between $B=1$ and the pitchfork 
bifurcation at $B\approx 1.748$. The profiles on a$\mathrm{2}$ are obtained as mirror 
images about $\theta = \pi$. Panel (b) gives profiles along a$\mathrm{1}$ between 
$B=0.19$ and $B=1$. Panel (c) shows profiles along the symmetric branches 
d$\mathrm{l}$ and n$\mathrm{l}$. The order of the legend corresponds to the order of 
solutions if one were to follow the branch. Panel (d) gives profiles along the symmetric 
branches d$\mathrm{u}$ and n$\mathrm{u}$. Thereby, label (i) gives the solution with 
$\|\delta h\|\approx0.57$ at $B=0$, then labels increase as one travels along the 
branches, until label (v) gives the solution with $\|\delta h\|\approx0.46$ at $B=0$.}
\label{Fig9}
\end{center}
\end{figure}
\renewcommand{\baselinestretch}{1.5}

In this Appendix, we give some further details of the behavior of the double-
drop solutions in Fig.~\ref{Fig6}.

We consider the left-shifted asymmetric drop profiles on the central a$\mathrm{1}$ 
branch and describe the changes in the drop profiles as $B$ varies. For any of the 
values of $B$ shown, the right-shifted asymmetric drop on the a$\mathrm{2}$ branch 
is obtained by considering the mirror image about $\theta = \pi$. For $B>1$ 
(Fig.~\ref{Fig9}(a)) the drop underneath the cylinder decreases in height and volume, 
while the larger drop on top of the cylinder increases its volume as liquid is transferred. 
Simultaneously, the short film between the drops decreases in length, because the 
volume of the larger drop increases but its height decreases. At $B\approx 1.748$, the 
gap between the drops disappears, the solution becomes symmetric and 
a$\mathrm{1}$ and a$\mathrm{2}$ terminate on n$\mathrm{u}$ in a 
symmetry-breaking pitchfork bifurcation slightly before n$\mathrm{u}$ and 
d$\mathrm{u}$ annihilate in the saddle-node bifurcation at $B\approx 1.749$. Moving 
on a$\mathrm{1}$ from $B=1$ in the opposite direction (Fig.~\ref{Fig9}(b)), we 
observe that the drop beneath the cylinder increases in height and volume, while the 
larger drop on top becomes smaller. The sizes of the gaps between the drops remain 
nearly the same, but the drops shift. At $B=0.19$ the solution itself has recovered the 
mirror symmetry at $\theta = \pi$ and the branches a$\mathrm{1}$ and a$\mathrm{2}$ 
end in a pitchfork bifurcation on d$\mathrm{l}$.

The steady profiles on the symmetric branches d$\mathrm{l}$ and n$\mathrm{l}$ 
(Fig.~\ref{Fig6}) show two drops of equal height arranged in a symmetric way about 
$\theta = \pi$ that are positioned on the sides of the cylinder, i.e., the holes are on top 
and underneath. The steady-state solution with the larger \bfuwe{$L^2$-norm} (on 
d$\mathrm{l}$), features at $B=0$ a flat film underneath the cylinder separating the 
drops, while the edges of the drops meet above the cylinder (Fig.~\ref{Fig9}(c)). 
Following this branch of steady-state solutions first along d$\mathrm{l}$ then along 
n$\mathrm{l}$ to the solution of smaller $L^2$-norm at $B=0$, the norm 
monotonically decreases since the separating flat film shrinks and the drops decrease 
in height. Selected profiles on the other symmetric branches d$\mathrm{u}$ and 
n$\mathrm{u}$ (Fig.~\ref{Fig6}) are given in Fig.~\ref{Fig9}(d). At $B=0$ on 
d$\mathrm{u}$ ($\|\delta h\|\approx 0.57$), there exist two identical drops on top and 
underneath the cylinder while the earlier discussed d$\mathrm{l}$ is the same solution 
shifted by $\pi/2$. They only start to differ once gravity is turned on, i.e., for $B\ge 0$. 
Moving along d$\mathrm{u}$ for increasing $B$, liquid is transferred to the drop on top 
and the films between the drops shrink until at $B\approx 1.749$ d$\mathrm{u}$ 
annihilates with n$\mathrm{u}$ in a saddle-node bifurcation.  Returning on 
n$\mathrm{u}$, for decreasing $B$, the fluid transfers back to the drop beneath the 
cylinder and the drops reach again equal heights at $B=0$, where they correspond to 
standard nucleation solutions (cf.~subcritical part of $n=3$ branch at $L=2\pi$ in 
Fig.~\ref{Fig8}).

\section{Asymptotic analysis for special cases}
\mylab{app:asym}

In Appendices~\ref{sec:B1} and \ref{sec:B2}, we analyze solutions in the limits of 
vanishing and infinite rotation numbers, respectively.

\subsection{Steady-state solutions without rotation}
\mylab{sec:B1}

The equation for the steady-state solutions when $\Omega=0$ is given by 
\begin{equation}
\pth \left\{ h^3 \pth\left[\pth^2 h + h - B\cos\theta + \Pi(h)\right] \right\}= 0.
\mylab{eq:A1}
\end{equation}
The equation can be integrated once, and the constant of integration can be removed 
because it represents the flux that is zero. We obtain
\begin{equation}
\pth\left[\pth^2 h + h - B\cos\theta + \Pi(h)\right]= 0.
\mylab{eq:A2}
\end{equation}
Integrating the equation again, we have
\begin{equation}
\pth^2 h + h - B\cos\theta + \Pi(h)= c,
\mylab{eq:A3}
\end{equation}
where $c=1+\frac{1}{2\pi}\int^{2\pi}_0 \Pi(h) \,d\theta$ is a constant. 

For partially wetting liquids, $\beta_0\ne 0$, we have numerically found that there 
exists a finite number of distinct (in the sense that the solutions are not obtained from 
each other by a shift in $\theta$) non-trivial solutions (e.g., three non-trivial solutions 
for $\beta_0=2$ and $B=0$ as shown in Figs.~\ref{Fig6} and~\ref{Fig8}), in addition to 
the trivial solution, $h_0\equiv 1$. Multiple non-trivial solutions exist up to a certain 
critical Bond number. Beyond the critical value (e.g., $B\approx 1.68$ for $\beta_0=2$) 
only the symmetric pendent drop solution exits. 

\begin{figure}
\begin{center}
{\includegraphics[width=0.49\hsize]{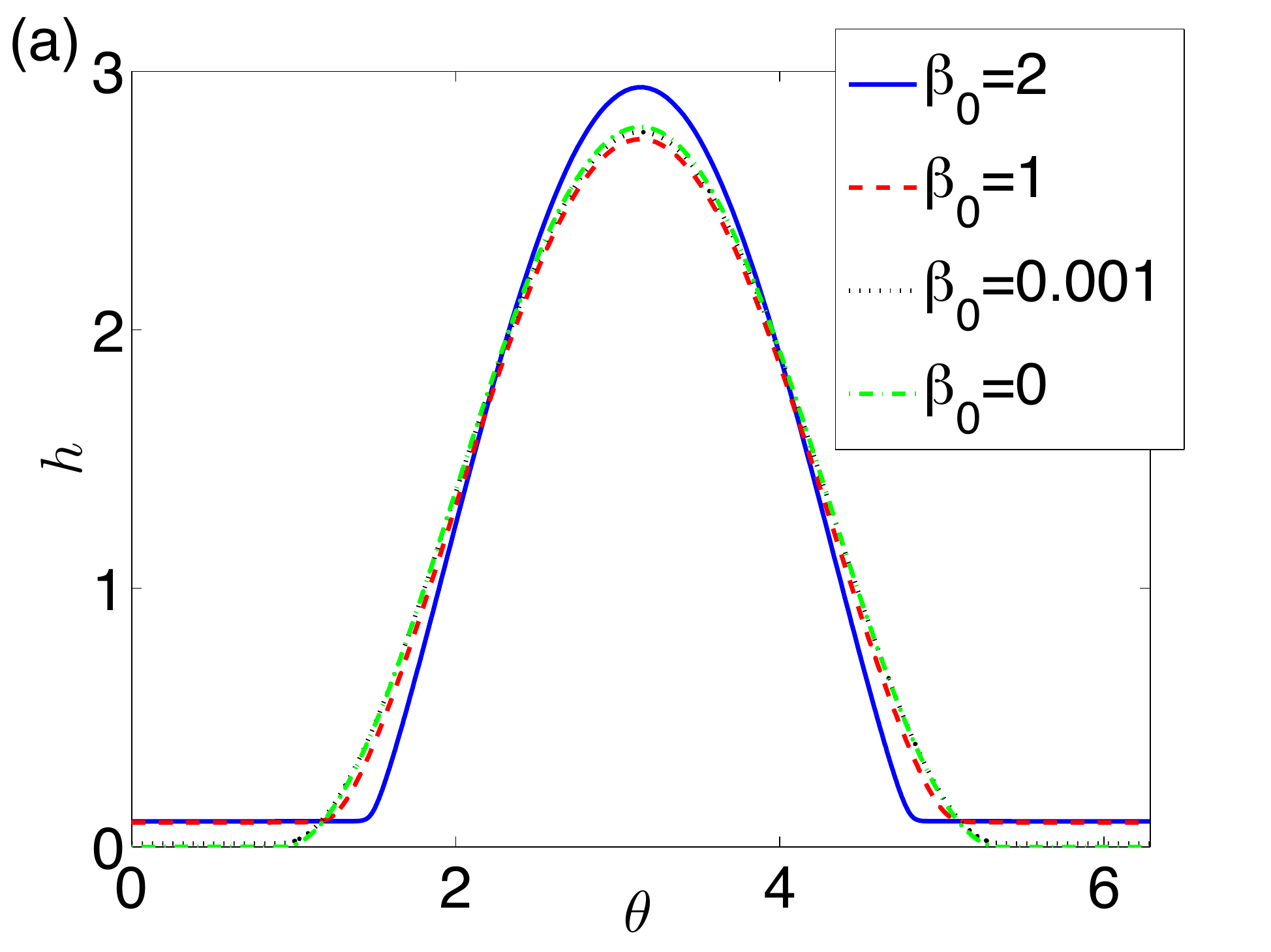}}
{\includegraphics[width=0.49\hsize]{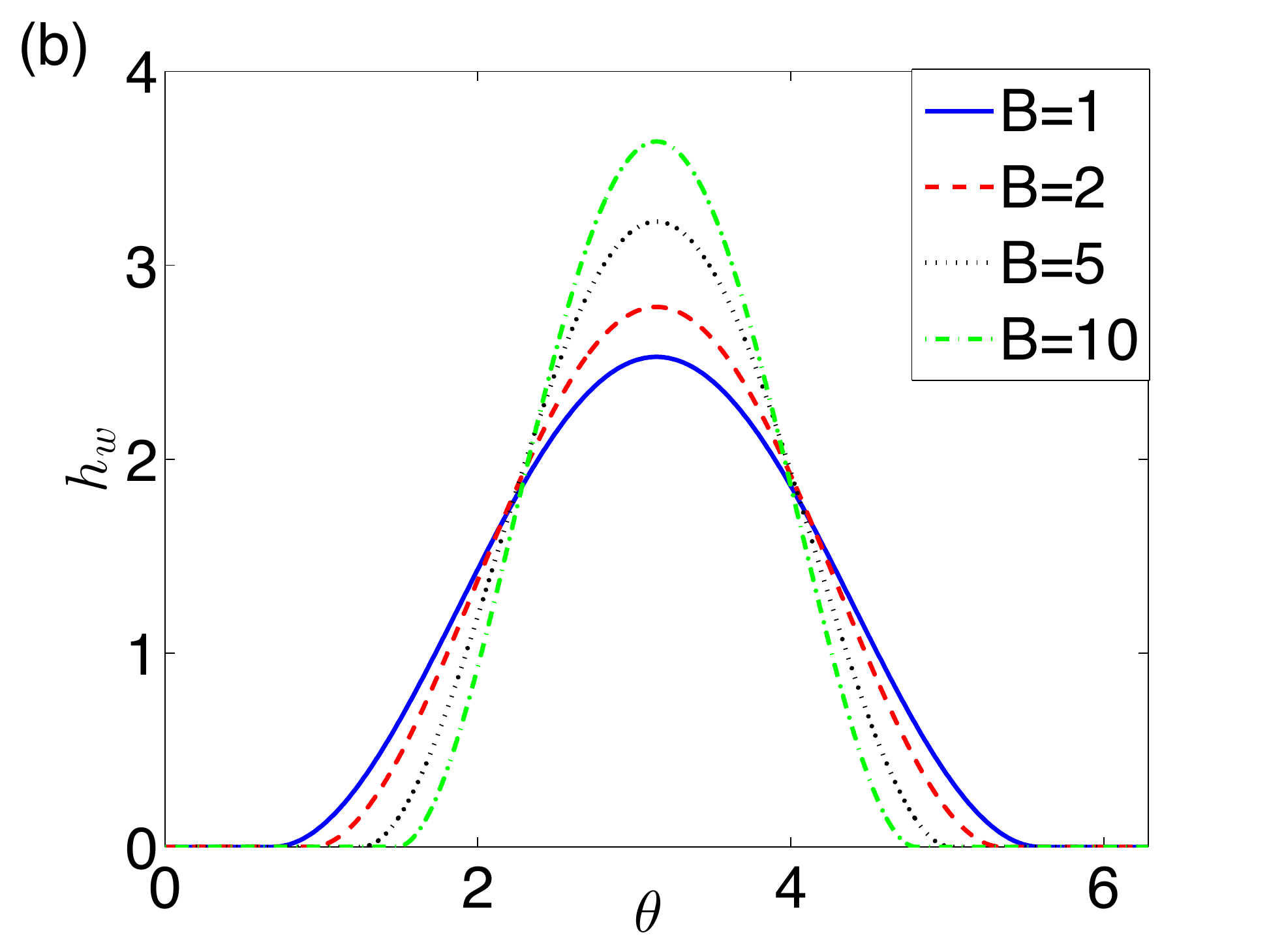}}
\caption{(Color online) 
(a) Steady-state solutions for partially wetting liquids at $B=2$ for various values of 
$\beta_0$. Note that the solution at $\beta_0=0$ is a weak solution given by 
Eq.~(\ref{eq:A5}). (b) Steady-state solutions for \bfuwe{completely} wetting liquids, 
$\beta_0=0$, for various values of $B$. }
\label{Fig21}
\end{center}
\end{figure}

On the other hand, for completely wetting liquids, the general solution of 
Eq.~(\ref{eq:A3}) can be written in the form
\begin{equation}
h_c = 1+\frac{B}{2}\left(1 + \theta \sin\theta\right)+A_0\sin\theta+A_1\cos\theta, 
\mylab{eq:A4}
\end{equation}
where $A_0$ and $A_1$ are constants. If $B=0$, we obtain an infinite number of 
$2\pi$-periodic solutions. However, for a non-zero $B$, $h_c$ is not a smooth 
$2\pi$-periodic function. That is, for a non-zero Bond number no classical solution 
exists~\cite{Reisfeld1992, Evans2004}. We, therefore, consider a weak formulation. 
Our numerical results indicate that as $\beta_0\rightarrow 0$, a solution of 
Eq.~(\ref{eq:A3}) approaches a shape with a compact support. This can be see in 
Fig.~\ref{Fig21}(a), where steady pendent drop solutions are shown for 
various values of $\beta_0$. We note that such compactly-supported weak solutions 
were also found in heated-film systems as ``dissipative 
compactons"~\cite{Shklyaev2010}. Self-similar compactly-supported solutions were 
studied~\cite{Witelski1998, King2001, Bernoff2002, Witelski2004} and the stability was 
analyzed by Laugesen et al.~\cite{Laugesen2002}.

We see that there exists a thin precursor layer for partially wetting liquids and the 
thickness of the thin layer decreases to zero as $\beta_0$ approaches zero. The fact 
that the precursor can go continuously to zero can be realized by a scaling argument. 
If $\beta_0$ is fixed, then in the asymptotic limit $h_0\rightarrow 0$, using the 
asymptotic expansion in $h_0$, i.e., $h=h_0f_0+h_0^2f_1+\cdots$, we find that the 
leading-order equation requires $f_0=1$, i.e., there exists a precursor film of thickness 
$h_0$ (plus higher-order corrections). On the other hand, if we assume that $h_0$ is 
small but fixed and consider the asymptotic limit $\beta_0\rightarrow 0$, we find that 
an appropriate scaling for the precursor thickness $h_p$ is 
$h_p\propto \beta_0^{1/3}$. Then, we find that in Eq.~(\ref{eq:A3}) $c=O(1)$ when 
$\beta_0\rightarrow 0$ and the leading-order balance gives the precursor film of 
thickness $h_p=\left(\frac{5}{3(c+B\cos\theta)}\right)^{1/6} h_0^{5/6}\beta_0^{1/3}+o(\beta_0^{1/3})$.
\renewcommand{\baselinestretch}{1}
\renewcommand{\baselinestretch}{1.5}

As a result, we use the following ansatz of a compactly-supported weak solution for 
the analytical expression of the asymptotic solution of completely wetting 
liquids~\cite{Jensen1997}: 
\begin{eqnarray}
h_w(\theta) = \left\{
\begin{array}{cc}
c_0 + \frac{B}{2}(\theta-\pi)\sin\theta + c_1\cos\theta, & 
\theta\in(\pi-\alpha,\pi+\alpha), \\
0, & \theta\notin(\pi-\alpha,\pi+\alpha), 
\end{array}
\right.
\mylab{eq:A5}
\end{eqnarray}
where $c_0$, $c_1$ and $\alpha$ are constants that should ensure that the solution 
$h_w$ satisfies the continuity and integral conditions:
\begin{equation}
h_w(\theta=\pi\pm \alpha) = \pth h_w(\theta=\pi\pm\alpha) = 0, \quad 
\frac{1}{2\pi}\int^{2\pi}_0 h_w \, d\theta = 1.
\mylab{eq:A6}
\end{equation}

Note also that for a purely macroscopic theory, i.e., without any disjoining pressure, 
weak solutions of partially wetting liquids can also be found by requiring 
$\pth h_w(\theta=\pi\pm\alpha) = \pm \beta_e$, where $\beta_e$ is the macroscopic 
equilibrium contact angle. For completely wetting liquids, $\beta_e=0$ and we obtain 
that the unknown constants $c_0$, $c_1$ and $\alpha$ should satisfy
\begin{eqnarray}
&& \frac{\alpha^2}{\sin\alpha} + \alpha\cos\alpha - 2\sin\alpha = \frac{2\pi}{B}, 
\mylab{eq:A7}\\
&& c_0 = \frac{B}{2}\left(\cos\alpha + \frac{\alpha}{\sin\alpha}\right), 
\mylab{eq:A8} \\
&& c_1 =  \frac{B}{2}\left(1 + \frac{\alpha\cos\alpha}{\sin\alpha}\right). 
\mylab{eq:A9}
\end{eqnarray}

Equations~(\ref{eq:A7})-(\ref{eq:A9}) can be solved numerically to obtain the constants 
of the asymptotic solution given by Eq.~(\ref{eq:A5}). The steady-state solutions for 
various Bond numbers are shown in Fig.~\ref{Fig21}(b). It is found that with 
decreasing $B$ the compact support of the solution becomes larger and the height of 
the central peak decreases. Besides, it can be shown that, in the limit 
$B\rightarrow 0$, we have $h_w=1-\cos\theta$. 

\subsection{Limit of large rotation number}
\mylab{sec:B2}

\subsubsection{Steady-state solutions}
\mylab{sec:B2a}

In the limit when $\Omega$ goes to infinity, it is appropriate to define the small 
parameter $\varepsilon=1/\Omega$ and rescale the time variable as 
$t=\varepsilon\tau$. Equation~(\ref{eq:timeevolution}) can then be rewritten as
\begin{equation}
\partial_{\tau}h = -\partial_{\theta} h - \varepsilon\,\partial_{\theta} 
\left\{ h^3 \pth\left[\pth^2 h + h - B\cos(\theta) + \Pi(h)\right] \right\}.
\mylab{eq:B1}
\end{equation}
As suggested by numerical experiments, \bfuwe{the $L^2$-norm of the steady-state 
solutions, as defined in Eq.~(\ref{eq:normsteady}),} approaches zero and, therefore, 
the solutions approach the flat-film state. Hence, we expand the steady-state 
solution using the following ansatz
\begin{equation}
h_s(\theta) = 1 + \varepsilon h_1(\theta) + \varepsilon^2 h_2(\theta) + \cdots. 
\mylab{eq:B2}
\end{equation}
\bfuwe{Substituting the expansion into} the equation for steady states, we can easily 
find the terms in the expansion. In particular, we find
\begin{equation}
h_1(\theta)=-B\sin(\theta),\quad 
h_2(\theta)=-\frac{3B^2}{2}\cos(2\theta)+\Pi'(1)B\cos(\theta),
\label{eq:h1_h2}
\end{equation}
where we use the prime to denote the derivative of $\Pi$ with respect to $h$.

Next, we analyze the linear stability of the steady-state solution. Assuming that 
the solution is the steady state with a small perturbation, i.e., 
$h(\theta, \tau) = h_s(\theta) + \delta \eta(\theta,\tau)$, $\delta\ll 1$, 
we obtain the following linearized equation:
\begin{equation}
\eta_{\tau} = -\eta_x + \varepsilon\ms{L}[\eta], 
\label{eq:eta}
\end{equation}
where
\begin{equation}
\ms{L}[\eta] = -\pth\left\{h_s^3\pth[\pth^2\eta+(1+\Pi'(h_s))\eta] 
+ 3h_s^2\pth[\pth^2h_s+h_s-B\cos(\theta)+\Pi(h_s)]\eta\right\}.
\label{eq:L_op}
\end{equation}
Assuming that $\eta = e^{\lambda \tau} g(\theta)$, we have 
\begin{equation}
\lambda g = -\partial_{\theta} g + \varepsilon\ms{L}[g].
\label{eq:eigvalue}
\end{equation}
By substituting the expansion (\ref{eq:B2}) in (\ref{eq:L_op}), we can find
\begin{equation}
\ms{L}[\eta]=\ms{L}_1[\eta]+\varepsilon \ms{L}_2[\eta]+\cdots,
\end{equation}
where
\begin{eqnarray}
\ms{L}_1[\eta] &=& -\pth\{ \pth^3\eta+(1+\Pi'(1))\pth\eta+3\sin(\theta)\eta \},\\
\ms{L}_2[\eta] &=& -\pth\{ [6Bh_1\sin(\theta)+(3+3\Pi'(1)+\Pi''(1))\pth h_1+\pth^3 h_1]
\eta\nonumber\\ 
&& +(3+3\Pi'(1)+\Pi''(1))\pth h_1 \pth\eta+3h_1\pth^3\eta \},
\end{eqnarray}
etc. We then expand $\lambda$ and $g$ in power series in $\varepsilon$, i.e., 
$\lambda=\lambda_0+\varepsilon\lambda_1+\cdots$ and 
$g=g_0+\varepsilon g_1+\cdots$, and substitute in (\ref{eq:eigvalue}). As a result, 
we obtain at $n^\text{th}$ order
\begin{equation}
\sum_{i=0}^n \lambda_ig_{n-i}=-\partial_\theta g_n+ \sum_{i=1}^n\ms{L}_i[g_{n-i}].
\label{eq:nth_order}
\end{equation}
At leading order, we have
\begin{equation}
\lambda_0 g_0 = -\partial_{\theta} g_0,
\end{equation}
and we obtain countably infinite number of solutions:
\begin{equation}
g_0^k(\theta) = e^{- \lambda_0^k\theta}, \quad \lambda_0^k =  i k, \quad 
k=0,\,\pm 1,\,\pm 2,\,\ldots.
\end{equation}
\bfuwe{We note that for each $k$, we can select an arbitrary amplitude for $g_0^k$, and we select it to be unity for convenience.}
At next order, we obtain the equation 
\begin{equation}
\partial_{\theta} g_1^k + \lambda_0^k g_1^k = \ms{L}_1[g_0^k] - \lambda_1^k g_0^k.
\mylab{eq:appe11}
\end{equation}
Denoting the operator on the left-hand side by $\ms{M}^k$, i.e., 
$\ms{M}^k[f]\equiv \partial_{\theta} f + \lambda_0^k f$, it is easy to verify that 
$g_0^k$ is the basis for the null space of the adjoint operator $\ms{M}^{k*}$, defined 
by
\begin{equation}
\ms{M}^{k*}[f]\equiv -\pth f + \overline{\lambda_0^k} f,
\end{equation} 
where the overline denotes complex conjugation. Then, the Fredholm alternative 
solvability condition for the equation for $g_1^k$ requires that the inner product of 
the right-hand side of Eq.~(\ref{eq:appe11}) and $g_0^k$ should vanish, which gives 
that
\begin{equation}
\lambda_1^k = 
\frac{\langle\ms{L}_1[g_0^k], g_0^k\rangle}{\langle g_0^k, g_0^k\rangle} = 
(\Pi'(1)+1)k^2 - k^4, \quad k=0,\,\pm 1,\,\pm 2,\,\ldots.
\end{equation}
Note that if 
\begin{equation}
\Pi'(1)>0,
\label{eq:instability_cond}
\end{equation}
then at least for $k=\pm1$, we obtain $\lambda_1^k>0$. Otherwise, all the 
$\lambda_1^k$'s are non-positive. Thus, if condition~(\ref{eq:instability_cond}) 
is satisfied, at least two eigenvalues (the ones corresponding to $k=\pm 1$) have 
positive real parts (assuming that $\varepsilon$ is sufficiently small). In such a 
case, the steady solution is unstable. For the disjoining pressure given by 
Eq.~(\ref{eq:djpress}), 
\begin{equation}
\Pi'(1)=-3H(1-2b). 
\end{equation}
Then, condition (\ref{eq:instability_cond}) is satisfied if $H>0$ and $b>1/2$ (which 
is an unphysical case), or $H<0$ and $b<1/2$. We note that our numerical findings 
agree with these theoretical conclusions. Namely, for a negative $H$ and sufficiently 
small $b$, we found in Sec.~IV that on the branch of steady-state solutions there 
exists a subcritical Hopf bifurcation as $\Omega$ increases, and the steady solution 
becomes unstable. However, there exists a stable time-periodic solution. The 
analysis of such time-periodic solutions is given in the next section.

We would also like to point out that for completely wetting liquids, when $H=0$ , 
we find that $\lambda_1^k=0$ for $k=\pm1$ and $\lambda_1^k<0$ for $|k|>1$. 
Hence, the stability cannot be determined at this order and higher-order terms 
are need to clarify the stability of the steady-state solutions for large values of 
$\Omega$. 

Equation (\ref{eq:nth_order}) for $n=2$ is
\begin{equation}
\pth g_2^k + \lambda_0^k g_2^k = 
\ms{L}_1[g_1^k] + \ms{L}_2[g_0^k] - \lambda_1^k g_1^k - \lambda_2^k g_0^k.
\label{eq:appe24}
\end{equation}
Then, the Fredholm alternative solvability condition implies that
\begin{equation}
\lambda_2^k = 
\frac{\langle\ms{L}_1[g_1^k]+\ms{L}_2[g_0^k]-\lambda_1^k g_1^k,g_0^k\rangle}
{\langle g_0^k, g_0^k\rangle}, \quad k=0,\,\pm 1,\,\pm 2,\,\ldots.
\end{equation}
For $k=\pm 1$ (note that it is sufficient to consider $k=1$ since 
$\lambda_n^{-1}=\overline{\lambda_n^1}\,$), we find
\begin{equation}
\lambda_2^{1} = \frac{\langle\ms{L}_1[g_1^{1}]+\ms{L}_2[g_0^{1}], g_0^{1}\rangle}
{\langle g_0^{1}, g_0^{1}\rangle}.
\label{eq:lam2}
\end{equation}
By solving (\ref{eq:appe11}), we can find that
\begin{equation}
g_1^{1}(\theta)=- 3 i B e^{- 2i\theta}+C_1e^{- i\theta},
\end{equation}
where $C_1$ is an arbitrary constant. Next, substituting the expressions for 
$g_0^{1}$ and $g_1^{1}$ in (\ref{eq:lam2}), we find 
\begin{equation}
\lambda_2^{1} =-\frac{15}{2}B^2 i.
\end{equation}
Since $\lambda_2^{\pm 1}$ are purely imaginary, we need to find the next order term.

Equation (\ref{eq:nth_order}) for $n=3$ is
\begin{equation}
\pth g_3^k + \lambda_0^k g_3^k = 
\ms{L}_1[g_2^k] + \ms{L}_2[g_1^k]+\ms{L}_3[g_0^k] 
- \lambda_1^k g_2^k - \lambda_2^k g_1^k-\lambda_3^k g_0^k.
\end{equation}
Then, the Fredholm alternative solvability condition implies that
\begin{equation}
\lambda_3^k = \frac{\langle\ms{L}_1[g_2^k]+\ms{L}_2[g_2^k]+\ms{L}_3[g_0^k]
-\lambda_1^k g_2^k-\lambda_2^k g_1^k,g_0^k\rangle}
{\langle g_0^k, g_0^k\rangle}, \quad k=0,\,\pm 1,\,\pm 2,\,\ldots.
\label{eq:lam3}
\end{equation}
By solving (\ref{eq:appe24}) for $k=1$, we can find that $ g_2^{1}(\theta)$. 
(Since the expression that we obtain is quite lengthy, we omit it here.) 
Next, substituting the expressions for $g_0^{1}$, $g_1^{1}$ and $g_2^{1}$ in 
(\ref{eq:lam3}) with $k=1$, we find 
\begin{equation}
\lambda_3^{1}=B^2\left[-81 +\frac{57}{2}\Pi'(1)+6 B^2\Pi''(1)+\frac{1}{4}\Pi'''(1)\right],
\end{equation}
which for $H=0$ becomes
\begin{equation}
\lambda_3^{1}=-81 B^2.
\end{equation}
This means that for completely wetting liquids and sufficiently large $\Omega$, 
$\mathrm{Re}(\lambda^k)<0$ for $k=\pm1,\pm 2,\,\ldots,$ i.e., the steady solution 
$h_s(\theta)$ is stable. 
Therefore, the behavior at large rotation numbers for completely wetting and 
partially wetting liquids is completely different -- for completely wetting liquids 
small-amplitude steady solutions are stable, whereas for even slightly non-wetting 
liquids such steady solutions are unstable and, instead, stable large-amplitude 
time-periodic solutions emerge (see the next section for the analysis of such 
time-periodic solutions).

For a general partially wetting liquid, we find 
\begin{equation}
\mathrm{Re}(\lambda^{1})= \varepsilon \Pi'(1) + 
\varepsilon^3B^2 \left[-81 +\frac{57}{2}\Pi'(1)+6 B^2\Pi''(1)+\frac{1}{4}\Pi'''(1)\right]
+O(\varepsilon^4),
\label{eq:lamda_expansion}
\end{equation}
which allows to estimate the value of the angular velocity $\Omega$ at which the 
steady-state solution looses stability and a time-periodic solution emerges. Indeed, 
using (\ref{eq:lamda_expansion}), we find that the condition $\mathrm{Re} 
(\lambda^{1})=0$ implies that the Hopf bifurcation occurs at 
\begin{equation}
\varepsilon\approx\frac{1}{B}
\sqrt{\frac{\Pi'(1)}{\left[81 -\frac{57}{2}\Pi'(1)-6 B^2\Pi''(1)-\frac{1}{4}\Pi'''(1)\right]}},
\label{eq:hopf_cond}
\end{equation}
(which is consistent with the assumption that $\varepsilon\ll 1$ when $\Pi'(1)\ll 1$ 
and/or $B\gg 1$).
Since $\varepsilon=1/\Omega$, condition (\ref{eq:hopf_cond}) is equivalent to
\begin{equation}
\Omega\approx B 
\sqrt{\frac{\left[81 -\frac{57}{2}\Pi'(1)-6 B^2\Pi''(1)-\frac{1}{4}\Pi'''(1)\right]}{\Pi'(1)}}.
\end{equation}
For the disjoining pressure given by (\ref{eq:djpress}), this condition becomes
\begin{equation}
\Omega\approx B\sqrt{\frac{54+19H-2Hb}{-2H(1-2b)}}.
\end{equation}
Considering the small-contact-angle limit, $\beta_0\ll 1$, and using that 
$H= -\frac{5}{3}\beta_0^2h_0^2$, we obtain the following scaling for the Hopf 
bifurcation point:
\begin{equation}
\Omega\approx \frac{9B}{\sqrt{5}h_0}{\beta_0^{-1}}.
\label{eq:hopf_cond2}
\end{equation}
We note, however, that this condition corresponds to the case when the first 
two terms in the asymptotic expansion for $\lambda^{1}$ become of the same 
order, i.e., the asymptotic expansion strictly speaking breaks down. Therefore, 
for a more rigorous analysis, the case $\beta_0= O(\varepsilon)$ should be 
treated separately, although it would lead to the same condition 
(\ref{eq:hopf_cond2}).

\subsubsection{Time-periodic solutions}
\mylab{sec:B2b}

To look for time-periodic solutions, we first impose a co-moving frame by defining 
$\xi=\theta-\tau$. Equation~(\ref{eq:B1}) is rewritten as
\begin{equation}
\partial_{\tau}h = - \varepsilon\,\partial_{\xi} 
\left\{ h^3 \partial_{\xi}\left[\partial_{\xi}^2 h + h - B\cos(\xi+\tau) + \Pi(h)\right] \right\}.
\mylab{eq:C1}
\end{equation}
Assuming that $h(\xi,\tau) = H_0(\xi, \tau) + \varepsilon H_1(\xi, \tau), + \cdots$, 
at leading order we have
\begin{equation}
\partial_{\tau} H_0 = 0.
\end{equation}
That is, $H_0(\xi, \tau) = \phi_0(\xi)$ for some function $\phi_0$. Periodicity and 
the mean film thickness condition require that $\phi_0$ is $2\pi$-periodic and that
$\int^{2\pi}_0 \phi_0(x) dx = 2\pi$, respectively. At next order, we obtain
\begin{equation}
\partial_{\tau} H_1 = -\partial_{\xi}\left\{H_0^3\partial_\xi
\left[\partial^2_{\xi}H_0 + H_0 - B\cos(\xi+\tau)+\Pi(H_0)\right]\right\}.
\mylab{eq:C2}
\end{equation}
Integration with respect to $\tau$ over $[0, 2\pi]$ gives
\begin{equation}
\partial_{\xi}\left\{H_0^3\partial_\xi 
\left[\partial^2_{\xi}H_0 + H_0 +\Pi(H_0)\right]\right\} =0.
\end{equation}
By integrating this equation with respect to $\xi$, we find
\begin{equation}
H_0^3\partial_\xi 
\left[\partial^2_{\xi}H_0 + H_0 +\Pi(H_0)\right] =c_2,
\end{equation}
where $c_2$ is a constant. One more integration, after dividing by $H_0^3$, 
implies that $c_2$ should be zero. Therefore, we obtain
\begin{equation}
\partial_\xi 
\left[\partial^2_{\xi}H_0 + H_0 +\Pi(H_0)\right] =0,
\end{equation}
which after one more integration implies
\begin{equation}
\partial_{\xi}^2 H_0 + H_0 + \Pi(H_0) = c,
\mylab{eq:C3}
\end{equation}
where $c$ is a constant. By integrating the latter equation once again and using 
the fact that the mean thickness is $1$, we actually find that 
$c=1+\frac{1}{2\pi}\int^{2\pi}_0 \Pi(H_0) d\xi$. Note that 
Eq.~(\ref{eq:C3}) is exactly the same as Eq.~(\ref{eq:A3}) for $B=0$. That is, 
in the limit of a large rotation number, the solution at leading order is \bfuwe{a steady 
solution} of zero Bond number co-rotating with the cylinder. 

As discussed in Sec.~\ref{Partial}, for partially wetting liquids there \bfuwe{exist three 
non-trivial large-amplitude solutions at zero Bond number. These are the solutions that 
can co-rotates with the cylinder at large rotation number.} For completely wetting 
liquids at zero Bond number, on the other hand, there exist infinitely many 
solutions given by Eq.~(\ref{eq:A4}) with $B=0$. However, we will show in the 
following that only the trivial solution $H_0\equiv 1$ exists at large rotation 
number. For this, we need to go to higher orders. 

From Eqs.~(\ref{eq:C2}) and (\ref{eq:C3}) we find that 
\begin{equation}
H_1 = B\partial_{\xi}\left[H^3_0 \cos(\xi+\tau)\right] + \phi_1(\xi),
\end{equation}
where $\phi_1$ is of zero mean. To determine $\phi_1$, we consider the equation 
at next order:
\begin{equation}
\partial_{\tau} H_2 = -\partial_{\xi}\left\{H_0^3\partial_\xi
\left[\partial^2_{\xi}H_1 + H_1\right]+3BH^2_0\sin(\xi+\tau)H_1
\right\}.
\mylab{eq:C4}
\end{equation}
Integration with respect to $\tau$ over $[0, 2\pi]$ gives
\begin{equation}
\partial_{\xi}\left\{H_0^3\partial_\xi 
\left[\partial^2_{\xi} \phi_1 + \phi_1\right]\right\}
= \partial_{\xi}\left\{\frac{3}{2}B^2H^5_0\right\}.
\mylab{eq:C5}
\end{equation}
We denote the linear operator on the left-hand side by $\ms{K}$, and find that 
its adjoint operator is
\begin{equation}
\ms{K}^*[g] = 
\partial^3_{\xi}\left\{H_0^3\partial_\xi g\right\}
+ 
\partial_{\xi}\left\{H_0^3\partial_\xi g\right\}.
\end{equation}
It can be easily shown that 
\bfuwe{$\int \frac{\sin\xi}{H_0^3}\,d\xi$ and $\int \frac{\cos\xi}{H_0^3}\,d\xi$} are in the null space 
of $\ms{K}$. Then, by the Fredholm alternative solvability condition, the 
right-hand side of Eq.~(\ref{eq:C5}) should be 
orthogonal to these two functions. This leads to
\begin{equation}
\int^{2\pi}_0 H^2_0\sin\xi \,d\xi = \int^{2\pi}_0 H^2_0\cos\xi \,d\xi = 0.
\end{equation}
For a completely wetting liquid, the general form of $H_0$ is 
$H_0=1+A_0\sin(\xi)+A_1\cos(\xi)$. It can be easily seen that the latter 
integral conditions imply that $A_0=A_1=0$. Therefore, \bfuwe{$H_0\equiv 1$.} 
That is, for completely wetting liquids, there do not exist large-amplitude time-periodic 
solutions. Moreover, we find that for a completely wetting liquid
\begin{equation}
H_1 = -B\sin(\xi+\tau)=-B\sin(\theta),
\end{equation}
i.e., $H_1$ is in fact time-independent and is exactly the same as the order 
$\varepsilon$ term in the expansion of a small-amplitude steady solution, 
see Eq.~(\ref{eq:h1_h2}). In fact, it can be shown that at all the orders
the \bfuwe{$H_k$'s} 
are time-independent, and we recover the small-amplitude steady solution 
discussed in the previous section. We conclude that for the completely 
wetting case, time-periodic solutions do not exist for large values of 
$\Omega$, which is consistent with our numerical observations in 
Sec.~\ref{sec:timetransition}.

\bibliography{RC}

\end{document}